\documentclass[useAMS,usenatbib]{mn2e}
\usepackage{graphicx}
\usepackage{longtable}
\usepackage{multicol}
\pdfoutput=1

\title[Spectrophotometric Distances to Galactic H\,{\sc{ii}} regions]{Spectrophotometric Distances to Galactic H\,{\sc{ii}} regions}
\author[A. P. Mois\'es et al.]
{A. P. Mois\'es$^{1}$\thanks{E-mail: apmoises@astro.iag.usp.br}, 
A. Damineli$^{1}$
E. Figuer\^edo$^{1}$
R. D. Blum$^{2}$
P. S. Conti$^{3}$
\newauthor and C. L. Barbosa$^{4}$ \\
$^{1}$IAG-USP, Rua do Mat\~ao, 1226, 05508-090, S\~ao Paulo, SP Brazil\\ 
$^{2}$NOAO, 950 N Cherry Ave., Tuczon, AZ 85719 USA\\
$^{3}$JILA, University of Colorado, Boulder, CO 80309-0440, USA\\
$^{4}$IP\&D-UNIVAP, Av. Shishima Hifumi, 2911, 12244-000 S. J. dos Campos, SP Brazil}
\date{Accepted 2010 September 14. Received 2010 September 13; in original form 2009 November 14}

\begin{document}

\pagerange{\pageref{firstpage}--\pageref{lastpage}}
\pubyear{2010}

\maketitle

\label{firstpage}


\begin{abstract}
We present a near infrared study of the stellar content of 35 H\,{\sc{ii}} regions in the Galactic plane, 24 of them have been classified as giant H\,{\sc{ii}} regions. We have selected these optically obscured star forming regions from the catalogs of \citet{Russeil03}, \citet{Conti04} and \citet{Bica03a}. In this work, we have used the near infrared domain $J-$, $H-$ and $K_{s}-$ band color images to visually inspect the sample. Also, color-color and color-magnitude diagrams were used to indicate ionizing star candidates, as well as, the presence of young stellar objects such as classical TTauri Stars (CTTS) and massive young stellar objects (MYSOs). We have obtained {\it Spitzer} IRAC images for each region to help further characterize them. {\it Spitzer} and near infrared morphology to place each cluster in an evolutionary phase of development. {\it Spitzer} photometry was also used to classify the MYSOs. Comparison of the main sequence in color-magnitude diagrams to each observed cluster was used to infer whether or not the cluster kinematic distance is consistent with brightnesses of the stellar sources. We find qualitative agreement for a dozen of the regions, but about half the regions have near infrared photometry that suggests they may be closer than the kinematic distance. A significant fraction of these already have spectrophotometric parallaxes which support smaller distances. These discrepancies between kinematic and spectrophotometric distances are not due to the spectrophotometric methodologies, since independent non-kinematic measurements are in agreement with the spectrophotometric results. For instance, trigonometric parallaxes of star-forming regions were collected from the literature and show the same effect of smaller distances when compared to the kinematic results. In our sample of H\,{\sc{ii}} regions, most of the clusters are evident in the near infrared images. Finally, it is possible to distinguish among qualitative evolutionary stages for these objects. 

\end{abstract}


\begin{keywords}
HII regions -- infrared: massive stars.
\end{keywords}


\section{INTRODUCTION}

Massive stars play an important role in the evolution of galaxies. They have strong winds and emit a large fraction of their radiation as UV photons; at the end of their evolution they explode as supernova, recycling enriched material into the interstellar medium. Indeed, during their short lives, they are responsible for a large amount of the momentum and kinetic energy input into the interstellar gas. 

Thus, the formation of these massive stars, as well as their interaction with their natal environment, is one of the most important subjects in astrophysics. These massive stars are formed in molecular clouds, at places where local agglomerations of matter \citep{Blitz91}, made up of dense gas and appreciable concentrations of dust, may undergo quasi-static gravitational contraction \citep{McKee03}, forming the so called pre-stellar core (T $\approx$ $10$ - $30$ K). This phase presents the youngest epoch in which one can identify a high mass star in the process of formation. Due to their low temperatures, they are detectable at $\lambda$ $\approx$ $4-8$ $\mu$m as absorption sources when seen against the bright Galactic plane, and are detectable in the far infrared (FIR) and submilimeter (sub-mm) in emission \citep{Ward98}. This phase does not last more than $10^{6}$ years \citep{Ward94}.

The subsequent phase is the hot core \citep{Kurtz00} phase. Hot cores (HC) have T $>$ $100$ K and are dense ($n_{H_{2}}$ = $10^7$ $cm^{-3}$). A rapidly accreting massive star is located inside the core. The massive star acquires most of its mass in this phase and, due to this accretion, becomes sufficiently hot, and substantial UV photons are produced. The surrounding hydrogen is rapidly ionized forming a hyper compact H\,{\sc{ii}} region (HCH\,{\sc{ii}}), but this hot gas is not typically detectable in the optical. HCH\,{\sc{ii}} regions are defined as being smaller than $0.01$ pc \citep{Kurtz02} and are very faint or undetectable even at $cm$ wavelengths \citep{Churchwell02} due to their small emission measure. A few of these regions were observed in the hydrogen recombination lines H42$\alpha$-H66$\alpha$ with FWHM $\approx$ $50$-$180$ $kms^{-1}$ \citep*{Johnson98}. Little is known about HCH\,{\sc{ii}} regions, but they are treated as an intermediate stage between HCs and the ultra compact H\,{\sc{ii}} regions (UCH\,{\sc{ii}}).

UCH\,{\sc{ii}} regions represent the earliest phase in which the newly born massive star can be detected by its ionizing radiation. This detection is not direct yet. The natal dust cocoon that surrounds the ionized hydrogen radiates in the mid and far infrared. Differently from low mass stars, massive stars start to burn hydrogen well before the accretion phase finishes \citep{Bernasconi96}. Aided by its wind, the intense radiation from the massive star dissipates and evacuates the surrounding gas and dust that gradually expands \citep{Wood89}. As it does so, its optical depth diminishes and the OB-type exciting star becomes revealed, first in the near infrared and as the gas expands it becomes revealed also in the optical domain. The UCH\,{\sc{ii}} region also becomes larger, forming a compact H\,{\sc{ii}} and finally a normal H\,{\sc{ii}} region, when the OB star exhibits a naked photosphere.

A complete knowledge about the formation and evolution of the massive ionizing stars is fundamental to understanding the evolution of the H\,{\sc{ii}} regions as a whole and their influence on Galactic structure. To further this goal, we have made detailed studies of the stellar content of 35 Galactic H\,{\sc{ii}} regions, where 24 of them have been classified as giant H\,{\sc{ii}} regions (GH\,{\sc{ii}}, $N_{LyC}$ $>$ $10^{50}$ photons per second). These GH\,{\sc{ii}} regions are the best tracers of the spiral structure of the Milky Way, and we argue that some distances to these objects, derived by kinematic techniques are systematically overestimated. 

In this work, we have made a study of the stellar content in the near infrared domain of each H\,{\sc{ii}} region from our sample, indicating, when it is possible, the ionizing sources as well as massive young stellar object  (MYSO) candidates. The presence (or not) of a cluster of stars (which is typically defined as a clear overdensity in the stellar counts), young stellar objects, and nebular emission were used to establish an evolutionary stage for each star-forming region. Here, we have adopted an evolutionary scale from the youngest ($stage$ $A$) to the most evolved ($stage$ $D$). 

In many H\,{\sc{ii}} regions \citep{Blum99,Blum00,Blum01,Figueredo05,Figueredo08}, the spectral type of the ionzing sources were identified, as well as massive objects still surrounded by disks or circumstellar envelopes, MYSOs, which typically do not yet reveal their photospheric features due to the emission from hot circumstellar dust. The disks of MYSOs may be identified by modelling the Keplerian velocities from the CO band head emission profile \citep[e.g.,][]{Blum04} seen toward some of these objects. These studies used the Spectral Atlas of Hot, Luminous Stars at 2 $\mu$m \citep*{Hanson96} to determine the spectral type of the massive stars in giant H\,{\sc{ii}} regions. In many cases significant differences from kinematic distances were found using spectroscopic parallaxes. An important kinematic discrepancy was pointed out by \citet{Xu06}. They showed that the distance to the massive star-forming region W3OH, in the Perseus spiral arm, derived from trigonometric parallax is smaller than that obtained from radio kinematic techniques by a factor of $2$. This difference from the kinematic distance to W3 (by a factor of 2) is similar to that found by Navarete et al. (in preparation) using $K$-band spectrophotometric results. Also, classical T-Tauri Stars (CTTS), objects that exhibit long-wavelength dust emission, generally atributed to a circumnstellar disk, are identified, when present, through near infrared color excess. 

The procedure used to analyse the presence of MYSOs, ionizing stars and the evolutionary stage for each H\,{\sc{ii}} region is discussed in the section 4. The individual study of the stellar content of each H\,{\sc{ii}} region is given in the section 5. In the section 6, we present the MYSOs found in our sample and their classifications from near- and mid-infrared photometry. Another difficulty with such regions, is to determine their distances. The most common manner to obtain a distance of a H\,{\sc{ii}} region is using kinematic methodologies. In this work, we compare these kinematic distances with that from non-kinematic techniques. These non-kinematic distances are derived from trigonometric parallax as well as spectrophotometric parallax. In the section 7, we have collected trigonometric distances from the literature, as well as, the spectral type of the ionizing sources of some H\,{\sc{ii}} regions (when they exist in the literature) to derive spectrophotometric distances. Both distances (from trigonometric and spectrophotometric parallax), show discrepancies with kinematic distances. 

\begin{table*}
\caption{H\,{\sc{ii}} regions used in the present work. Names and Galactic coordinates are given in columns 1, 2 and 3, respectively. Kinematic distances are presented in columm 4. $N(LyC)$ are presented in columm 5; most of them are Giant H\,{\sc{ii}} regions, $N(LyC)$ $>$ $10^{50}$ $s^{-1}$. The seeing for each region ($K_s$-band) is given in column 6. In column 7, we show the $evolutionary$ $stage$ derived in this work. In column 8 we indicate if the cluster is closer (CL), further away (FW), agrees (AG) with the adopted kinematic distance or if the data are not conclusive (unknown-UN).}
\begin{tabular}{cccccccc}
\hline
 Name & $l$ & $b$ & $d_{Kin}^{1}$ & $N(LyC)$ & Seeing$^{2}$ & Evolutionary & Distance \\ 
      &     &     & $(kpc)$   & $log(s^{-1})$ & $(")$ $K_s$-band & Stage & Classification \\ \hline
 M8 & $5.97$ & $-1.18$ & $2.8$ & $50.19$ & $0.84^a$ & B & AG \\ 
 W31-South$^{1}$ & $10.2$ & $-0.3$ & $4.5$ & $50.66$ & $0.56^b$ & B-C & CL$^{5}$\\ 
 W31-North$^{1}$ & $10.3$ & $-0.1$ & $15.1$ & $50.90$ & $0.87^a$ & B & CL \\
 W33$^{4}$ & $12.8$ & $-0.2$ & $3.9$ & $50.01$ & $0.69^c$ & A & UN \\
 M17 & $15.0$ & $-0.7$ & $2.4$ & $51.22$ & $0.61^b$ & B & CL \\
 (4) & $22.7$ & $-0.4$ & $10.6$ & $49.73$ & $0.77^b$ & C-D & CL \\ 
 W42 & $25.4$ & $-0.2$ & $11.5$ & $50.93$ & $0.59^b$ & B & CL$^{5}$ \\ 
 W43 & $30.8$ & $-0.2$ & $6.2$ & $50.83$ & $0.78^c$ & C & CL$^{5}$ \\ 
 K47$^{4}$ & $45.5$ & $+0.1$ & $7.0$ & $49.67$ & $0.78^b$ & A & UN \\
 W51 & $48.9$ & $-0.3$ & $5.5$ & $50.03$ & $1.20^a$ & B & CL \\ 
 W51A & $49.5$ & $-0.4$ & $5.5$ & $50.94$ & $0.99^a$ & B & CL$^{5}$ \\ 
 W3$^{3}$ & $133.7$ & $+1.2$ & $4.2$ & $50.25$ & $0.86^6$ & C & CL$^{5}$ \\ 
 RCW42 & $274.0$ & $-1.1$ & $6.4$ & $50.36$ & $0.72^a$ & B & AG \\ 
 RCW46 & $282.0$ & $-1.2$ & $5.9$ & $50.32$ & $1.80^a$ & B & AG \\ 
 NGC3247 & $284.3$ & $-0.3$ & $4.7$ & $50.96$ & $0.81^a$ & B-C & CL \\ 
 NGC3372 & $287.4$ & $-0.6$ & $2.5$ & $50.11$ & $0.69^a$ & C & CL \\ 
 NGC3603 & $291.6$ & $-0.5$ & $7.9$ & $51.50$ & $0.75^a$ & B-C & CL \\
 -- & $298.2$ & $-0.3$ & $10.4$ & $50.87$ & $0.53^b$ & B & UN \\
 -- & $298.9$ & $-0.4$ & $10.4$ & $50.87$ & $0.81^a$ & A & UN \\ 
 (4) & $305.2$ & $+0.0$ & $3.5$ & $49.53$ & $0.75^a$ & A & UN \\ 
 (4) & $305.2$ & $+0.2$ & $3.5$ & $49.64$ & $0.81^a$ & B-C & AG \\ 
 (4) & $308.7$ & $+0.6$ & $4.8$ & $48.59$ & $1.02^a$ & D & AG \\ 
 RCW87$^{4}$ & $320.1$ & $+0.8$ & $2.7$ & $48.85$ & $1.17^a$ & B & UN \\ 
 -- & $320.3$ & $-0.2$ & $12.6$ & $50.11$ & $1.41^a$ & A-B & UN \\ 
 RCW92$^{4}$ & $322.2$ & $+0.6$ & $4.0$ & $49.52$ & $1.20^b$ & A-B & UN \\ 
 RCW97 & $327.3$ & $-0.5$ & $3.0$ & $50.14$ & $0.81^a$ & A & AG \\ 
 -- & $331.5$ & $-0.1$ & $10.8$ & $51.16$ & $0.84^a$ & A & CL \\
 -- & $333.1$ & $-0.4$ & $3.5$ & $50.08$ & $0.59^b$ & B & CL$^{5}$ \\ 
 -- & $333.3$ & $-0.4$ & $3.5$ & $50.04$ & $0.84^a$ & A & UN \\ 
 -- & $333.6$ & $-0.2$ & $3.1$ & $50.43$ & $1.14^a$ & A & UN \\ 
 RCW108$^{4}$ & $336.5$ & $-1.5$ & $1.5$ & $48.29$ & $0.99^a$ & A-B & AG \\ 
 (4) & $336.8$ & $-0.0$ & $10.9$ & $50.48$ & $0.77^b$ & A & AG \\ 
 RCW122$^{4}$ & $348.7$ & $-1.0$ & $2.7$ & $48.41$ & $0.88^b$ & A-B & AG \\ 
 (4) & $351.2$ & $+0.7$ & $1.2$ & $49.67$ & $0.99^a$ & B & FW \\ 
 RCW131$^{4}$ & $353.2$ & $+0.6$ & $1.0$ & $49.32$ & $0.99^a$ & B & FW \\ \hline
\label{table1}
\end{tabular}
\\
\footnotesize{
{\parbox{06.7in}{(1) Kinematic distances adopted here are from \citet{Russeil03}; exceptions are W31-South and W31-North for which we have used distances from \citet{Corbel04};
(2) Except for W33 and W43, which are based on CIRIM data, all the regions above have data from CTIO Blanco-4 meter telescope (ISPI or OSIRIS). Instruments used are denoted a - ISPI, b - OSIRIS and c - CIRIM;
(3) For W3 we have used 2MASS photometric data;
(4) These regions aren't in the sample of Conti \& Crowther (2004), but we have derived the $N(LyC)$ following their work.
(5) These regions have spectrophotometric distances which differ from kinematic results.
}}}
\end{table*}


\section{SELECTION OF THE SAMPLE}

In this work, we present a near infrared study of the stellar content of 35 Galactic H\,{\sc{ii}} regions (Table \ref{table1}). Our sample encompasses that of \citet{Conti04}. In that paper, they conducted a Galactic census of Galactic Giant H\,{\sc{ii}} regions, based on the all-sky 6-cm data set of \citet{Kuchar97}, in connection with the kinematic distances obtained by \citet{Russeil03}. Some H\,{\sc{ii}} regions of our sample were based on the \citet{Dutra03} and \citet{Bica03a} catalogs, who discovered new infrared clusters in the southern and northern hemispheres based on the 2MASS catalog. 

In the following sections, we will present near infrared photometric data of each Galactic H\,{\sc{ii}} region of our sample. We present color-color and color-magnitude diagrams (C-C and C-M diagrams, respectively) of each of them. Also, we present false color images of $JHK_s$ ($J$ is blue, $H$ is green and $K_s$ is red), and false color images of 4.5, 5.8 and 8.0 $\mu$m IRAC-{\it Spitzer} images (blue, green and red, respectively).


\section{NEAR- AND MID-INFRARED OBSERVATIONS}

The $J$ ($\lambda$ $\approx$ 1.28 $\mu$$m$, $\Delta$$\lambda$ $\approx$ 0.3 $\mu$$m$), $H$ ($\lambda$ $\approx$ 1.63 $\mu$$m$, $\Delta$$\lambda$ $\approx$ 0.3 $\mu$$m$) and $K_s$ ($\lambda$ $\approx$ 2.19 $\mu$$m$, $\Delta$$\lambda$ $\approx$ 0.4 $\mu$$m$) band images were obtained on the nights of 1, 4 and 20 May 1999; 19 and 21 May 2000; 10 and 12 July 2001, at the Cerro Tololo Interamerican Observatory (CTIO) 4-m Blanco telescope, using the facility infrared imager OSIRIS (FOV of 93$\times$93 arcsec and pixel scale of 0.161''/pixel). On the nights of 3, 4, 5, 6 and 11 July 2005 and 3, 4, 5, 6 and 7 June 2006 we obtained images using the facility infrared imager ISPI (FOV of 10.25 $\times$ 10.25 arcmin and pixel scale of 0.3''/pixel), also at Blanco 4m telescope. Also, on the nights of 28 and 29 August 1998 we obtained images on the CTIO 4-m telescope using the infrared facility CIRIM (FOV of 102 $\times$ 102 arcsec and pixel scale of 0.40``/pixel). OSIRIS, ISPI and CIRIM are described in instrument manuals, found on the CTIO web pages (http://www.ctio.noao.edu). 

The data were processed with standard methodology for near infrared images: the images were linearized and corrected for bad pixels, flatfielded and sky subtracted using a blank sky image. The fluxes were extracted using the IDL code Starfinder \citep{Diolaiti00}, except for the regions W31-South and G333.1-0.4 for which we have used the published photometry (\citet{Blum00} and \citet{Figueredo05}, respectively). The fluxes were calibrated according to the 2MASS photometric system \citep{Skrutskie06} to produce a self consistent set of magnitudes, including that from published data. Also, for the W3 H\,{\sc{ii}} region we have used 2MASS $JHK_s$ images and photometric data \citep{Skrutskie06}.

For saturated objects, we have adopted 2MASS photometric data. For non-detections in the $J-$band, we have adopted a limiting magnitude based on $90$th percentile of detected objects. This procedure was based on a test where we have added 9000 artificial stars randomly in our images. These stars had magnitudes varying from $J$ = 15.0 to $J$ = 19 mag and in intervals of 0.5 mag. In some situations, we also needed to use this procedure in $H$-band. These objcets, where we have adopted the $90\%$ limiting magnitudes, are represented by arrows insted of points. The inclination of the arrows follows the interstellar reddening.

Also, we present IRAC-{\it Spitzer} color images. IRAC (Infrared Array Camera) is the mid-infrared camera on the {\it Spitzer} Space Telescope, with four arrays observing at 3.6, 4.5, 5.8 and 8.0 $\mu$m \citep{Fazio04}. The images were obtained using the software leopard (http://archive.spitzer.caltech.edu/) and the {\it Spitzer} program ID for each H\,{\sc{ii}} region is indicated in its respective subsection. The mosaic images were constructed from bcd IRAC images using Mopex software and the IRAC photometry of MYSO candidates was realized using the IDL code Starfinder on the mosaic images following the photometric calibration manual (http://ssc.spitzer.caltech.edu/irac/iracinstrumenthandbook/).

\section{\bf ANALYSES}

\subsection{Reddening Vectors}

There are several interstellar extinction laws in the literature, e.g. \citet{Mathis90,Indebetouw05,Nishiyama06}, but the photometric system plays a very important role in such a choice when we are dealing with color-color diagrams. We have chosen the reddening vector from \citet{Straizys08b} which is derived by fitting a large number of Red Clump (RC) stars along the Galactic plane. RC stars are the metal rich equivalents of the horizontal branch stars and are assumed to have absolute luminosities weakly dependent on ages and chemical composition, and thus are used as standard candles. 

The \citet{Stead09} interstellar extinction law ($A_{\lambda} \propto \lambda^{-\alpha}$) has an exponent of $\alpha$ = 2.14, which is one of the largest values derived so far. On the other hand, \citet{Mathis90} has one of the smallest values ($\alpha$ = 1.70).  These laws are extreme situations and should cover the range of reddening expected in the Galaxy. 

With the choice of standard candles (RC stars) and the availability of deeper datasets \citep[$e.g.$, 2MASS, UKIDSS, see:][respectively]{Skrutskie06,Hewett06}, which cover a large portion of the Galactic plane (and therefore higher values of extinction), the value of $\alpha$ has increased from \citet{Mathis90} where $\alpha$ = 1.70, \citet{Indebetouw05} with $\alpha$ = 1.86 and \citet{Nishiyama06} with $\alpha$ = 1.99. Using $\alpha$ = 2.14 and the central wavelengths for the 2MASS system we derive a slope  $E_{J-H}/E_{H-K_s}$ = 2.07, which is in excellent agreement with the results from \citet{Straizys08b}; based on 2MASS data they find a slope of $E_{J-H}/E_{H-K_s}$ = 2.00 (from their slope we rederived the exponent, $\alpha$ = 2.02). \citet{Stead09} also have pointed out that their results are in agreement with the results from \citet{Indebetouw05} if one uses the same effective wavelengths of the 2MASS filters as \citet{Stead09} have used. We have used \citet{Straizys08b} reddening lines in the color-color diagrams, since their results are based on the same photometric system as ours (2MASS system), and their results (slope of $E_{J-H}/E_{H-K_s}$ and $\alpha$) lie between the two extreme interstellar laws illustrated above \citep{Mathis90,Stead09}.

\subsection{C-C and C-M Diagrams}

\begin{figure*}
\begin{minipage}[b]{0.49\linewidth}
\includegraphics[height=8.0cm,width=\linewidth]{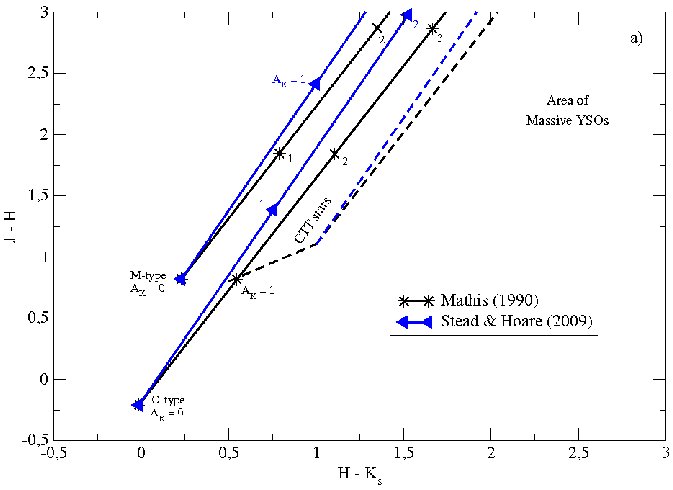}
\end{minipage}
\begin{minipage}[b]{0.49\linewidth}
\includegraphics[height=8.0cm,width=\linewidth]{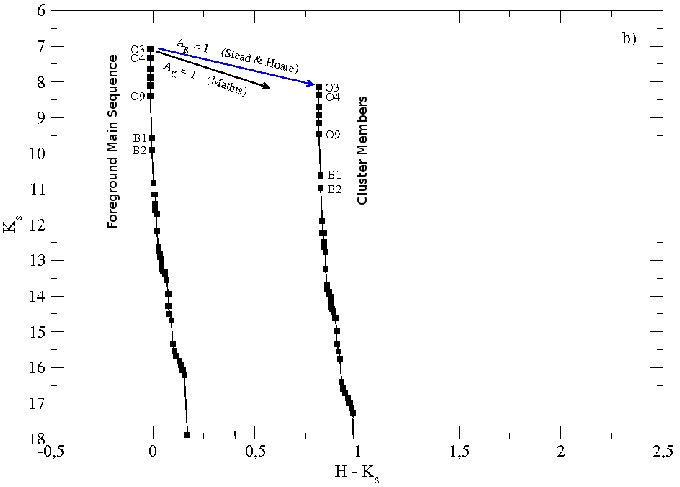}
\end{minipage}
\caption{{ Example of C-C and C-M diagrams. 1a): Color-color diagram (C-C), with reddening lines for a M-type star (upper) and for an O-type star (lower). Here, we compare two different interstellar reddening laws. Black lines have a slope of $E_{J-H}/E_{H-K_s}$ =  1.83 \citep{Mathis90}. Blue lines have a slope of 2.07 \citep{Stead09}. For both cases, we have marked the $A_K$ values. 1b): Color-magnitude diagram (C-M) with the location of the main sequence only affected by distance (left sequence) and the main sequence affected also by the interstellar reddening (line to the right). The two extreme cases of reddening vectors are also displayed.}}
\label{fig:CMD-template}
\end{figure*}

We use the C-C and C-M (Color-Color and Color-Magnitude, respectively) diagrams (Fig.~\ref{fig:CMD-template}, left and right, respectively) to select candidates to ionizing sources in each H\,{\sc{ii}} region, and follow up $K_s$-band spectroscopy with the aim of deriving the spectrophotometric distances. Some regions are deeply embedded in nebulosity and their stellar content is not detectable. In others, the H\,{\sc{ii}} region does not seem to be associated with a clear over-density of point sources. But, for most of our sample of H\,{\sc{ii}} regions, we find embedded stellar clusters and clear candidates for ionizing sources. 

In the C-C diagram (Fig.~\ref{fig:CMD-template}a), we can see several lines in black and blue, where four are continuous and two are dashed. Each color represents an interstellar reddening slope ($E_{J-H}/E_{H-K_s}$) for color-color diagrams. The upper lines (black and blue) are the reddening lines for a M-type star, while the lower lines are for an O-type star. The intrinsic colors for the M-type star were obtained from \citet{Frogel78} and the intrinsec colors for the O-type star are from \citet{Koornneef83}. These intrinsic colors were corrected for the 2MASS photometric system using the relations from \citet{Carpenter01}. The dashed lines show the location of the CTTS sequence and the expected reddening lines. T--Tauri stars are low-mass young stellar objects stars and they can be separated in two subclasses: CTTS and weak-line T--Tauri stars (WTTS). CTTS are thought to evolve first into WTTS, where they become virtually disk-less and no longer shows signs of significant accretion, and eventually into solar-like stars on the main sequence \citep{Robrade07}. 
Objects to the right of the CTTS region can be more embedded (younger) YSOs. The brighter of which are MYSOs.
It should be noted that some MYSOs will have excess emission in all the near infrared bands due to the reprocessing of their intense radiation. Thus, the CC diagram does not show a unique distinction between the effects of extinction and excess emission. Nevertheless, deeply embedded objects are often found to the right in the near infrared C-C diagram due to stronf $K-$band excess.
Finally, there are others types of young objects, such as Herbig Haro stars \citep[e.g.][]{Nishiyama07,Subramaniam06}, but we do not attempt to identify them specifically in this work.

In the C-M diagram (Fig.~\ref{fig:CMD-template}b) we have labeled the position of the brightest candidate members. In this plot we see two lines that represent the main sequence stars at the adopted (kinematic) distance for each HII region. This main sequence line is constructed using $M_V$ for O-type stars from \citet*{Vacca96} and for the others stars we have used $M_V$ from \citet{Wegner07}, for all stars the $M_{V}-M_{K}$ and $M_{H}-M_{K}$ colors used here are from \citet{Koornneef83}. All these magnitudes and colors were corrected to the 2MASS system. The first line, to the left, represents the main sequence for the foreground objects without reddening (only the inverse square law with distance was considered) and a second line, to the right, represents the main sequence for the members of a cluster (interstellar reddening also included). Two reddening vectors, with $A_{K}$ = 1.0 mag, are also plotted. They show the effect on the main sequence stars of interstellar extinction.

In the diagrams, we compare the effect of two different interstellar extinction laws discussed above. In the C-C diagram, the black lines have a slope of $E_{J-H}/E_{H-K_s}$ = 1.83 \citep{Mathis90} while the blue lines have a slope of 2.07 \citep{Stead09}. Neither slope is related to any particular photometric system, they were derived from their respective universal interstellar extinction laws \citep{Mathis90,Stead09}. However, there are various types of filters with different effective wavelengths and the reddening  measured will depend on which filters are used. 

In most H\,{\sc{ii}} regions two groups of objects are displayed in these (C-C and C-M) diagrams. The first one is the foreground objects, and the second group is formed by the members of the clusters themselves (e.g. Fig.~\ref{fig:G305-CMD}) projected along the same line of sight. Foreground objects can be distinguished from cluster objects by using their position in the diagrams together with qualitative information in the images. In some situations, there are different groups of objects belonging to the same H\,{\sc{ii}} region. This may occur when a bright cluster has swept away it's natal material from the central region, and triggered star formation at its periphery (or where stars are independently forming in the nearby molecular cloud) producing both main sequence cluster stars and young stellar objects. Also, differential reddening may scatter the distribution of objects in these diagrams. 

In most of the C-C and C-M diagrams, objects with excess emission in the $K_{s}$-band (evidenced by large $H - K_{s}$ color) are seen. The brightest of these objects are the MYSOs, recently formed massive stars in the earliest phases of their the life. Such objects are stars in which nuclear fusion has most likely started in the core, but they have not yet begun to ionise their surroundings to form an HII region \citep{Urquhart09} and are surrounded by warm dust and/or disks and so often do not show photospheric features. Many of these objects are likely late O-type or early B-type stars, so-called OB stars of second rank whose more massive neighbors have already shed their natal envelopes and disks. 

\subsection{NIR and MIR Images, Evolutionary Stages}

The present sample contains star clusters in different stages of formation. Using our $JHK_{s}$ photometric data, together with {\it Spitzer} images, we can estimate the evolutionary stage of each region by making several assumptions.  An evolutionary stage can be inferred with the adoption of the following criteria: In the first stage ($stage$ $A$), the image is dominated by nebulosity in the $K_{s}$-band (mainly Br$\gamma$ at 2.167 $\micron$), in the {\it Spitzer} image the PAH emission (mainly in the 5.8 and 8.0 $\micron$, green and red, respectively) is dominant and there are few stars; In the second ($stage$ $B$), we can see a cluster of stars with some `naked' star candidates, a large number of CTTS and some MYSOs; nebulosity in both images is not so dominant. In the next stage ($stage$ $C$), we detect only minor nebulosity ($K_{s}$-band) and some emission in the {\it Spitzer} image mainly due to gas warm dust (red), with a well-defined cluster of 'naked' stars and a few CTTS and no MYSOs. In the fourth stage ($stage$ $D$), we just see a cluster of stars and no nebulosity in the region. In each region of our sample, we have used these assumptions to estimate an evolutionary stage, which goes from the younger ($A$) to older ($D$) regions.


\section{\bf NEAR- and MID-IFRARED IMAGES WITH COLOR-COLOR AND COLOR-MAGNITUDE DIAGRAMS}


\subsection{G5.97-1.18 (M8)}

A few stars possibly associated with a stellar cluster were detected at R.A.: 18h03m40s and Dec.: -24d22m40s (J2000). Nebular emission (mainly $Brackett$ $gamma$) is strong (Fig.~\ref{fig:G5-2-color}, which makes it very difficult to study the stellar content. This object is also the well known Hourglass region of the Lagoon Nebula (M8) and it is home to the O7 star Herschel 36 \citep{Thompson06}. Due to this nebular emission, we can see few objects associated with this region. 

The {\it Spitzer} image ({\it Spitzer} program ID: 30570) shows a central bright region, associated with the embedded objects \#01 and \#41, and a nebulosity with main contributions from the 5.8 and 8.0 $\micron$ bands (green and red, respectively and mainly associated to PAHs) dominating all the field. The stars present in the $JHK_{s}$ color image are very embedded in the bright central nebula. Inside that nebula, we can distinguish two sources, likely the ionizing sources of this region, objects \#01, which is also called Hershel 36, and \#41 (Fig.~\ref{fig:G5-2-color}, left side). Unfortunately, object \#01 is saturated in the $JHK_{s}$ image, and we could not obtain good photometry for it. But, its coordinates, centered on the nebula, suggest it is Herschel 36. We find 2MASS $J$, $H$ and $K_{s}$-band photometry ($J$ = 7.94; $H$ = 7.45 and $K_{s}$ = 6.91 mag). However, \citet{Goto06} show, with better resolution data, that this is not a single object. Near our object \#01 we also detect Her 36 SE, which is a red extension 0".25 southeast of Herschel 36. Object \#41 is also in the center of the nebulosity. Its position in the diagrams (C-C and C-M diagram, Fig.~\ref{fig:G5-2-CMD}) show that it is also likely to be an ionizing source of this region.

Object \#01 is indicated in the C-M and C-C diagrams (Fig.~\ref{fig:G5-2-CMD}) based on its 2MASS photometry. We can see, in the C-C diagram, it displays some color excess. Object \#41 is located in the C-C diagram in a region of infrared excess. It is a region between the CTTS region and the YSO area. Other objects, \#26, \#37, \#49, \#63, \#66, \#71, \#78, \#82 and \#108 (with $H - K_{s}$ $\approx$ 1.3 mag) are located well between the reddening lines. Objects, \#151 and \#115 are to the right of the O-type reddening line, in the CTTS region, and in the {\it Spitzer} image they present small MIR emission. Object \#432 is outside the CTTS range, and is probably a YSO. In the {\it Spitzer} image we can see it as a red object. The number of CTTS is notably larger than the number of YSOs. 

This information suggests an evolutionary $stage$ $B$. The size of both images is $\approx$ 3 arcmin on a side. The adopted distance is 2.8 kpc \citep{Russeil03} and its Lyman continuum flux from \citet{Conti04} is $1.55$ $\times$ $10^{50}$ photons per second. Looking at the position of the brightest objects of this region and the reddening vector, the kinematic distance seems to be in agreement with our data.


\subsection{\bf G10.2-0.3 (W31 - South)}

The Galactic GH\,{\sc{ii}} region G10.2-0.3 is part of the W31 complex \citep{Shaver70}. It is one of the largest H\,{\sc{ii}} complexes in the Galaxy with intense star-forming regions. \citet{Wilson74} show that no optical nebulosity appears to be associated with this region, and that this complex is actually formed by three H\,{\sc{ii}} regions: (i) G10.2-0.3 (W31 - South; RA=18:09:21.0, Dec=-20:19:30.9 (J2000)); (ii) G10.3-0.1 (W31 - North; RA=18:08:52.2, Dec=-20:05:53.4 (J2000)) and (iii) G10.6-0.4 (W31B; RA=18:10:28.7, Dec=-19:55:51.7 (J2000)).

Here, we discuss the GH\,{\sc{ii}} region: W31-South (G10.2-0.3), where a stellar cluster was detected. \citet{Kim02} classified this region as a shell morphological type. \citet{Wilson72} derived the kinematic distance as $d_{kin}$ $>$ 4.4 $\pm$ 0.9 kpc (corrected for $R_{\odot}$ $=$ 8.5 kpc), and \citet{Corbel04} derived 4.5 $\pm$ 0.6 kpc ($R_{\odot}$ $=$ 8.5 kpc). Using this distance, \citet{Conti04} derived a Lyman continuum flux of $4.57$ $\times$ $10^{50}$ $s^{-1}$.

\citet*{Blum01} made a detailed study in the near infrared domain of this region. In the Fig.~\ref{fig:W31-color} we reproduce their $JHK_{s}$ image and in the Fig.~\ref{fig:W31-CMD} we reproduce their photometric data transformed to the 2MASS photometric system. In their work, they identified YSOs and four O-type stars. \citet{furness09} recently observed these O-stars with the {\it Spitzer} IRS.

The {\it Spitzer} image ({\it Spitzer} program ID: 3337) at the right shows nebular emission that did not appear in the optical domain \citep{Wilson74}, and it is a little faint in the near infrared image (left hand). The O-type stars \citep[\#2, \#3, \#4 and \#5,][]{Blum01} are faint in this {\it Spitzer} image. This color image points to the presence of nebular material, mainly strong PAH (Polycyclic Aromatic Hidrocarbons) emission (shown in red and not detected at $4.5$ $\mu$m, which is more intense at $\approx$ 6 $\mu$m). The $4.5$ $\mu$m (blue), on the other hand, contains one potentially strong emission feature from H\,{\sc{ii}} regions, the free-free $Br\alpha$ recombination line at 4.05 $\mu$m. Dust is present in all bands, but is strongly present at 8.0 $\mu$m (red).

In the Fig.~\ref{fig:W31-CMD}, we see the C-C and C-M diagrams (color-color and color-magnitude diagrams, respectively). There, we can see two groups of points: (i) one at $H - K_{s}$ $\approx$ 0.4 mag representing the foreground objects and (ii) another at $H - K_{s}$ $\approx$ 1.5 mag representing the cluster members.

In the C-M diagram (Fig.~\ref{fig:W31-CMD}, right), we see some bright objects in the second group: \#2, \#3, \#4 and \#5. Since objects \#1, \#9, \#15, \#26 and \#30 are to the right of the CTTS loci, they are classified as YSOs. In this region we can identify a cluster of stars associated with a nebulosity. The main sequence plotted there is for the kinematic distance, d = 4.5 kpc. Looking at this line and the reddening vector ($A_{K}$ = 1 line), we can see the brightest cluster members seem to be brighter than a reddened O3-type star. This suggests that this kinematic distance is larger than expected, as found by \citet{Blum01} and confirmed by \citet{furness09}.

In the C-C diagram (Fig.~\ref{fig:W31-CMD}, left), we see the O-type stars \citep{Blum01}, cited above, between the lines of natural interstellar reddening. The YSO candidates are at the right of the O-type line, which indicates an excess in $K_{s}$-band emission. This excess comes from circumstellar material that does not allow us to see their photospheric features. Using these diagrams to identify the O-type stars, it was possible to select them for follow up $K$-band spectroscopy. \citet{Blum01} determined a spectrophotometric distance to this region. They showed that objects: \#2, \#3, \#4 and \#5 are, in fact, O-type stars (O5.5 V) by comparing their spectra with that of a $K$-band catalogue of hot stars \citep{Hanson96}. In this way, they found a distance of $d$ $\approx$ 3.4 $\pm$ 0.3 kpc; \citep[see also][]{furness09}. This distance is smaller than the lower limit of the kinematic distances of \citet{Wilson74} and \citet{Corbel04} cited above. Since this region has some objects with naked photospheres, several CTTS and some YSOs, we classify it as $stage$ $B-C$.


\subsection{G10.3-0.1 (W31 - North)}

A stellar cluster was detected at R.A.: 18h08m59s and Dec.: -20d04m50s (J2000). \citet{Wilson74} pointed out that this region is part of the W31 complex. As discussed above, \citet{Corbel04} have shown this region is just in the line of sight of W31, but it is much farther. They have derived a distance of 15.1 kpc. This distance may be too large for the region, as can be seen in the effect of inverse square law in the main sequence location when using this value. The main sequence indicates the ionizing sources of this region (OB-type stars) should be fainter than our data for that distance (Fig.~\ref{fig:G10-CMD},right side). The distance to this cluster might be smaller, which would provide a better fit between the apparent main sequence and the bright stellar content (i.e., the stars clustered near \#82 in Figure A6). The brightest objects may be evolved massive stars in the cluster (\#4, \#6, \#7, \#10 and \#31) given the significant gap between them and the next brightest stars. Using the distance of 15.1 kpc, \citet{Conti04} indicate that this region has $N(LyC)$ of $7.94$ $\times$ $10^{50}$ $s^{-1}$.

In the color images (mainly the $JHK_{s}$ image, Fig.~\ref{fig:G10-color}) we can see a small cluster of stars. In the {\it Spitzer} image ({\it Spitzer} program ID: 146), we see the nebulosity, mainly, in the SE direction with a bright object (\#1032, a massive YSO candidate). The images have size of $\approx$ 2.0 $\times$ 1.5 arcmin. 

The white box shows the area used to obtain the photometry. Looking at the diagrams (C-C and C-M, Fig.~\ref{fig:G10-CMD}), we note that objects numbers \#4, \#6, \#7, \#10 and \#31 seem to be evolved stars and are not very close on the expected main sequence location for the cluster of stars. Object \#48 is a foreground object. Objects \#72, \#82 and \#92, are very close to the line of reddening for O-type stars in the C-C diagram (Fig.~\ref{fig:G10-CMD}). Objects \#81 and \#106 are in the CTTS $loci$, and objects \#96 and \#1032 (which is very bright in the {\it Spitzer} image) are at the right of the region for CTTS, indicating they are YSOs. Object \#1032 was not detected at $J$-band, so we have used a limiting magnitude of $J$ = 17.0 mag for $90\%$ detectability. 

With this information, two YSOs, several CTTS and a well-defined cluster of stars we can put this region in the evolutionary $stage$ $B$. The kinematic distance of 15.1 kpc may be too large as can be seen in the C-M diagram. If objects \#72, \#82 and \#92 belong to the cluster and are on the main sequence, then a smaller distance is indicated. Objects \#10 and \#31 may be background giant stars since they are apparently bright but lie along the reddening line at large extinction. Alternatively, they could be luminous evolved stars in the cluster seen behind a higher column of dust. 


\subsection{G12.8-0.2 (W33)}

No stellar cluster was detected at R.A.: 18h14m14s and Dec.: -17d55m47s (J2000). The distance to this region is 3.9 kpc \citep{Russeil03}. This region belongs to a more extended H\,{\sc{ii}} region, the W33 complex \citep*{Beck98}. Following the work of \citet{Conti04} we derived the Lyman continuum flux, using $T_{e}$ from \citet{Downes80} and $S_{\nu}$ from \citet{Kuchar97}. The derived $N(LyC)$ is $1.02$ $\times$ $10^{50}$ $s^{-1}$, which tells us this is a GH\,{\sc{ii}} region. \citet{Keto89} have observed an expanding shell of gas around the H\,{\sc{ii}} region with $NH_{3}$ and derived a dynamical time scale of $\sim$ $10^{5}$ years for the complex.

In the color images we can see a bipolar structure. But we can not see a well-defined cluster of stars. The image sizes are $\sim$ 1.0 arcmin on a side. In the {\it Spitzer} image ({\it Spitzer} program ID: 146) we can see PAHs emission (green and red) as well as the dark cloud also visible in the $JHK_{s}$ image.

Objects \#3, \#4 and \#6 are in the foreground. Objects \#5, \#17 and \#27 are following the interstellar reddening lines. Objects \#1, \#2, \#7, \#8 and \#10 have excess emission and are in the region of massive YSOs. The tip of the arrows indicate the positions of the objects considering the limiting magnitude of $J$ = $H$ = 17.5 mag (detections above $>$ 90$\%$ completeness) and their inclination is due to the reddening effect. Object \#8 was detected only in $K_{s}$-band, its inclination follows the interstellar law adopted. The objects with vertical lines were detected in $H$ and $K_{s}$-band, so their $H - K_{s}$ is well determined.

The absence of a cluster makes a distance determination impossible. This absence of a cluster, no CTTS and only a few MYSOs (C-C and C-M diagrams, Fig.~\ref{fig:W33-CMD}) indicate this region is at $stage$ $A$. 


\subsection{G15.0-0.7 (M17)}

A stellar cluster was detected at R.A.: 18h20m30s and Dec.: -16d10m48s (J2000). \citet{Conti04} have derived a Lyman continuum flux of $N(LyC)$ = $1.66$ $\times$ $10^{51}$ $s^{-1}$ using a distance of 2.4 kpc (from \citet{Russeil03}.

\citet*{Hanson97} have performed a near infrared study of this region, and have identified nine O-type stars using a $K$-band spectral classification scheme. These stars were used to derive a spectrophotometric distance of 1.3 kpc, smaller than the results obtained from kinematic tehcniques \citep[2.4 kpc from][]{Russeil03}. \citet*{Chini80} have made a multicolor study (UBVRI) in the stellar content of M17, and in subsequent works, \citep*[e.g.][]{Chini98,Chini04} have shown the presence of YSOs in this young region. 

The M17 H\,{\sc{ii}} region is larger than that shown in the color images (Fig.~\ref{fig:M17-color}). Our data are focused on the central cluster, but with better spatial resolution. Both images show a dark cloud to the East, while in the {\it Spitzer} image ({\it Spitzer} program ID: 107) we can see nebular emission in the SW direction. In the \citet{Hanson97} work, object \#189 is resolved into our objects \#100 and \#200; this effect does not explain the shorter distance obtained by them, since their spectrophotometric distance is based on many OB stars. Unfortunately, these objects are saturated, and the 2MASS images do not have sufficient spatial resolution to separate them.

Objects \#1, \#2, \#3, \#4 and \#7 are stars that belong to the M17 star cluster, since they have similar colors (Fig.~\ref{fig:M17-CMD}). Objects \#8 and \#17 follow the interstellar reddening lines, while objects \#10 and \#24 are MYSOs candidates. Object \#24 was not detected in $J$-band, so we have used a limiting magnitude of $J$ = 16.0 mag (see above for definition of limiting mag). Objects \#5, \#6 and \#23 are in the CTTS $loci$. 

The presence of a cluster of stars, several CTTS and some YSOs indicate this is a region in an evolutionary $stage$ $B$. The discrepancy between the main sequence line and the bright stars shows the kinematic distance is larger than observed in our data \citep[consistent with][]{Hanson97}, since the tip of the main sequence line at this distance is fainter than some of our objects.


\subsection{G22.7-0.4}

A stellar cluster was detected at R.A.: 18h34m09s and Dec.: -09d14m26s (J2000). This region appears old, as can be seen from the images which lack strong nebulosity. In larger {\it Spitzer} images ({\it Spitzer} program ID: 146), we can see this region lies on the line of sight to the W41 complex, which has radio coordinates at $\approx$ 7 arcmin NE but has a diameter of $\theta$ = 18.93 arcmin \citep*{Smith78}. This cluster of stars ([MCM2005b]9) is included in the Glimpse catalog of new star clusters \citep{Mercer05}. The star cluster is easily seen and we can see some nebulosity in the {\it Spitzer} image (Fig.~\ref{fig:W41-color}, right side). The size of both images is $\approx$ 2.5 arcmin on a side. If this cluster belongs to the W41 complex, which is not obvious, we can assume its distance is 10.6 kpc \citep{Russeil03} and has a Lyman continuum flux of $N(LyC)$ = $5.37$ $\times$ $10^{49}$ $s^{-1}$. \citet{Leahy08} derived a distance of 4.9 kpc to the region SNR W41 (G23.3-0.3) and overlapping H\,{\sc{ii}} regions. This cluster of stars, which seems to be in projection in the line of sight, was not considered in their work. \citet{Messineo10} found a spectrophotometric distance of 4.2 $\pm$ 0.4 kpc, using two identified O9-B2 supergiants (our objects \#1 and \#6).

Looking at the diagrams (C-C and C-M, Fig.~\ref{fig:W41-CMD}), we can note two groups of objects. The first group of stars with $H - K_{s}$ $\approx$ 0.8 mag and a second with $H - K_{s}$ $\approx$ 1.0 mag. The diagrams show us that objects \#1, \#2, \#3, \#4 and \#6 are on the expected location for stars affected only by the interstellar reddening. These bright objects are saturated in our data and the adopted magnitude values are from 2MASS.

The main sequence line for this distance does not match the observed data well. This cluster of stars is likely closer than what is expected from kinematic results and, in fact, it probably does not belong to the W41 complex. It is more likely a foreground cluster of (evolved) stars. Although the cluster appears evolved, we see some nebulosity in the MIR with a few CTTS. We thus place it in an evolutionary $stage$ $C-D$


\subsection{G25.4-0.2 (W42)}

A few stars associated with an embedded stellar cluster were detected at R.A.: 18h38m15s and Dec.: -06d47m58s (J2000). This region is located in the fourth Galactic quadrant and \citet{Lester85} determined that W42 is at the `near' kinematic distance (3.7 kpc for $R_{\odot}$ = 8 kpc). \citet{Conti04} derived a $N(LyC)$ of $8.51$ $\times$ $10^{50}$ photons per second, using the adopted kinematic distance of 11.5 kpc from \citet{Russeil03}, if using the `near' distance from \citet{Lester85} it would not be giant ($0.9$ $\times$ $10^{50}$ photons). \citet*{Blum00} have made a detailed study of this region. They presented high-spatial resolution $J$, $H$ and $K_{s}$-band images of this massive star cluster. In the Fig.~\ref{fig:W42-color}, to the left, we can see a color image reproduced from \citet{Blum00}. The respective diagrams with the near infrared photometry are presented at the Fig.~\ref{fig:W42-CMD}. The {\it Spitzer} image ({\it Spitzer} program ID: 186) shows nebular emission (mainly in red, $8.0$ $\micron$), which indicates the presence of young embedded stars.

\citet{Blum00} obtained $K_{s}$-band spectra of three of the brightest four stars in the center of the cluster (objects \#1, \#2 and \#3). Object W42 No. \#1, the brightest star, was classified as kO5-O6. With these spectra, \citet{Blum00} derived a ZAMS distance of 2.2 kpc, almost half of the `near' kinematic value \citep{Lester85}. Objects, \#2 and \#3 show no stellar absorption features. This fact, combined with their position in the C-C diagram showing excess emission, lead us to classify them as MYSOs. Object \#57 is very bright in the {\it Spitzer} image but almost invisible in the near infrared image. Since it was not detected at $J$-band, we have used a magnitude limit of $J$ = 16.5 mag, and we suggest it is an excellent YSO candidate.

The presence of nebulosity, CTTS and some MYSOs indicate this region is at $stage$ $B$. The images have $\approx$ 1.5 arcmin on a side. The main sequence line also does not fit our data; as can be seen, the kinematic distance is much larger than that which would be expected to a good fit.


\subsection{G30.8-0.2 (W43)}

A stellar cluster was detected at R.A.: 18h47m37s and Dec.: -01d56m42s (J2000). \citet*{Blum99} have made a detailed study of the stellar content of this region. They have presented $J$, $H$ and $K_{s}$-band data and a new distance to this region, based on $K$-band spectrophotometric parallax.

In the near infrared color image, we can see a small and very crowded, cluster of stars. This cluster is surrounded by a dark lane with some foreground objects in the line of sight. In the {\it Spitzer} image ({\it Spitzer} program ID: 186), we can see the presence of modest  nebular emission (Fig.~\ref{fig:W43-color}). \citet{Blum99} have obtained $K$-band spectra of three of the brightest stars in the center of the cluster. Objects \#1, \#2 and \#3 are in the CTTS region (Fig.~\ref{fig:W43-CMD}), but their spectra show photospheric features.

\citet{Blum99} find that W43 No. \#1, the brightest star, has a spectrum similar to the optically classified WN7 star WR 131 \citep*{Figer97} and W43 Nos. \#2 and \#3 are O-type stars. The distance to this region was determined to be 4.3 kpc, while \citet{Russeil03} derived a distance of 6.2 kpc. \citet{Conti04}, using the kinematic distance, have derived a $N(LyC)$ of $6.76$ $\times$ $10^{50}$ photons per second. Object \#9 is very bright in the {\it Spitzer} image and is very faint in the near infrared image. The limiting magnitude used for objects not detected in $J$-band is $J$ = 17.0 mag. 

The presence of a cluster of stars with most objects in the CTTS $loci$, and a few YSOs, together with a Wolf-Rayet star, indicate this is a slightly evolved star-forming region, and we classify it in the evolutionary $stage$ $C$.


\subsection{G45.5+0.1 (K47)}

G45.5+0.1 (K47) is located at R.A.: 15h09m59s and Dec.: -58d17m26s (J2000) and no stellar cluster was detected. The adopted kinematic distance to this region is 7.0 kpc \citep{Russeil03}. For this distance, we have obtained a $N(LyC)$ of $4.68$ $\times$ $10^{49}$ $s^{-1}$ following the work of \citet{Conti04}. This is a small region with only a few (detected) stars associated with it. In the {\it Spitzer} image ({\it Spitzer} program ID: 187) the nebulosity dominates all the field of view (Fig.~\ref{fig:K47-color}, right side), and we can see the central bright region. The image sizes are $\approx$ 1.5 arcmin on each side.

Looking at the diagrams (C-C and C-M, Fig.~\ref{fig:K47-CMD}), we can see that objects \#1, \#10 and \#12, with $H - K_{s}$ $\approx$ 0.2 mag, are likely in the foreground. Objects \#2, \#5, \#8 and \#9 are on the expected main sequence location for O-type stars (Fig.~\ref{fig:K47-CMD}, left side). Object \#20 was not detected in $J$-band, so we assumed the value derived from the completeness limit $J$ = 16.5 mag. Its real $J - H$ color will follow the arrow. Object \#6 is in the CTTS region, but its photometry may be contaminated with nebular emission since it appears as bright as a de-reddened O3 star (C-M diagram). Several objects with larger excess are also seen. 

The comparative amount of $K_{s}$-band excess objects and the color images indicates that this is a region in the evolutionary $stage$ $A$. The analyses of the kinematic distance, in this case, is inconclusive due to the absence of a cluster.


\subsection{G48.9-0.3 (W51)}

A stellar cluster was detected at R.A.: 19:22:15.0s and Dec.: +14:04:20s (J2000). This is one of the most luminous complexes of massive star-forming regions in the Galaxy \citep{Goldader94} with multiple H\,{\sc{ii}} regions \citep{Wilson70} with at least six regions hosting embedded clusters, all of them optically obscured \citep*{Kumar04}. \citet{Sato10} derived a trigonometric parallax distance of $5.41^{+0.31}_{-0.28}$ kpc to the Main/South part of this compelx, using $H_{2}O$ maser. In the $JHK_{s}$ image, we can see a well-defined cluster of stars as well as nebulosity associated with it. In the {\it Spitzer} image ({\it Spitzer} program ID: 187) the nebular pattern is easily seen. The bright red object in the central part is a contamination of object \#63 by an image artefact. The adopted distance to this region is 5.5 kpc \citep{Russeil03}. Using that distance, \citet{Conti04} derived a $N(LyC)$ of $1.07$ $\times$ $10^{50}$ photons per second (i.e., a GH\,{\sc{ii}} region).

The most prominent stars present a $H - K_{s}$ $\approx$ 0.5 mag, but we can find objects, associated with the nebulosity, with smaller values: \#7, \#15, \#49, \#62, \#63, \#65 and \#202 at $H - K_{s}$ $\approx$ 0.25 mag, as well as objects more reddened $H - K_{s}$ $\approx$ 1.0 mag (\#31, \#40 and \#144). Objects \#91, \#240 and \#540 are in the CTTS loci. In the $JHK_{s}$ color image we can see a cometary shape in the nebulosity (Fig.~\ref{fig:W51-color}, left side), while in the {\it Spitzer} image this shape is more complex (Fig.~\ref{fig:W51-color}, right side). The arrows indicate the location of the (not detected in $J$-band) YSOs: objects \#203, \#238, \#526 and \#1063. Their position in the C-C diagram follow the limiting arrows (based on a magnitude limit of $J$ = $H$ = 17.0 mag). \citet{Kang09} have made a study of embedded young stellar object candidates in the W51 complex and objects \#526 and \#1063 were also indicated as YSOs. Object \#63 is the brightest in the cluster. In the diagrams, its position suggests it may be an unobscured O-type star, while in the {\it Spitzer} image it is, still, very bright. 

The presence of some CTTS and a few MYSOs, nebulosity, and a well-defined cluster suggest this region is at evolutionary $stage$ $B$. In this region, we see that the tip of the main sequence is fainter than objects \#7 and \#15 if they are assumed not to be evolved or foreground stars, which indicates the adopted kinematic distance may be larger than the real distance, i.e. W51 may be  closer than what is given by kinematic results. This is consistent with the low value of reddening to the cluster too. Alternately, if \#7 and \#15 are not part of the cluster, then the kinematic distance may be accurate.


\subsection{G49.5-0.4 (W51A)}

A few stars possibly associated with an embedded stellar cluster were detected at R.A.: 19h23m42s and Dec.: +14d30m33s (J2000). This is one of the most luminous regions in the W51 complex, which is divided into eight smaller radio sources: W51A to W51H. The W51 complex is located at a kinematic distance of 5.5 kpc (near distance), adopting the value derived by \citet{Russeil03}. For this distance, \citet{Conti04} derived for W51A a $N(LyC)$ of $8.71 \times 10^{50}$ photons per second, indeed a GH\,{\sc{ii}} region. Nebulosity (Fig.~\ref{fig:W51A-color}) is well distributed through the field of view of the image with bright and dark components.

In the C-C diagram there are several objects in the CTTS region (objects \#7, \#17, \#24, \#25, \#44, \#50, and \#103). Also we see YSO candidates (objects \#45, \#61, \#62, \#73 and \#98). Objects \#21, \#32, \#57 and \#60 are quite close to the line of reddening for O-type stars. Objects \#52 and \#59 are foreground sources.

\citet{Figueredo08} have made a detailed study of the stellar content of this region. They have used spectrophotometric parallax of 4 O-type stars (\#44, \#50, \#57 and \#61; O5, O6.5, O4 and O7.5, respectively) to derive a distance of 2.0 $\pm$ 0.3 kpc. The arrows in the C-C diagram are based on the magnitude limit of $J$ = 16.5 mag, and indicate objects not detected in $J$-band. The presence of a few cluster members, associated with the color images which shows strong nebular emission (mainly $Br\gamma$) and the absence of stellar objects on the {\it Spitzer} image ({\it Spitzer} program ID: 187), indicates this region is very young.

Also, \citet{Barbosa08} have presented high spatial resolution spectroscopy of two very massive young stars in early formation stages, W51 IRS 2E and IRS 2W, (Fig.~\ref{fig:W51-color} left side). Both of them are embedded sources in the Galactic compact H\,{\sc{ii}} region W51 IRS2. Barbosa et al. find a distance of 5.1 kpc based on their spectrum of the source associated with W51d in IRS2.

Moreover, \citet{Xu09} have derived a trigonometric parallax to IRS2W using 12 GHz methanol masers and obtained a distance of $5.1^{+2.9}_{-1.4}$ kpc, close to the adopted kinematic value and that of \citet{Barbosa08}. \citep{Sato10} using $H_{2}O$ maser parallax, in the W51 Main/South region, found a distance of $5.41^{+0.31}_{-0.28}$ kpc. The discrepancy on the distances of W51A and IRS2 indicates that these two regions may not be physically connected and that the stars observed by \citet{Figueredo08} are closer along the line of site and projected onto W51A. 

Objects IRS2E and IRS2W are associated with star forming regions of evolutionary $stage$ $A$, while the others objects are associated with type $B$. There are some objects brighter than the tip of the main sequence line, but they are likely foreground objects.


\subsection{G133.7+1.2 (W3)}

A stellar cluster was detected at R.A.: 02h26m34.4s and Dec.: +62d00m45s (J2000). It is at the Perseus spiral arm, and its adopted kinematic distance is 4.2 kpc \citep{Russeil03}. Included in the sample of GH\,{\sc{ii}} regions of \citet{Conti04}, it has a Lyman continuum flux of $N(LyC)$ = $1.78 \times 10^{50}$ photons per second. The $JHK_{s}$ results presented here (images and photometry) are from 2MASS.

In the $JHK_{s}$ color image, we see a cluster of stars in the center of the field, and some nebular emission to the NW and to the SE, surrounding the cluster. Each of the $J$, $H$ and $K_{s}$ images is a 18$'$ $\times$ 18$'$ mosaic, constructed from several 2MASS images. In the {\it Spitzer} image ({\it Spitzer} program ID: 127), we see the nebulosity of this field in detail, and it appears that this nebulosity belongs to a unique region, which is not so obvious in the $JHK_{s}$ image.

The brightest star of the central cluster (\#159) was used by \citet{Humphreys78} to derive a spectrophotometric distance (in the optical domain) to this region, and they found a distance of 2.2 kpc. However, the adopted kinematic distance is 4.2 kpc \citep{Russeil03}. Using trigonometric parallax, \citet{Xu06} derived a distance of 1.95 kpc to the star-forming W3OH. W3OH is a region that belongs to the W3 complex, and it is seen in the $JHK_{s}$ and {\it Spitzer} images indicated by the star \#248 (see Fig.~\ref{fig:W3-color}). The distances obtained by parallaxes (spectrophotometric and trigonometric) are in a good agreement with each other and both smaller than the kinematic result by a factor of 2.

Furthermore, Navarete et al. (in preparation), derived a distance of 1.85 $\pm$ 0.92 kpc to W3. They have used $K$-band spectrophotometric parallax (to the O-type stars \#159, \#390 and \#559) and their results are in accordance with that from \citet{Xu06} and \citet{Humphreys78}. 

We can see in the C-C diagram (Fig.~\ref{fig:W3-CMD}, on the left) that objects \#252, \#347, \#390 and \#559, with $H - K_{s}$ $\approx$ 0.5 mag, are near the O-type reddening line, and objects \#444 and \#248 are in the CTTS region.

The tip of the main sequence is fainter than the brighest object of this region (\#159), as is expected since the kinematic distance appears to be too large. This region has a well-defined cluster, nebulosity is seen in both images, especially in the {\it Spitzer} image. There are several CTTS (e.g.: \#444 and \#248) and some massive YSOs. So, we can classify this region as evolutionary $stage$ $B$, while the central cluster is in a evolutionary $stage$ $C$.


\subsection{G274.0-1.1 (RCW42)}

The Galactic GH\,{\sc{ii}} region G274.0-1.1 is also known as RCW42 and a stellar cluster was detected at R.A.: 09h24m30.1s and Dec.: -51d59m07s (J2000). It belongs to a larger structure, a shell called $GSH 277+00+36$ that is at a distance of 6.5 kpc \citet{Macclureetal03}, is 600 pc in diameter, and extends above and below 1 kpc of the Galactic midplane. In the $JHK_{s}$ color image (Fig.~\ref{fig:G274-color}, left side), we see a crowded cluster of stars surrounded by a reddish nebula. We can see, in the NE part of this region, a dark cloud obscuring most of the background stars, possibly precluding the detection of other cluster members. regions like this are very difficult to analyse for cluster membership due to their crowded fields and embedded stars. The {\it Spitzer} image ({\it Spitzer} program ID: 40791) shows a field dominated by weak nebular emission, with the region surrounding the cluster emitting mostly at 8.0 $\micron$ (red).

The distance of G274.0-1.1 used by \citet{Conti04} is 6.4 kpc \citep{Russeil03}. That distance leads to a Lymann continuum luminosity of $N(LyC)$ $\approx$ 2 $\times$ $10^{50}$ photons per second. This implies, at least, a dozen early O-type stars associated within the region. 

Looking at the diagrams (C-C and C-M), the objects numbers \#20, \#24, \#30 and \#32, with $H - K_{s}$ $\approx$ 0.5 mag, are on the expected main sequence location and affected only by the interstellar reddening. Moreover, these stars are close to sites of nebular emission, some of them near the center of the cluster. This suggests these objects may be the ionizing sources of the H\,{\sc{ii}} region. Object \#6 is in the foreground. Also, object \#14, which is a bright star and less reddened than the others cited above, could be an O3-O4V star. On the other hand, objects \#21, \#36, \#40 and \#42 show a color excess, objects \#31 and \#33 are bright in $K_{s}$-band and are at the right of the CTTS region with $H - K_{s}$ $\approx$ 1.8 mag (see the CCD in Fig.~\ref{fig:G274-CMD}). 

The cluster members present a large range of colors, indicating they are very embedded, and we also see a large amount of CTTS as well some massive YSOs, but the nebular component does not emit strongly (it is mostly 'dark') Thus we suggest  an evolutionary $stage$ $B$. In this region, it is not clear if the main sequence line (adopted kinematic distance) is in agreement with the observed data. However, the tip of this main sequence is brighter (as one would expect) than our data, which indicates the adopted kinematic distance may be correct.


\subsection{G282.0-1.2 (RCW46)}

A stellar cluster was detected at R.A.: 10h06m38.1s and Dec.: -57d12m28s (J2000) toward the GH\,{\sc{ii}} region also known as RCW46. In the $JHK_{s}$ color image (Fig.~\ref{fig:G282-color}, left side), we see a small crowded cluster of stars. Nebulosity, in the central part, is visible in both images and in the {\it Spitzer} color image ({\it Spitzer} program ID: 30734)  we can see a surrounding shell nebulosity with heated dust emitting at 8.0 $\micron$ at the central part. In the southeast part of this region, there is a dark cloud obscuring most of the background stars. 

The distance of G282.0-1.2 used by \citet{Conti04} was 5.9 kpc \citep{Russeil03}. That distance leads to a Lymann continuum luminosity of $N(LyC)$ $\approx$ $2.09$ $\times$ $10^{50}$ photons per second (GH\,{\sc{ii}} region).

Looking at the diagrams (C-C and C-M, both at Fig.~\ref{fig:G282-CMD}), the objects \#9, \#10, \#11, \#12, \#13, \#18 and \#20, with $H - K_{s}$ $\approx$ 1.0 mag, are near the expected main sequence location and affected only by the interstellar reddening. These stars are close to the center of the cluster. This suggests they may be the ionizing sources of this H\,{\sc{ii}} region. Also, the object \#5, which is a bright star and less reddened, is in the line of reddening of a M-type star (C-C diagram). Objects \#8, \#15 and \#16 seem to be foreground stars. On the other hand, object \#10 seems to be a highly reddened late O-type star (see the C-C diagram in Fig.~\ref{fig:G282-CMD}), possibly on the far side of the cluster. In the C-M diagram, object \#31 and \#46 appear like very reddened O-type stars, and in the C-C diagram, we can see they present a $K_{s}$-band excess, and are in the YSO area.

The {\it Spitzer} image shows a shell-like structure with some stars well inside the shell. There are 2 YSOs, several CTTS and almost no nebulosity in the near infrared image, mostly visible in the {\it Spitzer} image and a well-defined cluster. We thus put it in the evolutionary $stage$ $B$. The kinematic distance may be correct here since the tip of the main sequence line is brighter than the observed data. Except object \#5, which, if de-reddened, may be brighter than the tip of this main sequence, but it is not clear if this object belongs to the H\,{\sc{ii}} region.


\subsection{G284.3-0.3 (NGC3247)}

A stellar cluster was detected at R.A.: 10h24m17.3s and Dec.: -57d45m36s (J2000). This is a typical H\,{\sc{ii}} region. A well defined cluster with a strong nebulosity surrounding it. The distance to this region is 4.7 kpc \citep{Russeil03}. At this distance, its Lymann continuum luminosity is $N(LyC)$ $\approx$ $9.12$ $\times$ $10^{50}$ photons per second \citep{Conti04}. In the cluster we find some stars as O-type candidates. In the {\it Spitzer} image ({\it Spitzer} program ID: 195, Fig.~\ref{fig:G284-color}, right side), we also see the few brightest stars. Object \#13 is remarkable since it shows a jet above it. Emission in 5.8 $\micron$ (green) dominates the field at NW of the central cluster, while in the SE direction it is emission at 8.0 $\micron$ that dominates.

Actually, this region extends over a larger area, and in this work we depicted only the central part. The larger (not shown) image measures $\approx$ 10 arcmin, but the region reproduced at Fig.~\ref{fig:G284-color} covers only $\approx$ 3.35 arcmin on a side.

Looking at the diagrams (C-C and C-M, Fig.~\ref{fig:G284-CMD}), we can easily seen two sets of objects. The first group of stars with $H - K_{s}$ $\approx$ 0.5 mag and a second group with $H - K_{s}$ $\approx$ 1.0. In the {\it Spitzer} image (Fig.~\ref{fig:G284-color}, right side) we can see the objects with infrared excess and the nebulosity. The main candidates to be ionizing sources (\#3, \#4, \#7, \#8, \#10, \#11, \#12, \#13, \#14, \#20, \#21 and \#25) are in the first group of objects. Unfortunately, objects \#3 and \#4 are saturated in our images, so we have used 2MASS data. Objects numbers \#54, \#60 and \#70 are YSO candidates. 

Since, there are many CTTS, and several MYSOs, and a well-defined cluster with nebulosity surrounding it (as can be seen in the near and mid infrared images), we can associate this region with $stage$ $B-C$. It is clear the O-type stars at the main sequence are fainter (more distant) than our data suggesting the cluster is closer to us than given by kinematic distance. This indicates the kinematic distance may be in error, and the real distance may be closer.


\subsection{G287.4-0.6 (NGC3372)}

G287.4-0.6 is also known as the Carina nebula (NGC3372) and a stellar cluster was detected at R.A.: 10h43m50.1s and Dec.: -59d32m47s (J2000). The distance of G287.4-0.6 used by \citet{Conti04} was 2.5 kpc \citep{Russeil03}. That distance leads to a Lymann continuum luminosity of $N(LyC)$ $\approx$ $1.29$ $\times$ $10^{50}$ photons per second, indicating this is a GH\,{\sc{ii}} region.

In the $JHK_{s}$ color image (Fig.~\ref{fig:G287-color}, left side), we have focused on the crowded cluster of stars. In a larger area (not shown here), we note the well-known strong nebula surrounding this cluster of stars. In the {\it Spitzer} image ({\it Spitzer} program ID: 30734) the 5.8 $\micron$ (green, and not strongly) dominates the field. Looking at the image, we can see that this region is not very distant from the Sun, since the stars can easily be distinguished and they are not strongly reddened by interstellar extinction (C-M and C-C diagrams). 

Looking at the diagrams (C-C and C-M), we note objects \#100, \#1, \#2, \#4, \#5, \#8, \#10, \#14, \#17 and \#45, with $H - K_{s}$ $\approx$ 0.1 mag, are on the expected main sequence and affected only by the interstellar reddening. These stars are close to the center of the region (except objects \#1, \#2 and \#5, which may be not connected to the main cluster), indicating they may be the ionizing sources of this region. Object \#45 is remarkable due the presence of a bow shock above it. 

We can see (color image, C-C and C-M diagrams) some reddened objects, but they appear to be objects behind the dust lane of this H\,{\sc{ii}} region. Object \#138, with $H - K_{s}$ $\approx$ 2.5 mag, is located at a position of an O-type star with a strong infrared excess. In the C-C diagram, this object is located to the right of the position of the CTTS, indicating it is a YSO.

The brightest objects (\#100, \#1, \#2, \#3, \#4, \#5 and \#8) are saturated in our data, so we have used 2MASS values for $J$, $H$ and $K_{s}$ for each of them, as the resolution of 2MASS is poorer than our images, these fluxes may be affected by nearby objects. But, assuming these fluxes are are well determined in the 2MASS catalog, the brightest stars of the cluster are above the tip of the main sequence line. This indicates that the kinematic distance might be too large, and the real distance could be smaller.

We classify this region as evolutionary $stage$ $C$, since we can see the presence of a central cluster cleared of nebular emission. In the central area, we see little nebulosity in both images, there are several CTTS and just one YSO, which is not located near the central cluster.


\subsection{G291.6-0.5 (NGC3603)}

A stellar cluster was detected at R.A.: 11h15m07.1s and Dec.: -61d15m37s (J2000). The adopted distance used by \citet{Conti04} was 7.9 kpc \citep{Russeil03}. That distance leads to a Lymann continuum luminosity of $N(LyC)$ $\approx$ $3.16$ $\times$ $10^{51}$ photons per second. \citet{Melena08} have derived a spectrophotometric distance of 7.6 kpc, but in the optical domain. \citet{Moffat02} have found a large number of X-ray sources, these sources were found with greater frequency toward the cluster center and with no obvious optical counterparts.

In both color images (Fig.~\ref{fig:NGC3603-color}), $JHK_{s}$ and {\it Spitzer} ({\it Spitzer} program ID: 40791), we see nebulosity surrounding a well defined cluster of stars. Except objects \#6, \#19 and \#20, which are labeled in the $JHK_{s}$ image, all the other objects are located inside the black box.

Using isochrone fitting, \citet{Stolte04} derived a distance of 6.0 kpc to this region, which is significantly smaller than that derived by kinematic technique, 7.9 kpc, \citet{Russeil03}. \citet{Melena08} have made a detailed study in the stellar content of NGC3603 and found several O-type stars.

The crowded cluster (\#6, \#10, \#25, \#29, \#34, \#49, \#51, \#55, \#68 and \#71) in the center of the image provides the ionizing sources of this region. All the objects in the central bright part are located at the tip of the C-M diagram, with $H - K_{s}$ $\approx$ 0.3 mag, and are inside the red square in the C-C diagram (Fig.~\ref{fig:NGC3603-CMD}). Object \#6 is saturated in our images, so we have used 2MASS values ($J$ = 8.60; $H$ = 8.03 and $K_{s}$ = 7.72 mag). Object \#20 is in the CTTS region at the C-C diagram, but if de-reddened (C-M diagram) it would be a `naked' O-type star. Object \#19 is in the YSO region. 

The presence of this well-defined cluster, some nebulosity in the neighbourhood of the star cluster (as can be seen in both images, the near infrared and {\it Spitzer}), several CTTS and some MYSOs leads to a region between $stages$ $B$ or $C$. This is also another case of strong disagreement between the main sequence line (kinematic distance) and the observed data: the cluster appears to be closer than what is predicted by kinematic techniques. However, the tip of the observed sequence includes evolved stars, and this makes it brighter than the main sequence.


\subsection{G298.2-0.3}

A few stars associated with an embedded stellar cluster were detected at R.A.: 12h09m58.1s and Dec.: -62d50m00s (J2000). The adopted distance used by \citet{Conti04} was 10.4 kpc \citep{Russeil03}. That distance leads to a Lymann continuum luminosity of $N(LyC)$ $\approx$ $7.41$ $\times$ $10^{50}$ photons per second.

In the $JHK_{s}$ color image (Fig.~\ref{fig:G298-2-color}, left side), we see nebulosity that spans the image. We also see a very embedded cluster and, at the central part of this region, some bright stars. Since the nebulosity dominates this field, we have detected few objects (Fig.~\ref{fig:G298-2-CMD}) that might be associated with the embedded cluster. The {\it Spitzer} image ({\it Spitzer} program ID: 189) shows us that this region is very young with an embedded cluster of stars in the bright area.

The best candidates to be an ionizing source of this region is object \#4. This object seems to be less affected by the nebulosity than the others (C-C diagram, Fig.~\ref{fig:G298-2-CMD}). 

Looking at the diagrams, we note objects number \#5, \#10, \#21, \#58 and \#26 are the brightest objects and have an $H - K_{s}$ around 0.5 mag. Object \#32 is near the O-type reddening line in the C-C diagram. Object \#23 is a bright object located in the CTTS region. 

The presence of an embedded cluster indicates this is a region at $stage$ $B$. As can be seen in the $JHK_{s}$ color image, the cluster of stars is  compact and very crowded; this may indicate the real distance to this region is large, as suggested by the kinematic results. The cluster members (\#21, \#23 and \#32), if de-reddened would be brighter than the tip of the main sequence line, but they can be foreground objects. The analyses of the kinematic distance is inconclusive.


\subsection{G298.9-0.4}

The Galactic GH\,{\sc{ii}} region G298.9-0.4 is located at R.A.: 12h15m25.1s and Dec.: -63d01m13s (J2000) and no stellar cluster was detected. The adopted distance used by \citet{Conti04} was 10.4 kpc \citep{Russeil03}. That distance leads to a Lymann continuum luminosity of $N(LyC)$ $\approx$ $7.41$ $\times$ $10^{50}$ photons per second.

In the $JHK_{s}$ color image (Fig.~\ref{fig:G298-color}, left side), we see small nebulosity, and several field stars. In the line of sight of the small nebulosity, we see only a few stars. The {\it Spitzer} image ({\it Spitzer} program ID: 189) shows a bright and red (8.0 $\micron$) area that coincides with the nebulosity and a few embedded objects in the $JHK_{s}$ image. Object \#21 is very bright in the {\it Spitzer} image.

The identification of ionizing source candidates in this region is not so obvious. In the C-M diagram (Fig.~\ref{fig:G298-CMD}), objects \#12 and \#13 are in the area expected for an ionizing source. However, looking at the C-C diagram, object number \#13 is near the reddening line for M-type stars. Object \#12 is near the line of reddening for O-type stars, without infrared excess, but its connection with the nebulosity is not so direct when we look at the images. On the other hand, object \#21, which is very bright and shows a large $K_{s}$-band excess is a MYSO.

The presence of nebular emission, and the presence of several CTTS with one MYSO indicate this region is at $stage$ $A$. In this case, it is very difficult to point to stars associated with nebulosity. The brightest objects (\#12 and \#13) may be foreground stars. So, it is not obvious if the adopted kinematic distance is correct or not.


\subsection{G305.2+0.0}

The H\,{\sc{ii}} region G305.2+0.0 is located at R.A.: 13h11m15s and Dec.: -62d45m20s (J2000), and a few stars associated with an embedded stellar cluster were detected. It is at a distance of 3.5 kpc \citep{Russeil03}, and has $N(LyC)$ = $3.39$ $\times$ $10^{49}$ $s^{-1}$. This region was divided in two parts (Fig.~\ref{fig:G305-0-color}), and the photometry was carried out on the objects within both white rectangles (Fig.~\ref{fig:G305-0-color}, on the left). Strong nebular emission can be seen in both regions, mainly at longer wavelenths. But, we can not see a cluster of stars. The nebulosity becomes more evident in the {\it Spitzer} image ({\it Spitzer} program ID: 189), where in some places the image is saturated. Also, we can see shells and a cavity in the lower left (SE direction, also evident in the near infrared).

Looking at the diagrams (Fig.~\ref{fig:G305-0-CMD}), we note that objects numbers \#134, \#266 present very similar colors ($H - K_{s}$ $\approx$ 0.8 mag), while \#126, with $H - K_{s}$ $\approx$ 0.2 mag, seems to be a foreground star in the line of sight of the nebulosity. \#86, \#125 and \#252 seem to be background stars with a large amount of nebular material in front of them. In the C-C, object \#873 is in the YSOs region.

In the diagrams (Fig.~\ref{fig:G305-0-CMD}), we can see two sets of objects. The first group of stars, with $H - K_{s}$ $\approx$ 0.30 mag, are likely foreground objects, while the sparse group of stars, with $H - K_{s}$ $\approx$ 0.75 mag, are likely members of the cluster.

Since we don't see a well defined cluster, the analyses of the kinematic distance is inconclusive. The Brackett gamma emission is strong, there is at least one YSO (object \#873) and some stars in the CTTS region; we find this region is at $stage$ $A$.


\subsection{G305.2+0.2}

A stellar cluster was detected at R.A.: 13h11m40s and Dec.: -62d33m09s (J2000). Its distance is 3.5 kpc \citep{Russeil03} and its Lyman continuum flux is $N(LyC)$ = $4.36$ $\times$ $10^{49}$ $s^{-1}$. The presence of a rich cluster of stars is evident. In the Fig.~\ref{fig:G305-color}, we show a 2'x2' portion of the ISPI image, centered on the cluster of stars. A faint nebular emission appears to surround the cluster, and is better seen in the {\it Spitzer} image ({\it Spitzer} program ID: 189). This faint nebulosity indicates this cluster is very evolved.

Looking at the diagrams (C-C and C-M, Fig.~\ref{fig:G305-CMD}), we suggest that objects numbers \#38, \#39, \#40, \#59 and \#65 are on the expected main sequence location for O-type stars and are affected only by interstellar reddening. These stars are close to the nebulosity, as seen in the {\it Spitzer} image (Fig.~\ref{fig:G305-color}, right side), which indicates they may be the ionizing sources of the H\,{\sc{ii}} region. 

In the diagrams (Fig.~\ref{fig:G305-CMD}), we can see two groups of objects. The first group of stars with $H - K_{s}$ $\approx$ 0.3 mag and a second group with $H - K_{s}$ $\approx$ 0.75. The cluster members are in the second group of points, and the first group are likely foreground objects. There are some objects with $K_{s}$-band excess. Objects \#129 and \#174 follow the reddening vectors of main sequence stars, while the object \#134 has an excess emission in $K_{s}$-band and it is near the region of the CTTS. 

The well-defined cluster together with surrounding nebular emission, several CTTS and some YSOs indicate that this is a region between $stages$ $B$ and $C$. The agreement between the kinematic distance and our observed data, also, seems to be valid in this region.


\subsection{G308.7+0.6}

A stellar cluster was detected at R.A.: 13h40m12.1s and Dec.: -61d43m46s (J2000). It is at a distance of 4.8 kpc \citep{Russeil03}, and we derived a $N(LyC)$ = $3.89$ $\times$ $10^{48}$ $s^{-1}$. This seems to be an evolved H\,{\sc{ii}} region, since the cluster members are well distinguished, and we can not see any nebulosity surrounding them in the $JHK_{s}$ color image. In the same way, the {\it Spitzer} image ({\it Spitzer} program ID: 190) does not show strong nebulosity, only a tiny amount of emission at 8.0 $\micron$. 

In the $JHK_{s}$ color image (Fig.~\ref{fig:G308-color}, left side), we see a cluster of stars, which is evident in the diagrams at $H - K_{s}$ around 0.70 mag.

Looking at the diagrams (C-C and C-M), objects \#28, \#31, \#44 and \#51 are on the expected main sequence location and affected only by interstellar reddening. These stars are close to the center of the cluster. These stars are located in a first group of objects with $H - K_{s}$ $\approx$ 0.2 mag. A second group of stars is also seen with $H - K_{s}$ $\approx$ 0.7 mag. In this second group we find objects \#4, \#9 and \#22. Objects \#4 and \#22 are only affected by interstellar reddening, but object \#9 has an excess in $K_{s}$-band as can be seen in the C-C diagram (Fig.~\ref{fig:G308-CMD}, left size).

Also, there are no bright embedded objects. The absence of nebulosity is an indication that the winds from the massive stars have had time enough to sweep away the gas and intracluster dust. Also, we can see some objects redder than the others in the $JHK_{s}$ image. These objects may be surrounded by circumstellar material emiting  strongly in the $K_{s}$-band and in the IRAC channel 1 (e.g. \#4 and \#9). 

The objects \#1 and \#4 are saturated in our images. From 2MASS, their magnitudes are: \#1: $J$ = 10.36; $H$ = 7.84 and $K_{s}$ = 6.44 mag; and \#4: $J$ = 10.47; $H$ = 8.60 and $K_{s}$ = 7.83 mag. These data show object \#1 is, actually, a very bright object with infrared excess and located in the CTTS region ($H - K_{s}$ = $1.40$ mag and $J - H$ = $2.53$ mag).

In the C-M diagram, we see that the main sequence location is in good agreement with our observed data, which indicates that the adopted kinematic distance may be correct. This region has a well-defined cluster, there is no nebulosity in both images and a few CTTS. We thus assign it an evolutionary $stage$ $D$.


\subsection{G320.1+0.8 (RCW87)}

A stellar cluster was detected at R.A.: 15h05m25.1s and Dec.: -57d30m57s (J2000) toward G320.1$+$0.8, also called RCW87. Its distance is 2.7 kpc \citep{Russeil03} and has $N(LyC)$ = $3.55$ $\times$ $10^{48}$ $s^{-1}$. This seems to be a young H\,{\sc{ii}} region, since the cluster members are still surrounded by nebular emission (Fig.~\ref{fig:G320-color}).

In the $JHK_{s}$ color image (Fig.~\ref{fig:G320-color}, left side), we see a crowded cluster of stars and a bubble nebula is easily defined in the {\it Spitzer} image ({\it Spitzer} program ID: 190). In the {\it Spitzer} image, the contribution of the gas is more evident and stars \#3 and \#15 are obvious bright point sources in the IRAC channel 1 (centered at 3.5 $\mu$m).

Looking at the diagrams (C-C and C-M), objects \#9 and \#14 are on the expected main sequence location, but only object \#14 is close to the center of the H\,{\sc{ii}} region, which indicates it may be the ionizing source of the H\,{\sc{ii}} region. 

There are bright objects in the $K_{s}$-band with large infrared colors: \#10 and mainly \#3, \#15 and \#106. In the C-C diagram, we see that object \#10 is close to the line of reddening of a M-type star. But if we consider the normal scatter from the hot star line, it is possible it would be an ionizing source (in the CMD it is in the position for a reddened O-type star). However, object \#10 is not close to the center of the H\,{\sc{ii}} region. Also, in the C-C diagram we see that object \#106 is close to the line of reddening of an O-type star. Object \#3 is in the CTTS region. Object \#15 is very bright in the $K_{s}$-band and presents a large infrared excess emission. Since it was not detected in the $J$-band, we have adopted the magnitude limit $J$ = 17.0 mag. Its real position in the C-C diagram follows the arrow. The presence of a cluster, nebulosity in both images, one YSO and several CTTS indicate this is a region of evolutionary $stage$ $B$.

The agreement between the kinematic distance (main sequence line) and our observed data is not obvious in this case, since the $N_{LyC}$ and the kinematic distance means that there is only a single late O-type star. This is inconsistent with the C-M diagram that show three O-type candidates (\#9, \#10 and \#14). But if we consider that objects \#9 and \#10 do not belong to this region, the kinematic distance may be correct.


\subsection{G320.3-0.2}

The Galactic GH\,{\sc{ii}} region G320.3-0.2 is located at R.A.: 15h09m59s and Dec.: -58d17m26s (J2000), and no stellar cluster is evident. There is no strong nebular emission in the $JHK_{s}$ image. Also, we do not see a well-defined cluster. However, in the {\it Spitzer} image ({\it Spitzer} program ID: 190, Fig.~\ref{fig:G320-2-color}, right side), we find nebulosity mainly at 8.0 $\micron$ (red). \citet{Conti04} derived a $N(LyC)$ of $1.29$ $\times$ $10^{50}$ photons per second using a distance of $12.6$ kpc \citep{Russeil03}.

Looking at the diagrams (C-C and C-M, Fig.~\ref{fig:G320-2-CMD}),  objects numbers \#4 and \#6 seem to be on the expected main sequence location, but they don't seem to be O-type stars (C-C diagram), and are affected only by the interstellar reddening. However, both objects are saturated in our images, so we have used 2MASS photometry for them. These stars are close to the nebulosity, as seen in the {\it Spitzer} image (Fig.~\ref{fig:G320-2-color}, right side). Object \#13 presents a high reddening and is bright in the $K_{s}$-band, but looking at the C-C diagram (Fig.~\ref{fig:G320-2-CMD}, left side) this object does not show color excess. Actually, these objects (\#4, \#6, \#13, and also objects \#55 and \#142) may be in the foreground, projected in the direction of the nebulosity. Object \#90 seems to be associated with this region due the shell-like structure in the {\it Spitzer} image. Objects \#90 and \#203 are in the CTTS region and have aproximately the same $H - K_{s}$ color as object \#13.

The assignment of the evolutionary stage of this region is not easy, since we don't see a cluster, there is little nebulosity in both images and there aren't YSOs. However, the nebular emission, seen in the {\it Spitzer} image, may indicate an incipient cluster in the center of the field. We suggest this region is in a $stage$ $A-B$. The absence of an obvious cluster, together with nebular emission ({\it Spitzer} image), indicates this region may be at a larger distance, as predicted by kinematic results and the brightest objects are in the foreground.


\subsection{G322.2+0.6 (RCW92)}

G322.2+0.6 (RCW92) is located at R.A.: 15h18m39.1s and Dec.: -56d38m49s (J2000), and a few stars associated with an embedded stellar cluster were detected. \citet{Russeil03} derived a distance of 4.0 kpc and using this distance we obtained a Lyman continuum flux of $N(LyC)$ = $3.31$ $\times$ $10^{49}$ $s^{-1}$. In the near infrared color image (Fig.~\ref{fig:G322-color}), we see a cluster with embedded stars. And in the {\it Spitzer} image ({\it Spitzer} program ID: 146) the nebulosity dominates all the field, and shows that the cluster of stars seems to be in a cavity, or that a bubble of gas and dust is surrounding the cluster of stars.

We see in the $JHK_{s}$ image that the majority of the objects in this small field of view are that in the small cluster of embedded stars. Due to this strong nebulosity, outside this central cluster the stars are 'white' foreground or 'red' background objects. Objects \#1 and \#4, with $H - K_{s}$ $\approx$ 0.75 mag, seem to be associated with this region, due the near infrared color image and their location on the diagrams. Objects numbers \#2, \#3 and \#8, with $H - K_{s}$ $\approx$ 0.5 mag, are probably foreground stars projected onto this obscured region. Objects \#6 and \#9 present excess in the $K_{s}$-band and are in the CTTS region. Object \#7, also has a $K_{s}$-band excess, but more accentuated; it seems to be an YSO. 

The cluster, the presence of CTTS and a massive YSO with the strong nebulosity in the {\it Spitzer} image indicate this is a region in the evolutionary $stage$ $A-B$. 


\subsection{G327.3-0.5 (RCW97)}

G327.3-0.5 (RCW97) is located at R.A.: 15h53m02s and Dec.: -54d35m16s (J2000), and no stellar cluster was detected. This region does not seem to be very evolved, since we can not see an obvious cluster of stars. In the $JHK_{s}$ color image (Fig.~\ref{fig:RCW97-color}), we can see nebular emission and some foreground stars. However, this region is likely to be more complex than the near infrared data suggest. It could be a cluster with a very dark lane running through the middle, or two related ones. Indeed, the {\it Spitzer} image ({\it Spitzer} program ID: 191) shows a bubble of gas and dust to the NE and another one smaller near the center of the image, and a third at the position of object \#5. This may indicate the action of massive stars (or a cluster of massive stars) at different positions.

The kinematic  distance is 3.0 kpc \citep{Russeil03}. \citet{Conti04} derived its Lyman continuum flux, $N(LyC)$ = $1.38$ $\times$ $10^{50}$ photons per second (a GH\,{\sc{ii}} region).

Looking at the diagrams (C-C and C-M, Fig.~\ref{fig:RCW97-CMD}), objects \#2 and \#5, with $H - K_{s}$ $\approx$ 0.8 mag, are on the expected main sequence location for O-type stars. Objects \#13 and \#36, with $H - K_{s}$ $<$ 0.5 mag, are bluer than objects \#2 and \#5, suggesting that these objects are in the foreground. Objects \#2 and \#5 are close to the nebulosity, and in C-C diagram they do not show excess in $K_{s}$-band. Object \#2 is near the M-type reddening line, while \#5 and \#19 are near the O-type line. These facts indicate \#5 and \#19 may be the ionizing sources of the H\,{\sc{ii}} region. 

Object \#23 may be a background object, while objects \#16 and \#87 (not detected in $J$-band) present high infrared excess emission and are YSOs, though \#87 is not very bright in the $K_{s}$-band. The adopted $J$-band magnitude for a $90\%$ detectability is $J$ = 16.0 mag.  

The nebulosity in the near infrared and in the {\it Spitzer} images, the absence of a cluster, some CTTS and a few YSOs indicate this is a region in the evolutionary $stage$ $A$. In this region, the adopted kinematic distance and our observed data seems to agree. 


\subsection{G331.5-0.1}

The Galactic GH\,{\sc{ii}} region G331.5-0.1 is located at R.A.: 16h12m07s and Dec.: -51d27m03s (J2000), and a few stars possibly associated with a stellar cluster were detected. \citet{Russeil03} derived a distance of 10.8 kpc. At that distance, this region has a Lyman continuum luminosity of ($NLyC$) $1.45$ $\times$ $10^{51}$ photons $s^{-1}$ \citep{Conti04}.

Objects \#1, \#2 and \#3 are saturated in our data. Using 2MASS photometry we get: \#1: $J$ = 9.14; $H$ = 6.94 and $K_{s}$ = 5.88 mag. \#2: $J$ = 13.10; $H$ = 9.66 and $K_{s}$ = 7.84 mag. \#3: $J$ = 10.04; $H$ = 8.50 and $K_{s}$ = 7.96 mag. None of them has color excess. On the other hand, object number \#100 has a large excess emission in $K_{s}$-band (C-C diagram, Fig.~\ref{fig:G331-5-CMD}, left side), and it is very bright in the {\it Spitzer} image ({\it Spitzer} program ID: 191), suggesting that it is an YSO.

In the near infrared color image, we can see two regions of embedded stars. We can see a small cluster dominated by stars \#3, \#19, \#57, \#52, \#61, \#68, \#69 and \#71. In the bottom region we have objects \#1, \#2, \#99 and \#100. Most of the objects, in both regions, are foreground objects. We can see nebulosity in the near infrared and {\it Spitzer} images, but the presence of a cluster of stars is not so obvious. In the {\it Spitzer} image we can see some cavities, which may indicate the influence of massive stars over the nebular material.

This region has many CTTS, however, objects \#52 and \#100 are MYSOs. These characteristics indicate this is a region in the evolutionary stage $A$. The tip of the main sequence is fainter than the brightest suspected cluster members (e.g. \#57, \#58, and \#61), which indicates that the cluster maybe closer than the adopted kinematic distance.


\subsection{G333.1-0.4}

A stellar cluster was detected at R.A.: 16h21m03s and Dec.: -50d36m19s (J2000) and is a GH\,{\sc{ii}} region. \citet{Russeil03} derived a distance of 3.5 kpc. At that distance, this region has a Lyman continuum luminosity of $1.20$ $\times$ $10^{50}$ photons $s^{-1}$ \citep{Conti04}.

\citet{Figueredo05} have made a detailed study of the stellar content of this region. Object numbers \#1 and \#2, with $H - K_{s}$ $\approx$ 0.5 mag, were identified as O-type stars. Their $K_{s}$-band spectra were used to derive the spectroscopic parallax of this region. \citet{Figueredo05} derived a distance of $2.6$ $\pm$ $0.2$ kpc, which is smaller than that derived by the kinematic techniques.

Fig.~\ref{fig:G333-1-color} \citep[reproduced from][]{Figueredo05} shows a cluster of stars near the bottom of the image, with some objects still very embedded toward the top of the image. The {\it Spitzer} image ({\it Spitzer} program ID: 191) shows nebular emission and some bright objects (YSOs candidates).

Fig.~\ref{fig:G333-1-CMD} shows the photometric results as C-C and C-M diagrams. Objects \#10 and \#11 are in the CTTS region. Also, this region has several YSOs, for example, objects \#4, \#6, \#9, \#13, \#14, \#18, \#416, \#472, \#488 and \#598. Object \#18 (also an YSO) has a large infrared excess emission, it is very bright in the {\it Spitzer} image, and was not detected in $J$-band. Its adopted $J$-band magnitude is $J$ = 18.0 mag. 

The presence of nebulosity in the near infrared and {\it Spitzer} images, the well-defined cluster, some CTTS and a large percentage of YSOs indicate this region is at evolutionary $stage$ $B$. The main sequence line seems to fit our data, but the smaller distance derived from spectrophotometric results is more reliable \citep{Figueredo05}.


\subsection{G333.3-0.4}

G333.3-0.4 is located at R.A.: 16h21m31.7s and Dec.: -50d26m23s (J2000). In this GH\,{\sc{ii}} region we can see two regions of nebular emission. A small cluster is located at the position of the upper nebulosity. The identification of individual objects is very hard due to extinction, nebuar emission, and source crowding. In the {\it Spitzer} image ({\it Spitzer} program ID: 191, Fig.~\ref{fig:G333-color}, right side), we see the nebulosity and almost no stars. The adopted distance to this region is 3.5 kpc \citep{Russeil03}, with a Lyman continuum luminosity of $1.10$ $\times$ $10^{50}$ photons $s^{-1}$ \citep{Conti04}.

Looking at the C-M diagram (Fig.~\ref{fig:G333-CMD}b), object numbers \#1, \#2 and \#4, with $H - K_{s}$ $\approx$ 0.7 mag, are on the expected main sequence location for O-type stars. These stars are close to the nebulosity, as seen also in the {\it Spitzer} image (Fig.~\ref{fig:G333-color}, right side). These facts together, indicate they may be ionizing sources of the H\,{\sc{ii}} region. Also, object \#36 presents a large reddening, with a $H - K_{s}$ $\approx$ 2.2 mag, and is bright in the $K_{s}$-band, but looking at the C-C diagram (Fig.~\ref{fig:G333-CMD}, left side) we see this object does not look like an YSO. 

In this region, we do not see a rich star cluster, and in the diagrams we see some CTTS and YSOs. The region is best described as evolutionary $stage$ $A$. The projected size of the region may suggest it is at a large distance, but this analyses is inconclusive.


\subsection{G333.6-0.2}

G333.6-0.2 is located at R.A.: 16h22m11.9s and Dec.: -50d05m56s (J2000), and no stellar cluster was detected. The distance to this GH\,{\sc{ii}} region is $3.1$ kpc \citep{Russeil03} and using this distance \citet{Conti04} derived a Lyman continuum luminosity of $2.69$ $\times$ $10^{50}$ $s^{-1}$. \citet{Becklin73} noted this is the most luminous H\,{\sc{ii}} region in the wavelenght interval between 1 - 25 $\mu$m (radiating 5 $\times$ $10^{5}$ $L_{\odot}$ in this wavelength range). This region presents a very high obscuration \citep*[$A_{V}$ $\sim$ 21,][]{Rubin83} and it is difficult to associate it with a star cluster. \citet{Hyland80} have suggested that this region is, actually, a `blister' source, since it has a large intrinsic luminosity ($L \sim 3 \times 10^{6} L_{\odot})$ and presents a low degree of ionization. 

The $JHK_{s}$ image exhibits many field stars and bright nebulosity in the central region with some very embedded objects. In the {\it Spitzer} image ({\it Spitzer} program ID: 191), we see mainly the nebular material with some foreground objects (Fig.~\ref{fig:G333-2-color}, right side).

Looking at the images and diagrams (C-C and C-M, Fig.~\ref{fig:G333-2-CMD}), we identify two groups of objects. The first group of stars with $H - K_{s}$ = 0.5 mag are likely members of the cluster, while the second group with $H - K_{s}$ = 2.0 mag (\#2, \#12  and \#25) seems to be composed by background objects. Also, we note that object \#4 has excess emission and together with objects \#10 and \#29 is well inside the bright region. As we can see in the near infrared image, the nebular emission is very strong in this region. The objects \#4 and \#10 seem to be YSOs. It is difficult to identify other objects in this region, due to image crowding and intense nebular emission. Objects \#7, \#16 and \#24 are near the reddening line of M-type stars, and objects \#5, \#6, \#9, \#15, \#19, and \#55 are near the reddening line of O-type stars. 

The strong nebulosity in the near infrared and {\it Spitzer} images, the absence of a cluster, the presence of some CTTS and YSOs indicate this is a region at evolutionary $stage$ $A$. Also, due to the absence of a cluster, it is difficult to analyze the kinematic distance.


\subsection{G336.5-1.5 (RCW108)}

G336.5-1.5 (RCW108) is located at R.A.: 16h39m58.3s and Dec.: -48d52m38s (J2000), where a small stellar cluster was detected. We have adopted a distance of 1.5 kpc \citep{Russeil03} and the Lyman continuum flux for this distance is $N(LyC)$ = $1.95$ $\times$ $10^{48}$ photons per second, the least luminous source in our sample. This is an extremely obscured region, that shows strong nebular emission. Due to this strong nebulosity, it is difficult to identify its stellar content. 

Nevertheless, a careful examination of the images and diagrams (C-C and C-M diagrams, Fig.~\ref{fig:G336-2-CMD}) suggests that objects \#1 and \#18 are near reddening line for O-type stars. Object \#8 is near the M-type reddening line, and objects \#40 and \#71 seem to be background stars. Object \#27 is a foreground star. Objects \#3 and \#4 are close to the bright central region, as seen in the images (Fig.~\ref{fig:G336-2-color}). However, \#3 presents a high excess emission, is bright in the $K_{s}$-band and is in the YSOs region of the C-C diagram (Fig.~\ref{fig:G336-2-CMD}, left side). Objects \#4 and \#15 lie in the CTTS $region$. 

The presence of strong nebulosity in the near infrared and {\it Spitzer} images ({\it Spitzer} program ID: 112), the small cluster in the central area, some YSOs and several CTTS indicate this is a region at $stage$ $A-B$. If objects like source \#1 are cluster members, then the main sequence maybe in the correct position for the kinematic distance.


\subsection{G336.8-0.0}

The Galactic GH\,{\sc{ii}} region G336.8-0.0 is located at: 16h34m37s -47d36m47.8s (J2000) and is at a distance of 10.9 kpc \citep{Russeil03}. No stellar cluster was detected at these coordinates. Following the work of \citet{Conti04}, we derived a $N(LyC)$ = $3.02$ $\times$ $10^{50}$ photons per second. In the near infrared color image the nebulosity is not so obvious, while in the {\it Spitzer} color image ({\it Spitzer} program ID: 191), the nebulosity is stronger including a bright compact source which dominates the field. In neither image do we see a well defined cluster of stars (Fig.~\ref{fig:G336-color}). 

Looking at the diagrams (C-C and C-M diagrams, Fig.~\ref{fig:G336-CMD}), we find that objects \#4, \#5, \#6, \#10 and \#22, with $H - K_{s}$ $\approx$ 1.0 mag, appear to be on the expected main sequence location for O-type stars and are affected only by interstellar reddening. Also, objects \#8, \#23 and \#29 are in the line of sight of the small cluster of embedded stars, and they are on the CTTS $region$.

Object number \#7, with $H - K_{s}$ $\approx$ 0.1 mag, is likely a foreground object. Objects \#55 and \#68 are to the right of the CTTS line of reddening, indicating their $K_{s}$-band excess, but they don't appear in the {\it Spitzer} image. 

All the brightest objects are foreground candidates. There is not a well-defined cluster, there is some nebulosity in the images and the presence of some CTTS and two MYSO candidates indicate this is a region in an evolutionary $stage$ $A$. Due to the absence of a cluster, and a small nebulosity, it seems this region is very far away, and the adopted kinematic distance may be correct.


\subsection{G348.7-1.0 (RCW122)}

G348.7-1.0 (RCW122) is located at: 17h20m05.8s -38d57m37s (J2000), where a few stars possibly associated with an embedded stellar cluster were detected. The distance of this region is 2.7 kpc \citep{Russeil03}. Following the work of \citet{Conti04}, we derived a $N(LyC)$ = $2.57$ $\times$ $10^{48}$ photons per second. 

In this region, the presence of a stellar cluster is not so obvious. The nebular emission is very strong and the ionizing sources seem to be behind the nebulosity. Most of the stars present in the $JHK_{s}$ color image are foreground as can be seen in the C-C diagram (Fig.~\ref{fig:RCW122-CMD}, left side). The background component is very difficult to see here due to the strong obscuration. In the {\it Spitzer} image ({\it Spitzer} program ID: 192, Fig.~\ref{fig:RCW122-color}, right side), we see strong nebulosity associated with the region and the sources with infrared excess. In the dark region of the near infrared color image, we see point sources in the {\it Spitzer} image suggesting the cluster is hidden by a dark cloud along our line of site.

There is a group of stars with $H - K_{s}$ $\approx$ 0.3 mag (\#1, \#4, \#6, \#9, \#10, \#16 and \#22) which are likely in the foreground. Objects \#2, \#7, \#18 and \#21, with $H - K_{s}$ $\approx$ 0.7 mag, are close to the reddening line of O-type stars. Also, objects \#3, \#8, \#11 and \#20 are in the CTTS region. Object \#5 is an YSO. Object \#1 seems to be foreground and is not in the line of sight of the nebulosity. 

The diagrams show this region has several CTTS and YSOs, together with the presence of nebulosity and is lacking a well defined cluster of stars. We thus place this region in an evolutionary $stage$ $A-B$. The main sequence line seems to do a good fit in our photometric data, which may indicate the kinematic distance is correct.


\subsection{G351.2+0.7}

G351.2+0.7 is located at R.A.: 17h20m04.1s and Dec.: -35d56m10s (J2000), and a cluster of stars was detected. This is a very obscured H\,{\sc{ii}} region and is part of the large complex NGC6334 \citep[first identified by][]{Moran90}. This region is at a distance of 1.2 kpc; using this distance we have derived its Lyman continuum flux following \citet{Conti04}, $N(LyC)$ = $4.68$ $\times$ $10^{49}$ photons per second. 

G351.2+0.7 is described as a ring/shell of radio emission \citep{Jackson-Kraemer99} with a ring radius of about 1 arcmin (0.5 pc). No source was detected at the ring's center or as an ionizing source of the ring \citep{Jackson-Kraemer99}. It seems to be very young, and its members are still embedded in their natal clouds. A dark molecular cloud can be easily seen at the SW. This cluster is sweeping the material away and is eroding the surrounding material (Fig.~\ref{fig:G351-color}, lower right). The {\it Spitzer} image ({\it Spitzer} program ID: 20201, Fig.~\ref{fig:G351-color}, right side) shows nebulosity and several embedded sources.

Looking at the diagrams (C-C and C-M diagrams, Fig.~\ref{fig:G351-CMD}), we identify a group of stars with $H$ - $K_{s}$ around 0.30 mag: \#16, \#18, \#20, \#23, \#24, \#27, \#28, \#32 and \#59. However, these objects are sparsely distributed in the image which may indicate these stars are in the foreground. Objects \#18, \#24, \#23 and \#27 are very close to the reddening line for O-type stars. Still close to the O-type reddening line, but with a larger infrared color, we note objects: \#7, \#12  and \#213. Some combination of these could be the ionizing sources of the H\,{\sc{ii}} region. Object \#7 is saturated in our image, so we have used 2MASS photometry. Objects \#54, \#75 and \#82 seem to be background stars. Objects \#11, \#21, \#29, \#40, \#82 and \#83 are more close to the reddening line of M-type stars. Object \#111 presents a high $K_{s}$-band excess (C-C diagram, Fig.~\ref{fig:G351-CMD}, left side) and is likely a YSO with $H - K_{s}$ $\approx$ 1.9 mag. Objects \#17, \#42, \#58 and \#78 are in the CTTS $region$. 

The presence of clusters of stars, nebulosity in both images, some YSOs and several CTTS indicate this is a region in $stage$ $B$. The tip of the main sequence line is brighter than our brightest objects. This suggests that the adopted kinematic distance may be correct, or the region is a little further away than the kinematic distance.


\subsection{G353.2+0.6 (RCW131)}

G353.2+0.6 (RCW131) is located at R.A.: 17h25m37s and Dec.: -34d21m26s (J2000), where a cluster of stars was detected. This region is part of the NGC6357 star forming complex \citep*{Massi97} and its distance is 1.0 kpc \citep{Russeil03}. We have derived the Lyman continuum flux at this distance, following \citet{Conti04} we obtained $N(LyC)$ = $2.09$ $\times$ $10^{49}$ photons per second.

In the $JHK_{s}$ color image (Fig.~\ref{fig:G353-color}, left side), we see a large number of stars. Many of them are foreground objects. In the {\it Spitzer} image ({\it Spitzer} program ID: 20201, Fig.~\ref{fig:G353-color}, right side), we see, more easily, the nebulosity and some structures like pilars and filaments, a result of the action of stellar winds from massive stars.

Objects \#01, \#02, \#4 and \#7 are saturated in our images, so we have used 2MASS photometry. Looking at the diagrams (C-C and C-M diagrams, Fig.~\ref{fig:G353-CMD}), we identify two distinct group of objects. The first group of stars with $H - K_{s}$ = 0.3 mag are likely members of the cluster (\#01, \#02, \#4, \#7, \#37, \#41, \#44, \#54, \#60, \#65, \#66, \#76 and \#157) and are close to the O-type reddening line, except objects \#02 and \#54. These two objects are spatially (Fig.~\ref{fig:G353-color}) very close to each other, and their magnitudes could be affected by this proximity. The second group of stars with $H - K_{s}$ $\approx$ 1.5 mag are likely background objects. Objects \#68, \#74  and \#266 have larger infrared colors, but are close to the reddening lines of O-type stars. These stars are close to the nebulosity (Brackett gamma emission), this indicates they may be the ionizing sources of the H\,{\sc{ii}} region. Objects \#19 and \#33 are closer to the reddening line of M-type stars. Objects \#135 and \#185, among others, show reddened colors and are bright in the $K_{s}$-band, but their positions in the C-C diagram indicate they are in the CTTS region.

The presence of nebulosity in the NIR and MIR, several possibly `naked' O-stars, surrounding nebulosity, CTTS and YSOs indicate this is a region in evolutionary $stage$ $B$. The C-M diagram shows us the tip of the main sequence is brighter than the brightest stars. This may indicate that the kinematic distance may be correct, or a the region is a little further away than the kinematic distance.

\section{MYSOs in H\,{\sc{ii}} regions}

The objects selected as MYSOs candidates throughout this work are shown in Table 2. In some cases the IRAC images are crowded and do not have enough spatial resolution to resolve the cluster members. In other situations, the objects are so bright at longer wavelenghts that they become saturated, which might indicate they are MYSOs. Also, there are situations in which the nebular emission is so intense that it is not possible to determine the magnitude of the object.

In Figure 2, an IRAC-{\it Spitzer} color-color diagram is shown for the objects with measured magnitudes in the four IRAC channels (11 objects). Using the critera of \citet{Allen04} for IRAC color-color diagrams, it was possible to indicate five Class 0/I objects (green closed circles), two Class II objects (blue closed squares), one Class III object (red closed triangle) and two `naked photosphere' objects (black crosses). 

Only 11 objects (from a total of 65) have the 4 IRAC-channel magnitudes determined. For the remaining objects, that present at least the [3.6] and [4.5] $\mu$m measured magnitudes, it is possible to indicate if these objects are YSOs by using the identified YSOs from the Figure 2 in a $K_{s}$-[3.6] X [3.6]-[4.5] $\mu$m diagram (Figure 3). The identified YSOs from Figure 2 are represented by green closed circles and the YSO candidates (in the vicinity of the identified ones) are represented by black open circles. The `naked photosphere' objects are around [3.6]-[4.5] = 0.5 $\mu$m (Figure 3), together with the `naked photosphere' objects identified in Figure 2. In all, we identified 14 YSO candidates, and 21 `naked photosphere' candidates.

Table 2 also shows cases in which the objects were saturated in the IRAC-{\it Spitzer} images. In the Table 2, the `naked photosphere' objects are indicated by {\it NP}.

\begin{footnotesize}
\begin{table*}
\begin{center}
{\parbox{06.7in}{ \textbf{Table 2.} Massive YSOs identified using the color-color and color-magnitude diagrams (C-C and C-M, respectively) and with a counterpart in their respective {\it Spitzer} images. Columns 1 and 2 are the identifications of each region studied in this work. Column 3 gives the candidate number. Columns 4 and 5 are the coordinates (2000). Column 6 gives the MYSO $K_{s}$-band magnitude. Columns 7, 8, 9 and 10 give the IRAC magnitudes. Column 11 gives, when it is possible, the classification of the MYSO. Column 7--10 identification for the non detected objects: (1) Undetected object; (2) Saturated object; (3) Strong nebulosity obscurating the object and (4) Crowded cluster.}}
\begin{tabular}{ccccccccccc}
\hline
 region & Name  & Obj.& $R.A. (J2000)$ & $Dec. (J2000)$ & $K_s (mag)$&3.6$\mu$m&4.5$\mu$m&5.8$\mu$m&8.0$\mu$m&Class. \\ \hline
 G5.97-1.18    &M8       &\#01  &18:03:40.32&-24:22:42.70& $6.91$&   6.90 &  5.85  &  4.10  &  (2)   & YSO?     \\
 G5.97-1.18    &M8       &\#41  &18:03:40.37&-24:22:39.42& $9.18$&   7.10 &  6.14  &  4.66  &  (2)   & YSO?     \\ 
 G5.97-1.18    &M8       &\#432 &18:03:38.63&-24:22:24.20&$11.52$&   9.74 &  8.51  &  7.96  & 4.74   & YSO      \\
 G10.2-0.3$^a$ &W31-South&\#01  &18:09:27.64&-20:19:13.02& $9.45$&   7.42 &  6.17  &  (1)   &  (1)   & YSO?     \\ 
 G10.2-0.3$^a$ &W31-South&\#09  &18:09:26.98&-20:19:08.53&$10.67$&   7.97 &  6.65  &  (1)   &  (1)   & YSO?     \\
 G10.2-0.3$^a$ &W31-South&\#15  &18:09:27.28&-20:19:35.74&$11.02$&   9.70 &  (1)   &  (1)   &  (1)   & ?        \\
 G10.2-0.3$^a$ &W31-South&\#26  &18:09:26.28&-20:19:23.40&$11.49$&   8.10 &  6.70  &  (1)   &  (1)   & YSO?     \\
 G10.2-0.3$^a$ &W31-South&\#30  &18:09:25.80&-20:19:17.79&$11.83$&   8.83 &  7.50  &  (1)   &  (1)   & YSO?     \\
 G10.3-0.1     &W31-North&\#96  &18:08:58.20&-20:05:14.00&$11.49$&   9.85 &  9.74  &  9.32  &  (1)   & {\it NP}?\\
 G12.8-0.2     &W33      &\#01  &18:14:13.46&-17:55:38.95&$12.50$&   8.23 &  6.10  &  3.99  &  (1)   & YSO?     \\
 G12.8-0.2     &W33      &\#02  &18:14:12.53&-17:55:43.09&$12.63$&   (1)  &  (1)   &  (1)   &  (1)   & ?        \\
 G12.8-0.2     &W33      &\#07  &18:14:13.01&-17:55:27.94&$13.74$&   (1)  &  (1)   &  (1)   &  (1)   & ?        \\ 
 G12.8-0.2     &W33      &\#08  &18:14:14.58&-17:55:50.79&$13.95$&   (1)  &  (1)   &  (1)   &  (1)   & ?        \\
 G12.8-0.2     &W33      &\#10  &18:14:14.32&-17:55:56.71&$14.02$&   (1)  &  (1)   &  (1)   &  (1)   & ?        \\
 G15.0-0.7     &M17      &\#10  &18:20:30.55&-16:11:04.71&$10.08$&   8.05 &  6.88  &   6.28 & 5.81   & YSO      \\
 G15.0-0.7     &M17      &\#24  &18:20:30.73&-16:10:53.42&$10.97$&   (1)  &  (1)   &  (1)   &  (1)   & ?        \\
 G25.4-0.2$^b$ &W42      &\#03  &18:38:15.30&-06:47:51.88&$10.42$&   8.72 &  7.81  &  (2)   &  (3)   & YSO?     \\ 
 G25.4-0.2$^b$ &W42      &\#57  &18:38:14.57&-06:48:02.34&$12.96$&   7.60 &  6.78  &  (2)   &  (3)   & YSO?     \\ 
 G30.8-0.2$^c$ &W43      &\#09  &18:47:37.12&-01:56:42.54&$11.63$&   8.83 &  6.34  &  4.57  & 4.03   & YSO      \\
 G30.8-0.2$^c$ &W43      &\#10  &18:47:38.51&-01:56:43.17&$11.68$&   8.02 &  6.57  &  5.21  &  (1)   & YSO?     \\ 
 G45.5+0.1     &K47      &\#20  &19:14:22.09&+11:08:24.35&$13.28$&  10.12 &  9.14  &  (3)   &  (3)   & YSO?     \\
 G48.9-0.3     &W51      &\#203 &19:22:15.26&+14:04:27.88&$11.43$&  11.64 & 11.26  & 11.36  &  (1)   & {\it NP}?\\
 G48.9-0.3     &W51      &\#238 &19:22:11.64&+14:02:16.63&$11.61$&   9.39 &  9.10  &  8.39  &  (1)   & YSO?     \\  
 G48.9-0.3     &W51      &\#526 &19:22:07.82&+14:03:13.37&$12.66$&   8.63 &  7.19  &  6.05  & 4.86   & YSO      \\ 
 G48.9-0.3     &W51      &\#1063&19:22:19.02&+14:05:07.18&$13.59$&   9.58 &  8.26  &  7.14  & 6.26   & YSO      \\ 
 G49.5-0.4$^d$ &W51A     &\#45  &19:23:42.67&+14:30:27.56&$12.46$&   (3)  &  (3)   &  (3)   &  (3)   & ?        \\ 
 G49.5-0.4$^d$ &W51A     &\#61  &19:23:47.21&+14:29:43.69&$12.49$&   (3)  &  (3)   &  (3)   &  (3)   & ?        \\
 G49.5-0.4$^d$ &W51A     &\#62  &19:23:40.42&+14:29:32.22&$12.26$&   (3)  &  (3)   &  (3)   &  (3)   & ?        \\
 G49.5-0.4$^d$ &W51A     &\#73  &19:23:52.05&+14:28:50.30&$11.55$&   (3)  &  (3)   &  (3)   &  (3)   & ?        \\
 G49.5-0.4$^d$ &W51A     &\#98  &19:23:42.80&+14:30:29.70&$12.98$&   (3)  &  (3)   &  (3)   &  (3)   & ?        \\
 G274.0-1.1    &RCW42    &\#21  &09:24:25.76&-51:59:25.08&$11.01$&   (1)  &  (1)   &  (3)   &  (3)   & ?        \\ 
 G274.0-1.1    &RCW42    &\#31  &09:24:25.97&-51:59:23.59&$11.60$&   7.79 &  7.91  &  (3)   &  (3)   & {\it NP}?\\
 G274.0-1.1    &RCW42    &\#33  &09:24:26.39&-51:59:19.96&$11.78$&   8.71 &  8.66  &  (3)   &  (3)   & {\it NP}?\\
 G282.0-1.2    &RCW46    &\#31  &10:06:38.99&-57:11:58.35&$12.01$&   8.21 &  8.26  &  (1)   &  (3)   & {\it NP}?\\ 
 G282.0-1.2    &RCW46    &\#46  &10:06:36.89&-57:12:31.69&$12.56$&   9.55 &  8.88  &  8.60  &  (3)   & {\it NP}?\\
 G284.3-0.3    &NGC3247  &\#54  &10:24:01.13&-57:45:35.46&$10.25$&  10.55 &  9.87  &  9.57  &  (4)   & {\it NP}?\\ 
 G284.3-0.3    &NGC3247  &\#60  &10:23:55.74&-57:45:08.48&$10.39$&   (1)  &  (1)   &  (1)   &  (1)   & ?        \\
 G287.4-0.6    &NGC3372  &\#138 &10:43:35.12&-59:31:48.57&$11.17$&   9.28 &  8.76  &  8.52  & 8.26   & YSO      \\ 
 G291.6-0.5    &NGC3603  &\#19  &11:15:11.38&-61:16:44.91& $8.91$&   8.56 &  7.62  &  7.00  & 6.05   & YSO      \\
 G298.9-0.4    &--       &\#21  &12:15:20.01&-63:01:10.49&$10.70$&   6.85 &  6.24  &  (2)   &  (2)   & YSO?\\ 
 G305.2+0.0    &--       &\#873 &13:11:16.22&-62:46:21.57&$13.01$&   7.17 &  6.28  &  5.55  & 4.36   & YSO      \\ 
 G305.2+0.2    &--       &\#134 &13:11:31.72&-62:32:30.14&$11.25$&  10.56 & 10.05  &  (1)   &  (1)   & {\it NP}?\\
 G320.1+0.8    &RCW87    &\#15  &15:05:17.21&-57:30:02.31& $8.57$&   9.30 &  8.60  &  8.41  &  (2)   & {\it NP}?\\ 
 G322.2+0.6    &RCW92    &\#07  &15:18:38.78&-56:38:49.70&$11.17$&   (1)  &  (1)   &  (1)   &  (3)   & ?        \\ 
 G327.3-0.5    &RCW97    &\#16  &15:53:09.68&-54:34:31.13& $9.78$&   8.64 &  8.57  &  6.60  &  (1)   & {\it NP}?\\ 
 G327.3-0.5    &RCW97    &\#87  &15:53:03.15&-54:35:24.43&$11.82$&   7.20 &  6.71  &  6.19  & 6.46   & YSO      \\ 
 G331.5-0.1    &--       &\#100 &16:12:08.98&-51:28:02.93&$10.50$&   9.03 &  8.02  &  6.59  & 5.52   & YSO      \\
 G331.5-0.1    &--       &\#2758&16:12:10.01&-51:28:37.84&$14.52$&   6.79 &  5.85  &  7.41  &  (2)   & YSO?     \\ 
 G333.1-0.4$^e$&--       &\#04  &16:21:04.56&-50:35:42.00&$10.93$&   7.62 &  5.93  &  3.76  &  (3)   & YSO?     \\
 G333.1-0.4$^e$&--       &\#09  &16:21:02.47&-50:35:38.72&$12.12$&   9.15 &  8.68  &  8.05  &  (3)   & {\it NP}?\\
 G333.1-0.4$^e$&--       &\#06  &16:21:00.43&-50:35:08.37&$11.20$&   7.61 &  7.14  &  5.56  &  (3)   & {\it NP}?\\
 G333.1-0.4$^e$&--       &\#13  &16:20:59.70&-50:35:14.34&$12.28$&  10.50 & 10.03  &  (1)   &  (3)   & {\it NP}?\\ 
 G333.1-0.4$^e$&--       &\#14  &16:21:00.50&-50:35:09.37&$12.19$&   9.36 &  8.89  &  6.95  &  (3)   & {\it NP}?\\ 
 G333.1-0.4$^e$&--       &\#18  &16:21:02.62&-50:35:54.85&$12.46$&   (2)  &  (2)   &  (2)   &  (3)   & YSO?     \\ 
 G333.1-0.4$^e$&--       &\#416 &16:21:02.07&-50:35:16.02&$15.72$&   6.91 &  6.44  &  (1)   &  (3)   & {\it NP}?\\ 
 G333.1-0.4$^e$&--       &\#472 &16:21:04.02&-50:35:07.41&$15.05$&   (1)  &  (1)   &  (1)   &  (3)   & ?        \\ 
 G333.1-0.4$^e$&--       &\#488 &16:21:00.71&-50:35:05.39&$14.86$&   (1)  &  (1)   &  (1)   &  (3)   & ?        \\
 G333.1-0.4$^e$&--       &\#598 &16:21:06.68&-50:35:41.39&$10.50$&   (1)  &  (1)   &  (1)   &  (3)   & ?        \\ \hline
\end{tabular}
\label{table2}
\end{center}
\end{table*}
\end{footnotesize}

\begin{footnotesize}
\begin{table*}
\begin{center}
{\textbf{Table 2} - Continued}
\begin{tabular}{ccccccccccc}
\hline
 region & Name  & Obj.& $R.A. (J2000)$ & $Dec. (J2000)$ & $K_s (mag)$&3.6$\mu$m&4.5$\mu$m&5.8$\mu$m&8.0$\mu$m&Class. \\ \hline
 G333.6-0.2    &--       &\#04  &16:22:09.60&-50:05:59.13& $7.92$&   (2)  &  (2)   &  (2)   &  (2)   & YSO?     \\ 
 G333.6-0.2    &--       &\#10  &16:22:09.37&-50:06:00.59& $8.46$&   (2)  &  (2)   &  (2)   &  (2)   & YSO?     \\ 
 G336.5-1.5    &RCW108   &\#03  &16:40:01.06&-48:51:51.83& $8.93$&   6.93 &  6.00  &  4.79  & 5.10   & YSO      \\ 
 G336.8-0.0    &--       &\#55  &16:34:51.22&-47:33:16.11&$13.65$&  11.06 & 10.45  &  (1)   &  (1)   & {\it NP}?\\
 G336.8-0.0    &--       &\#68  &16:34:47.51&-47:32:10.16&$13.82$&  11.93 & 11.59  &  (1)   &  (1)   & {\it NP}?\\
 G348.7-1.0    &RCW122   &\#05  &17:20:06.65&-38:57:30.38&$10.83$&  10.82 &  8.61  &  (1)   &  (1)   & YSO?     \\
 G351.2+0.7    &--       &\#111 &17:19:57.87&-35:57:50.84&$10.42$&   7.79 &  7.12  &  6.07  &  (3)   & YSO?     \\ \hline
\end{tabular}
\label{table2b}
\end{center}
\begin{footnotesize}
{\parbox{06.7in}{References: (a) Blum et al. (2001); (b) Blum et al. (2000); (c) Blum et al. (1999); (d) Figuer\^edo et al. (2008); (e) Figuer\^edo et al. (2005).\\
Non detection objects: \\
(1) Undetected object; (2) Saturated object; (3) Strong nebulosity obscurating the object; (4) Crowded cluster.}}
\end{footnotesize}
\end{table*}
\end{footnotesize}

\begin{figure}
\includegraphics[height=08cm,width=08.5cm]{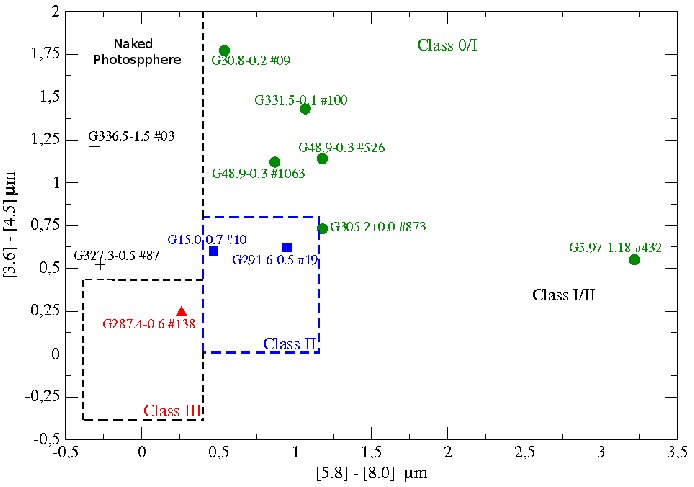}
\caption{{ Confirmed YSOs based on the four channel color-color diagram for the IRAC magnitudes according to \citet{Allen04} color classification.}}
\end{figure}

\begin{figure}
\includegraphics[height=08cm,width=08.5cm]{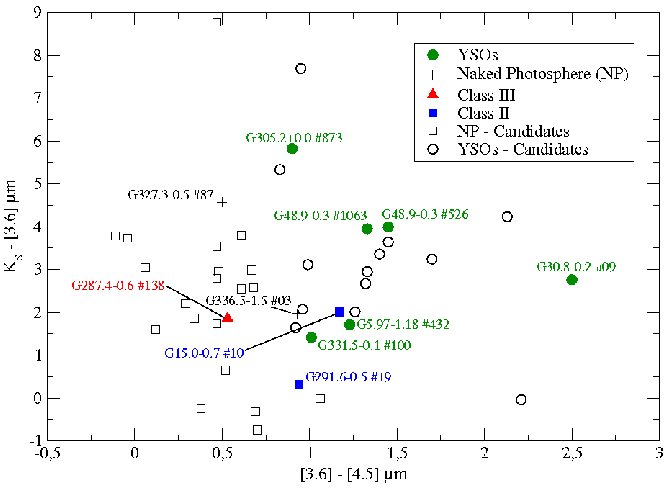}
\caption{{ Color-color plot combining $K_{s}$-band and 2 IRAC-{\it Spitzer} channels for objects without measured magnitude in the four IRAC-{\it Spitzer} channels, mainly at [8.0] $\mu$m. It is possible to identify `naked photosphere' stars (open black squares) and to suggest additional YSO candidates (open black circles).}}
\end{figure}

\section{Spectrophotometric and Trigonometric Parallaxes Distances}

In previous sections, we have used kinematic distances to the H\,{\sc{ii}} regions to check them with the photometry in C-M diagrams. Now, we compare these kinematic distances with that derived from trigonometric and from $K$-band spectrophotometric parallaxes. These non-kinematic results are useful to check the kinematic distances used to map the spiral pattern of the Milky Way. This sample of non-kinematic distances of H\,{\sc{ii}} regions encompass some of our objects, as well as that from other star-forming regions, and it is useful since it shows that there is an overall discrepancy between these two methodologies, kinematic and non-kinematic.

Using VLBI several authors \citep[see:][]{Reid09a} have derived trigonometric parallax distances to star forming regions. These results can be compared with the kinematic results in Table 3. In columns 1 and 2, we show the galactic coordinates. In column 3 we list the names of the regions. In column 4 the kinematic distances are given. In column 5 the published distances from trigonometric parallaxes are shown. In column 6, we give references.

\begin{footnotesize}
\begin{table*}
\begin{center}
{\parbox{06.7in}{\textbf{Table 3.} regions with distances derived from trigonometric parallax. In columns 1 and 2 are the galactic coordinates, in the column 3 names of the regions, and in column 4 the kinematic distances. In the column 5, we show the trigonometric parallax distances. References are given in column 6.}}

\begin{tabular}{|c|c|c|c|c|c|}
\hline
   $l$   & $b$      & \it{region}&$d_{Kin.}^{15}$&$d_{\pi}$         &\it{Ref.} \\ 
         &          &            & $kpc$  &  $kpc$                  &       \\  \hline
   0.67  &  $-0.03$ & Sgr B2	 & $8.50$ & $7.90^{+0.80}_{-0.70}$  & 14    \\ 
  23.01  &  $-0.41$ & G23.0-0.4  & $4.97$ & $4.59^{+0.38}_{-0.33}$  &  5    \\ 
  23.44  &  $-0.18$ & G23.4-0.2  & $5.60$ & $5.88^{+1.37}_{-0.93}$  &  5    \\ 
  23.66  &  $-0.13$ & G23.6-0.1  & $5.04$ & $3.19^{+0.46}_{-0.35}$  &  6    \\ 
  35.20  &  $-0.74$ & G35.2-0.7  & $1.98$ & $2.19^{+0.24}_{-0.20}$  &  4    \\ 
  35.20  &  $-1.74$ & G35.2-1.7  & $2.85$ & $3.27^{+0.56}_{-0.42}$  &  4    \\ 
  49.49  &  $-0.37$ & W51 (IRS2) & $5.52$ & $5.13^{+2.90}_{-1.40}$  &  3    \\ 
  59.78  &  $+0.06$ & G59.7+0.1  & $3.07$ & $2.16^{+0.10}_{-0.09}$  &  3    \\ 
 109.87  &  $+2.11$ & Cep A	 & $1.09$ & $0.70^{+0.04}_{-0.04}$  &  2    \\ 
 111.54  &  $+0.78$ & NGC7538	 & $5.61$ & $2.65^{+0.12}_{-0.11}$  &  2    \\ 
 122.02  &  $-7.07$ & IRAS00420  & $3.97$ & $2.17^{+0.05}_{-0.05}$  &  7    \\ 
 123.07  &  $-6.31$ & NGC281	 & $2.69$ & $2.82^{+0.24}_{-0.24}$  &  8    \\ 
 133.95  &  $+1.06$ & W3(OH)	 & $4.28$ & $1.95^{+0.04}_{-0.04}$  &  9    \\ 
 135.28  &  $+2.80$ & WB 89-437  & $8.68$ & $6.00^{+0.02}_{-0.02}$  & 10    \\ 
 188.95  &  $+0.89$ & S252	 & $4.06$ & $2.10^{+0.03}_{-0.03}$  &  1    \\ 
 196.45  &  $-1.68$ & S269	 & $3.98$ & $5.28^{+0.24}_{-0.22}$  & 11    \\ 
 209.01  & $-19.38$ & Orion	 & $0.99$ & $0.44^{+0.02}_{-0.02}$  & 12    \\ 
 232.62  &  $+1.00$ & G232.6+1.0 & $1.92$ & $1.68^{+0.11}_{-0.09}$  &  1    \\ 
 239.35  &  $-5.06$ & VY CMa	 & $1.56$ & $1.14^{+0.11}_{-0.09}$  & 13    \\ \hline
\end{tabular}
\label{table3}
\end{center}
\begin{footnotesize}
{\parbox{06.7in}{References: (1) - \citet[][a]{Reid09a}; (2) - \citet{Moscadelli09}; (3) - \citet{Xu09}; (4) - \citet{Zhang09}; (5) - \citet{Brunthaler09}; (6) - \citet{Bartkiewicz08}; (7) - \citet{Moellenbrock09}; (8) - \citet{Sato08}; (9) - \citet{Xu06,Hachisuka06}; (10) - \citet{Hachisuka09}; (11) - \citet{Honma07}; (12) - \citet{Hirota07,Menten07}; (13) - \citet{Choi08}; (14) - \citet[][c]{Reid09c}; (15) - \citet[][b]{Reid09b}.}}
\end{footnotesize}
\end{table*}
\end{footnotesize}

$K$-band spectrophotometric distances are derived from the distance modulus ({\it i.e.} $m_{K} - M_{K} = 5 \times log(d) - 5 + A_{K}$)  and the adoption of an interstellar extinction law. In this section, we have used H\,{\sc{ii}} regions with identifyied ionizing O-type stars found in the literature. Most of these regions are also in our sample. For each ionizing O-type star (for the spectral types of the ionizing sources, see the references that are listed in the notes of Table 4), we have used $M_{K}$ from \citet{Hanson96}, $M_{V}$ from \citet{Vacca96} and $V - K$ \citet{Koornneef83}. Also, two extreme interstellar extinction laws were used to analyse the effect of this parameter on the result. The two laws used are from \citet{Mathis90} and \citet{Stead09}. The law from \citet{Mathis90} gives an exponent of $\alpha$ = 1.70, while the law of \citet{Stead09} has an exponent of $\alpha$ = 2.14, which represent extreme situations of high and low interstellar extinctions, respectively. Indeed, since $A_{K} \propto \lambda_{K}^{-\alpha}$, \citet{Mathis90} give the largest values for $A_{K}$ compared to the values derived using \citet{Stead09}. This implies that the spectrophotometric distances derived using \citet{Mathis90} law, column 6 of Table 4, are smaller than those derived using \citet{Stead09}, column 7 of Table 4. When a H\,{\sc{ii}} region has more than one identified ionizing sources (column 4 in Table 4 shows the number of identified ionizing sources for each region), we consider the median of the spectrophotometric distances, for each interstellar extinction law. The median distance of both interstellar extinction laws are shown in column 8. These two extreme situations were used for 26 H\,{\sc{ii}} regions we have found in the literature and the results still show discrepancies with that from kinematic techniques. As can be seen in the Table 4, most of the regions have smaller distances than the kinematic results. The kinematic distances shown in the column 5 of Table 4 are, as throughout this work, from \citet{Russeil03}. Note that our spectrophotometric distances (columns 6, 7 and their median in column 8) are not the published values. Here we used only the published spectral types of the ionizing sources and computed distances based on adopting the two extreme interstellar laws.

\begin{footnotesize}
\begin{table*}
\begin{center}
{\parbox{06.7in}{\textbf{Table 4.} Spectrophotometric distances. Columns 1 and 2 give the galactic coordinates. Column 3 gives the names of the H\,{\sc{ii}} regions. In the column 4, we list the number of stars used in each region. In the column, the kinematic distances are given. Column 6 shows the spectrophotometric distance using the \citet{Mathis90} interstellar extinction law and column 7 the distance using the \citet{Stead09} interstellar extinction law. Column 8 is the average between both spectrophotometric distances.}}
\begin{tabular}{|c|c|c|c|c|c|c|c|c|}
\hline
   $l$  &$b$     &\it{region}    &$N^{\circ}$&$d_{kin}^{1}$&$d_{Mathis}$&$d_{Stead}$ & $\langle d_{spec} \rangle$\\ 
        &        &                      &   & kpc         & kpc          & kpc           \\ \hline
  $6.0$ & $-1.2$ & M8$^{2}$	    	& 1 & $2.80\pm1.0$& $0.90\pm0.35$& $0.99\pm0.38$ &  $0.95$\\ 
 $10.1$ & $-0.3$ & W31-South$^{3}$  	& 4 & $4.50\pm0.6$& $3.01\pm1.12$& $4.10\pm1.52$ &  $3.55$\\ 
 $10.3$ & $-0.1$ & W31-Norte$^{2}$  	& 2 &$15.10\pm1.3$& $2.02\pm0.77$& $2.76\pm1.08$ &  $2.39$\\ 
 $15.0$ & $-0.7$ & M17$^{4}$	    	& 3 & $2.40\pm0.5$& $2.01\pm0.75$& $2.19\pm0.82$ &  $2.10$\\ 
 $25.4$ & $-0.2$ & W42$^{5}$	    	& 1 &$11.50\pm0.3$& $2.46\pm0.90$& $2.89\pm1.07$ &  $2.67$\\ 
 $30.8$ & $-0.0$ & W43$^{6}$	    	& 3 & $6.20\pm0.6$& $3.64\pm1.37$& $6.16\pm2.32$ &  $4.90$\\ 
 $31.4$ & $+0.3$ & G31.4+0.3$^{2}$   	& 1 & $6.20\pm0.6$& $4.89\pm1.80$& $6.20\pm2.28$ &  $5.55$\\ 
 $34.3$ & $+0.1$ & G34.3+0.1$^{2}$   	& 3 &$10.50\pm0.3$& $1.68\pm0.66$& $2.55\pm1.00$ &  $2.11$\\ 
 $43.2$ & $+0.0$ & W49A$^{2}$	    	& 1 &$11.80\pm0.4$&$10.13\pm3.73$&$15.40\pm5.67$ & $12.76$\\ 
 $49.5$ & $-0.4$ & W51A$^{7}$	    	& 4 & $5.50\pm8.0$& $3.38\pm1.35$& $5.41\pm2.11$ &  $4.39$\\ 
$133.7$ & $+1.2$ & W3$^{8}$	    	& 3 & $4.20\pm0.7$& $1.97\pm0.73$& $2.19\pm0.82$ &  $2.08$\\ 
$189.0$ & $+0.8$ & G189.0+0.8$^{2}$  	& 2 & $0.80\pm1.9$& $3.62\pm1.51$& $4.33\pm1.82$ &  $3.97$\\ 
$217.4$ & $-0.1$ & BFS57$^{2}$	    	& 2 & $2.40\pm0.6$& $1.16\pm0.45$& $1.52\pm0.59$ &  $1.34$\\ 
$265.1$ & $+1.5$ & G265.1+1.5$^{2}$  	& 4 & $1.40\pm0.8$& $0.83\pm0.33$& $1.00\pm0.40$ &  $0.91$\\ 
$267.7$ & $-1.1$ & G267.7-1.1$^{2}$  	& 2 & $1.40\pm1.0$& $1.13\pm0.42$& $1.36\pm0.52$ &  $1.24$\\ 
$268.0$ & $-1.0$ & RCW38$^{2}$	    	& 1 & $1.40\pm1.0$& $1.05\pm0.42$& $1.75\pm0.70$ &  $1.40$\\ 
$282.0$ & $-1.2$ & G282.0-1.2$^{2}$  	& 2 & $5.90\pm0.5$& $6.00\pm2.32$& $7.94\pm3.07$ &  $6.97$\\ 
$291.3$ & $-0.7$ & NGC3576$^{2}$	& 1 & $3.10\pm9.8$& $0.87\pm0.32$& $1.09\pm0.41$ &  $0.98$\\ 
$298.2$ & $-0.3$ & G298.2-0.3$^{2}$  	& 2 &$10.40\pm0.5$& $4.18\pm1.56$& $5.29\pm1.97$ &  $4.73$\\ 
$326.6$ & $+0.6$ & RCW95$^{2}$       	& 2 & $2.80\pm0.3$& $1.63\pm0.63$& $2.02\pm0.78$ &  $1.82$\\ 
$328.3$ & $+0.4$ & G328.3+0.4$^{2}$  	& 2 & $6.10\pm0.6$& $5.27\pm2.18$& $6.34\pm2.59$ &  $5.80$\\ 
$332.6$ & $-0.6$ & RCW106$^{2}$	 	& 4 & $3.50\pm0.3$& $3.25\pm1.34$& $4.67\pm1.88$ &  $3.96$\\ 
$333.1$ & $-0.4$ & G333.1-0.4$^{9}$  	& 2 & $3.50\pm0.3$& $3.29\pm1.23$& $3.85\pm1.43$ &  $3.57$\\ 
$345.2$ & $+1.0$ & RCW116B$^{2}$	& 2 & $1.70\pm0.6$& $2.09\pm0.88$& $2.68\pm1.13$ &  $2.38$\\ 
$348.2$ & $-1.0$ & RCW121$^{2}$		& 1 & $2.70\pm0.5$& $2.74\pm1.01$& $2.95\pm1.08$ &  $2.84$\\ 
$351.2$ & $+0.7$ & NGC6334$^{2}$	& 1 & $1.20\pm1.1$& $1.30\pm0.50$& $2.34\pm0.90$ &  $1.82$\\
$351.6$ & $-1.3$ & G351.6-1.3$^{2}$  	& 2 &$14.30\pm0.8$& $2.28\pm0.94$& $2.90\pm1.17$ &  $2.59$\\ \hline
\end{tabular}
\label{table4}
\end{center}
\begin{footnotesize}
{\parbox{06.7in}{Notes: (1) All kinematic distances are from \citet{Russeil03}, except W31-South and W31-North with distances from \citet{Corbel04}; The adopted distance moduli are from: (2) \citet{Bik05}; (3) \citet{Blum01}; (4) \citet{Hanson96}; (5) \citet{Blum00}; (6) \citet{Blum99}; (7) \citet{Figueredo08}; (8) Navarete et al., {\it in preparation}; (9) \citet{Figueredo05}.}}
\end{footnotesize}
\end{table*}
\end{footnotesize}

\begin{figure*}
\begin{minipage}[b]{0.47\linewidth}
\includegraphics[height=09cm,width=09cm]{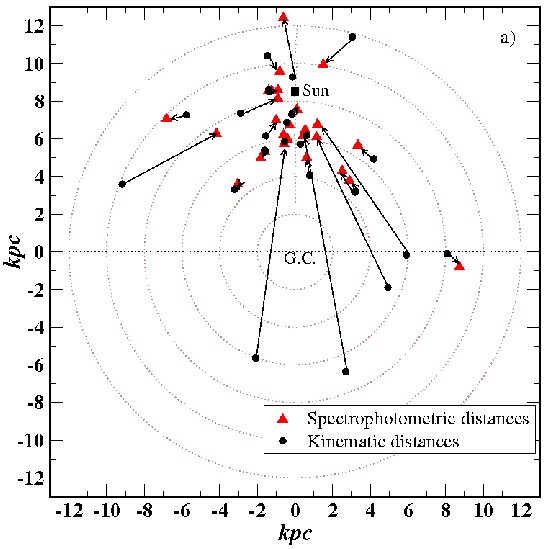}
\end{minipage} \hfill
\begin{minipage}[b]{0.47\linewidth}
\includegraphics[height=09cm,width=09cm]{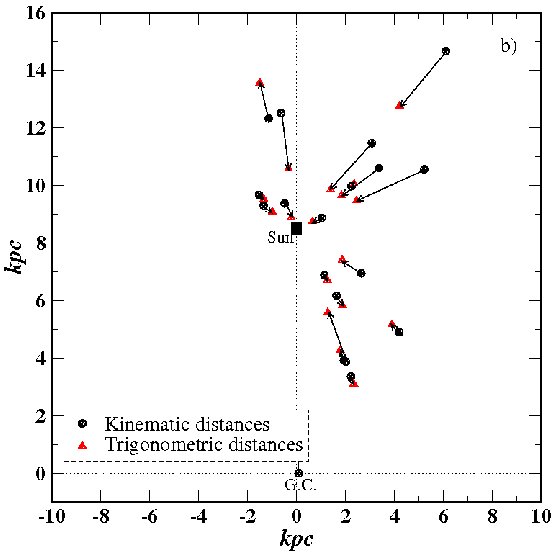}
\end{minipage}
\caption{{ Distribution of the H\,{\sc{ii}} regions in the Galactic plane. In the left (Fig. 4a) are the distances derived with $K$-band spectrophotometry, these distances are an average of the distances using \citet{Mathis90} and \citet{Stead09}. In the right (Fig. 4b) are the distances derived using trigonometric parallax. In both panels the kinematic distances are also shown. }}
\label{fig:gal-projec}
\end{figure*}

These effects, of discrepancies between the distances, can be seen in the Fig.~\ref{fig:gal-projec}, where the H\,{\sc{ii}} regions are displayed on the Galactic plane based on their distances. Black circles represent kinematic results, red triangles represent an average of the results using \citet{Mathis90} and \citet{Stead09} laws (Fig. 4a), and the trigonometric parallax distances (Fig. 4b). Discrepancies are clearly seen between the kinematic and non-kinematic results, where the second group has, in general, smaller distances. The arrows are proportional to the discrepancies.

\section{Conclusion}

In this work, we present a near infrared and {\it Spitzer} (mid infrared) study of 35 Galactic H\,{\sc{ii}} regions. These regions were chosen from the catalogs of \citet{Conti04}, \citet{Bica03a} and \citet{Dutra03}. \citet{Conti04} have carried out a complete census of 6-cm selected H\,{\sc{ii}} regions and identified 56 as Giant H\,{\sc{ii}} regions, based in part on mid- and far-IR fluxes from MSX and IRAS. \citet{Bica03a} and \citet{Dutra03} have carried out a 2MASS $J$, $H$, and $K_{s}$ survey of infrared star cluster across the Milky Way.

All the distances listed in the Table 1 are from kinematic techniques (rotational velocity plus a galactic rotation model) for consistency. Among our sample of 35 Galactic H\,{\sc{ii}} regions, we have defined 24 as GH\,{\sc{ii}} regions based on the kinematic distance and radio continuum luminosity. 

In this paper, we have focused on the $J$, $H$, and $K_{s}$-band photometric properties of the regions. Using morphological clues in near infrared and {\it Spitzer} IRAC images along with the C-C and C-M diagrams, we place each region into a qualitative evolutionary stage labeled {\it A}, {\it B}, {\it C}, or {\it D}. In the first ($stage$ $A$) we identify regions still very embedded in gas and dust, with little evidence of their emergent stellar component, like G12.8-0.2 (W33), G333.3-0.4, G333.6-0.2, for example. regions in the second $stage$ ($B$) have a well-defined cluster of stars, but with several objects with infrared excess (T-Tauri stars and YSOs) like G5.97-1.18 (M8) and G10.3-0.1 (W31-North). The third, $stage$ $C$, are those regions that we can distinguish a well-defined cluster of stars, a few objects with infrared excess and a nebulosity surrounding the naked cluster (i.e. on its periphery), like G30.8-0.2 (W43) and G287.4-0.6 (NGC3372). The fourth $stage$ ($D$) represents regions in which we do not see nebulosity, the cluster is well-defined and  stars are completely typified by colors for normal photospheres, like G308.7+0.6.

In this near and mid infrared study of the stellar content of these star-forming regions, we have also identified a sample of massive YSOs in our images and C-M and C-C diagrams, based on their large luminosities, as well as large infrared excess. As expected, the presence of YSOs, particularly the massive ones, is more prominent in the less evolved regions, where there is strong nebular emission. We present the list of the massive YSOs detected in this work in the Table 2. 

Qualitatively, we have used main sequence lines in the C-M diagrams to verify if the kinematic distance is consistent with the cluster members position in these diagrams. In some regions, we have shown large discrepancies, where the tip of the main sequence (O-type stars) is fainter than the brightest objects of the cluster. This implies that the real distance is smaller than that adopted from kinematic methodology. Other distance determinations, like spectrophotometric and trigonometric parallax have verified (typically) smaller distances compared to kinematic distances. In our sample of 35 Galactic H\,{\sc{ii}} regions, we suggest that roughly a third are consistent with a closer distance than is derived from kinematic techniques. These regions are marked as CL (closer) in the Table 1. We find nine H\,{\sc{ii}} regions have kinematic distances that are qualitatively consistent with main sequence locations in our C-M diagrams. These regions are marked as AG (agree) in the Table 1. We could not speculate on the agreement between the kinematic distances and the photometric data for ten of our sample of H\,{\sc{ii}} regions, due to small number of detected objects or the presence of known evolved stars. These regions are marked as UN (unknown) in the Table 1. Finally, two H\,{\sc{ii}} regions may be further away than their kinematic distances. Interestingly they are close to the Galactic center, where the determination of kinematic distances is not so easy. Our images and C-M and C-C diagrams provide excellent candidate sources to observe spectroscopically and so expand the sample of GH\,{\sc{ii}} regions with known distances. 

Quantitatively, we compared the kinematic distances with distances derived from $K$-band spectrophotometric and trigonometric parallax to 26 star forming regions found in the literature. We used two extreme interstellar extinction laws in the determination of the distance. In general, the distances derived by these two non-kinematic techniques are smaller than that derived by kinematic thecniques. The same discrepancies were found when we compare the results from trigonometric parallax with the kinematic thechniques.

There are three main conclusions in this work: 1) in most cases clusters are seen; 2) one can distinguish several evolutionary stages among these objects; and 3) the photometric distances are in many cases smaller than the kinematic values (similar to what is inferred from spectrophotometric and trigonometric parallaxes). Plans are underway to revisit the distance discrepancies among the kinematic, trigonometric and spectrophotometric determinations with the aim of better understanding our Galactic sprial structure.


\section*{Acknowledgments}

APM and AD are grateful to the Brazilian agency CNPq-MCT for continuous financial support. EF thanks L'Or\'eal-UNESCO-ABC for Brazil's 2009 For Women in Science grant. AD, EF and CLB acknowledge FAPESP for continuous financial support. PSC wishes to thank the NSF for continuous support. Based on observations obtained at the CTIO 1.5-m and Blanco 4-m telescopes, which are operated by the Association of Universities for Research in Astronomy Inc. (AURA), under a cooperative agreement with the National Science Foundation (NSF) as part of the National Optical Astronomy Observatories (NOAO).


\bibliographystyle{mn2e}
\bibliography{references}

\newpage

\appendix


\section[]{Figures}

All of the figures in this section have North pointing to the top and East to the left. For all regions, we present a $JHK_{s}$ color image, where $J$ is blue, $H$ is green and $K_{s}$ is red. We also present IRAC-{\it Spitzer} color images. The {\it Spitzer} images are combined with $4.6$, $5.8$ and $8.0$ $\mu$m as blue, green and red, respectively.

We show the photometric results in color-color and color-magnitude diagrams (C-C and C-M diagrams, respectively). We have used kinematic distances for consistency in all C-M diagrams when presenting the {\it un-reddened}  main sequence. The reddening vector is shown in each C-M diagram as well. In some regions the apparent discrepancy between the main sequence position and the observed photometry is used to suggest a different distance than the kinematic one. Details are given above for each cluster.

In the C-C diagram, we show the two lines of interstellar reddening for a M-type star \citep{Frogel78}, the top line, and for a O-type star \citep{Koornneef83}, the botton line. Ideally, all main senquence stars in the image, between these spectral types should be located between these two lines.  For objects with excess emission, they will be displaced toward larger colors (to the right in the diagram) into the CTTS region or even beyond.


\begin{figure*}
\begin{minipage}[b]{0.47\linewidth}
\includegraphics[height=09cm,width=\linewidth]{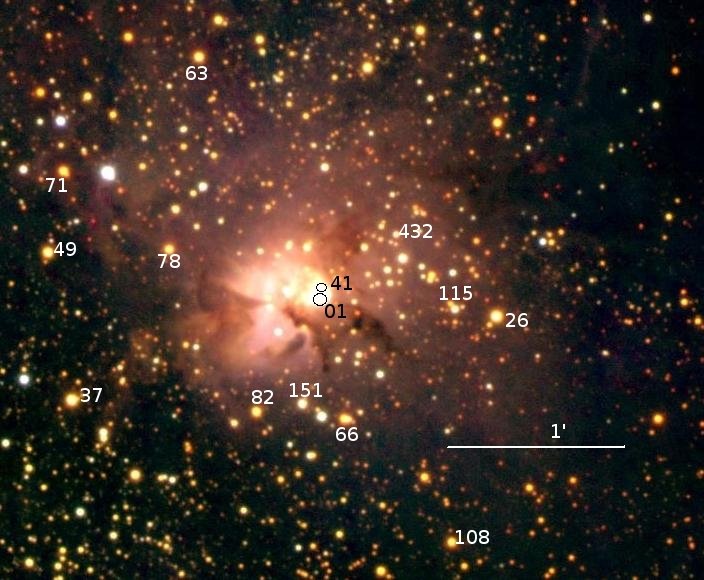}
\end{minipage} \hfill
\begin{minipage}[b]{0.47\linewidth}
\includegraphics[height=09cm,width=\linewidth]{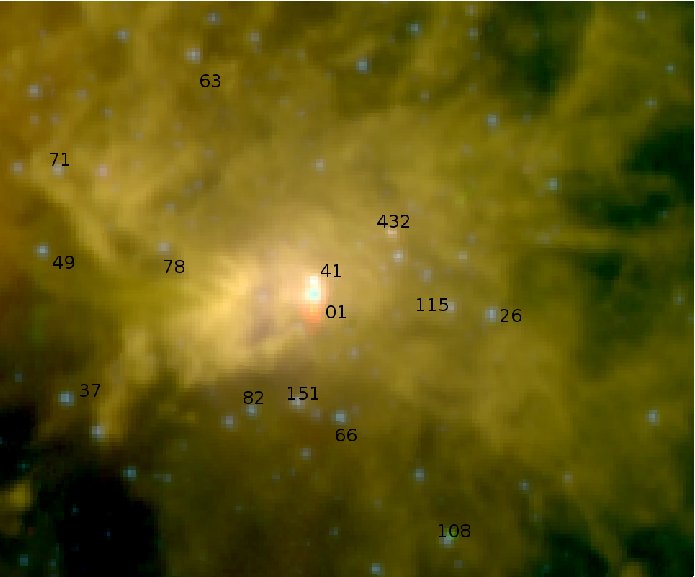}
\end{minipage}
\caption{{ Color images of G5.97-1.18 (M8). Left: $JHK_{s}$ color image. Right: IRAC-{\it Spitzer} color image. In both images, the size is $\approx$ 3.0 arcmin on a side.}}
\label{fig:G5-2-color}
\end{figure*}

\begin{figure*}
\begin{minipage}[b]{0.47\linewidth}
\includegraphics[height=10cm,width=\linewidth]{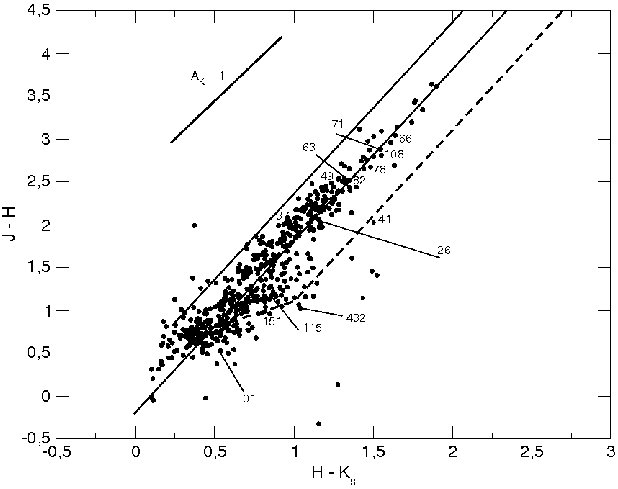}
\end{minipage} \hfill
\begin{minipage}[b]{0.47\linewidth}
\includegraphics[height=10cm,width=\linewidth]{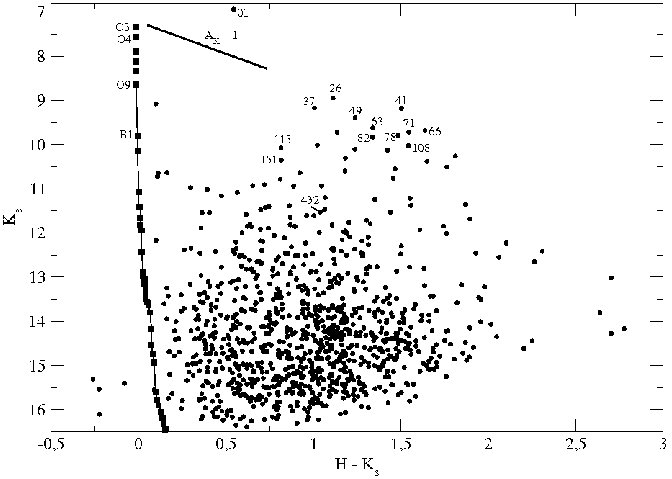}
\end{minipage}
\caption{{ H\,{\sc{ii}} region G5.97-1.18 (M8). Left: Color-color diagram (C-C). Right: Color-magnitude diagram (C-M). In the following figures, the main sequence (vertical line in the C-M diagram) position is affected only by the kinematic distance (see text). Extinction and reddening are given by the $A_{K}$ = 1 reddening vector according to the extinction law of \citet{Straizys08b}; see text. The kinematic distance to this cluster seems to agree with the photometry.}}
\label{fig:G5-2-CMD}
\end{figure*}


\begin{figure*}
\begin{minipage}[b]{0.49\linewidth}
\includegraphics[height=10cm,width=\linewidth]{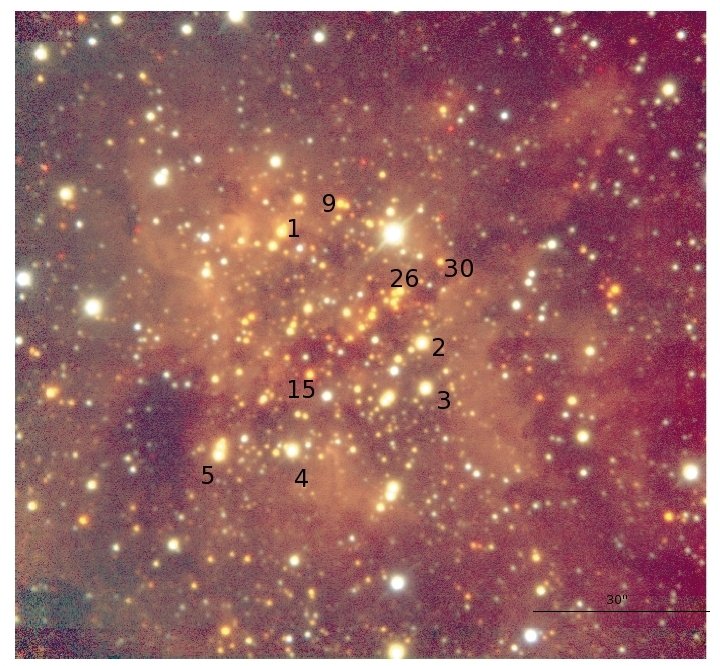}
\end{minipage} \hfill
\begin{minipage}[b]{0.49\linewidth}
\includegraphics[height=9.8cm,width=\linewidth]{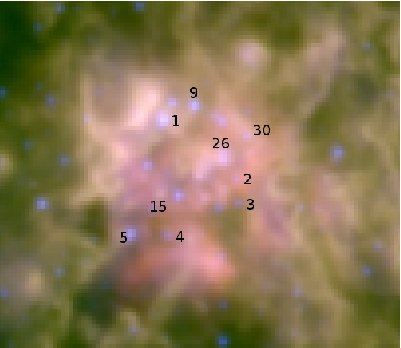}
\end{minipage}
\caption{{ Color images of G10.2-0.3 (W31-South), reproduced from Blum et al. (2001). Left: $JHK_{s}$. Right: IRAC-{\it Spitzer} color image. In both images the size is $\approx$ 1.8 arcmin on a side.}}
\label{fig:W31-color}
\end{figure*}

\vspace{4cm}
\begin{figure*}
\begin{minipage}[b]{0.47\linewidth}
\includegraphics[height=10cm,width=\linewidth]{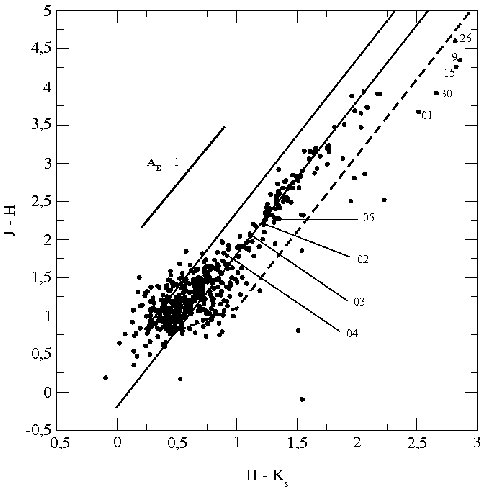}
\end{minipage} \hfill
\begin{minipage}[b]{0.47\linewidth}
\includegraphics[height=10cm,width=\linewidth]{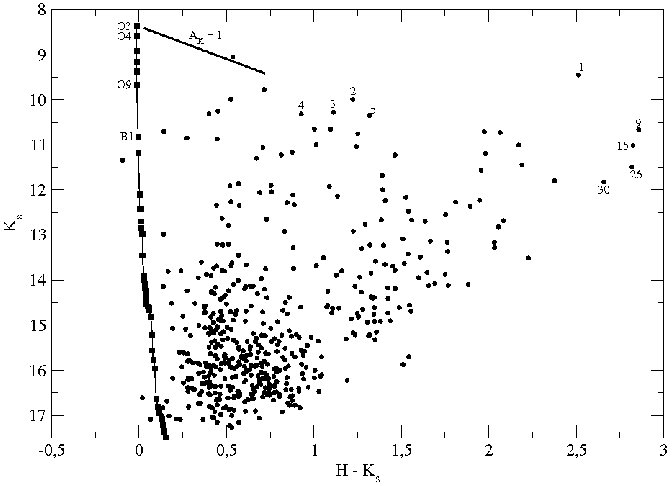}
\end{minipage}
\caption{{ H\,{\sc{ii}} region G10.2-0.3 (W31-South). Left: Color-color diagram (C-C). Right: Color-magnitude diagrama (C-M), the adopted kinematic distance is 4.5 kpc \citep{Russeil03}. The photometry is from \citet{Blum01} and corrected for 2MASS photometric system. Blum et al. used objects \#1, \#2 and \#3 to derive the spectrophotometric distance to this region ($d_{spec}$ = 3.4 kpc). As the spectrophotometric result shows, this cluster is closer to the Sun than the kinematic distance.}}.
\label{fig:W31-CMD}
\end{figure*}


\begin{figure*}
\begin{minipage}[b]{0.47\linewidth}
\includegraphics[height=08cm,width=09cm]{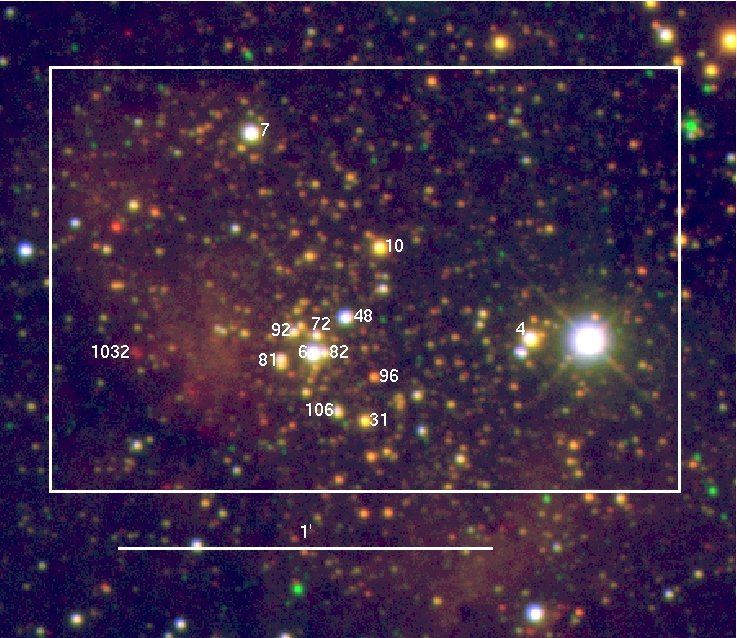}
\end{minipage} \hfill
\begin{minipage}[b]{0.47\linewidth}
\includegraphics[height=08cm,width=09cm]{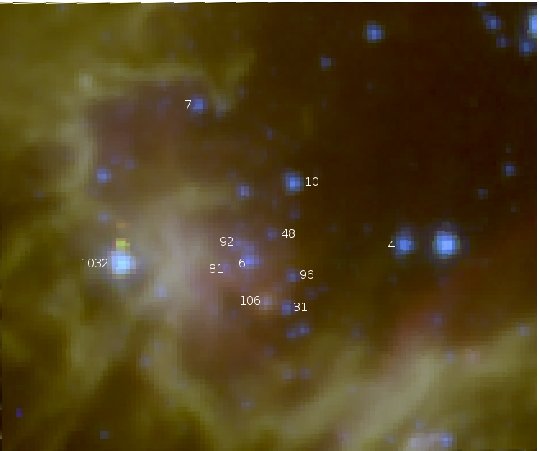}
\end{minipage}
\caption{{Color images of G10.3-0.1 (W31-North). Left: $JHK_{s}$ color image. Right: IRAC-{\it Spitzer} color image. In both images, the FOV is $\approx$ 2.0 x 1.5 arcmin.}}
\label{fig:G10-color}
\end{figure*}

\begin{figure*}
\begin{minipage}[b]{0.48\linewidth}
\includegraphics[height=09cm,width=\linewidth]{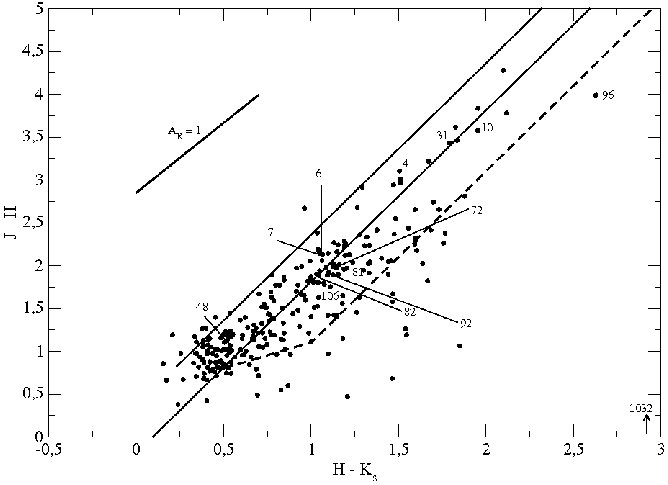}
\end{minipage} \hfill
\begin{minipage}[b]{0.49\linewidth}
\includegraphics[height=09cm,width=\linewidth]{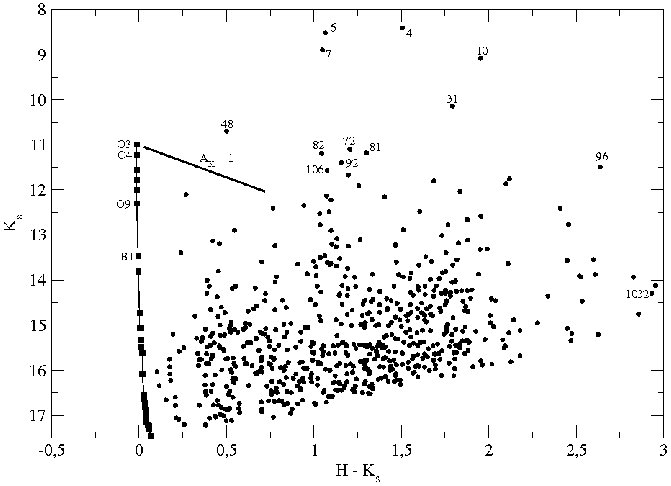}
\end{minipage}
\caption{{ H\,{\sc{ii}} region G10.3-0.1 (W31-North). Left: Color-color diagram (C-C). Right: Color-magnitude diagrama (C-M). The photometry was obtained from stars in the box shown in the near infrared color image. Object \#1032 has $J$-band magnitude $J$ = 17.0 mag (see text for limit determination). Clearly, the photometry is not in accordance with the main sequence line, which indicates to a smaller (closer) real distance.}}
\label{fig:G10-CMD}
\end{figure*}


\begin{figure*}
\begin{minipage}[b]{0.47\linewidth}
\includegraphics[height=10cm,width=\linewidth]{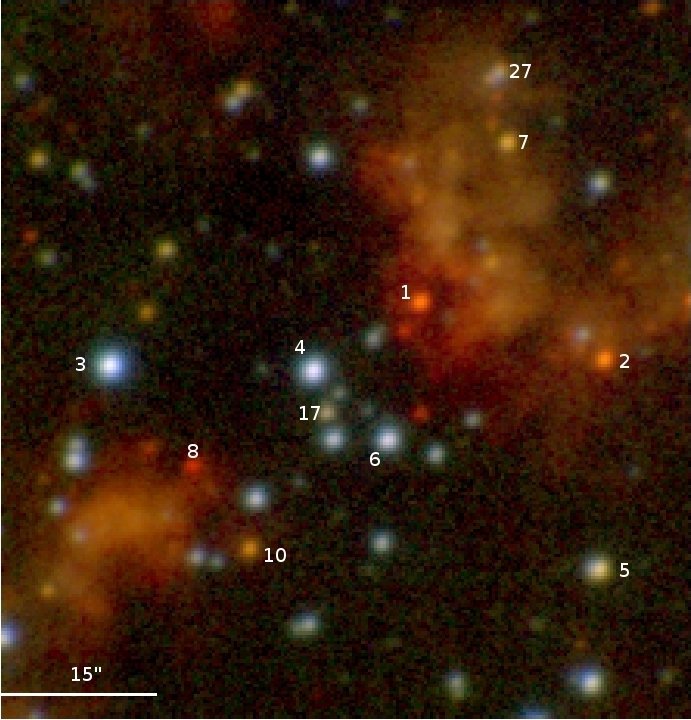}
\end{minipage} \hfill
\begin{minipage}[b]{0.47\linewidth}
\includegraphics[height=10cm,width=\linewidth]{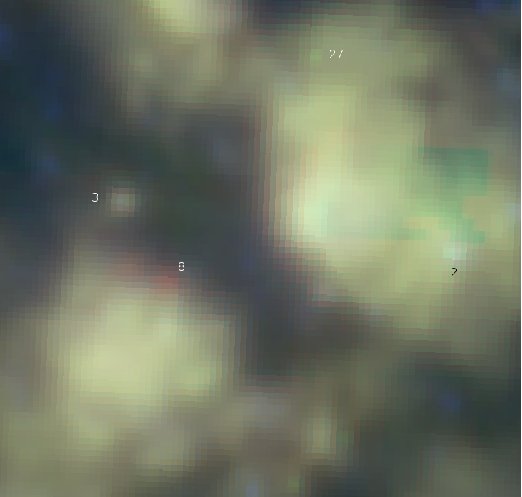}
\end{minipage}
\caption{{ Color images of G12.8-0.2 (W33). Left: $JHK_{s}$ color image. Right: IRAC-{\it Spitzer} color image. The size is $\approx$ 1.0 arcmin on a side.}}
\label{fig:W33-color}
\end{figure*}

\begin{figure*}
\begin{minipage}[b]{0.48\linewidth}
\includegraphics[height=10cm,width=\linewidth]{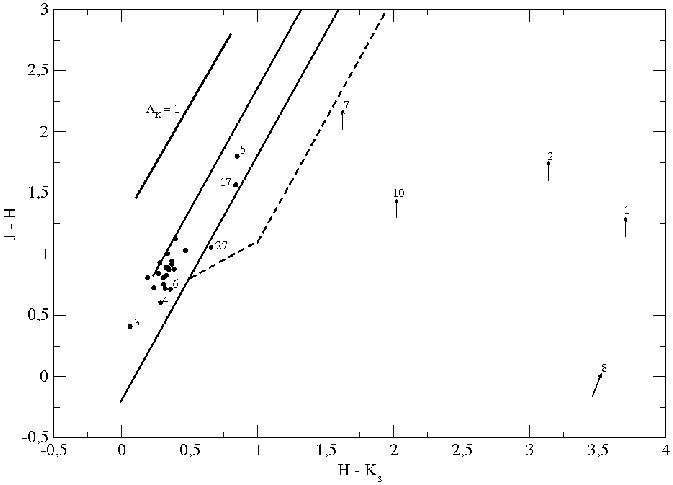}
\end{minipage} \hfill
\begin{minipage}[b]{0.49\linewidth}
\includegraphics[height=10cm,width=\linewidth]{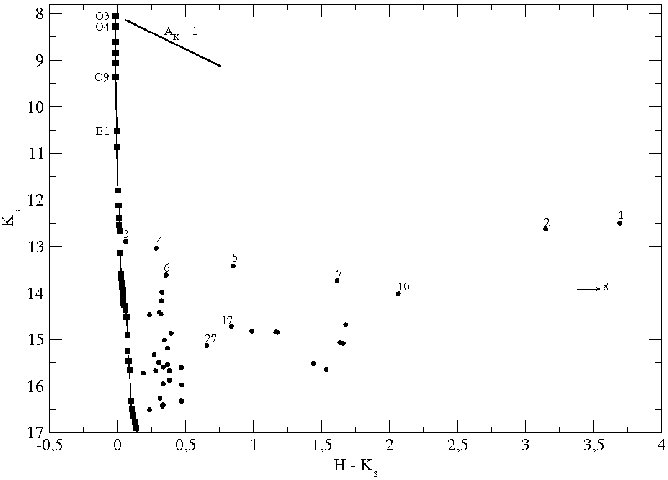}
\end{minipage}
\caption{{ H\,{\sc{ii}} region G12.8-0.2 (W33). Left: Color-color diagram (C-C). Right: Color-magnitude diagram (C-M). Objects not detected are shown (at the tip of the arrows) using limiting magnitudes above of $J$ = $H$ = 17.5 mag. The distance analyses is inconclusive in this case, due to the absence of a star cluster.}}
\label{fig:W33-CMD}
\end{figure*}


\begin{figure*}
\begin{minipage}[b]{0.47\linewidth}
\includegraphics[height=10cm,width=\linewidth]{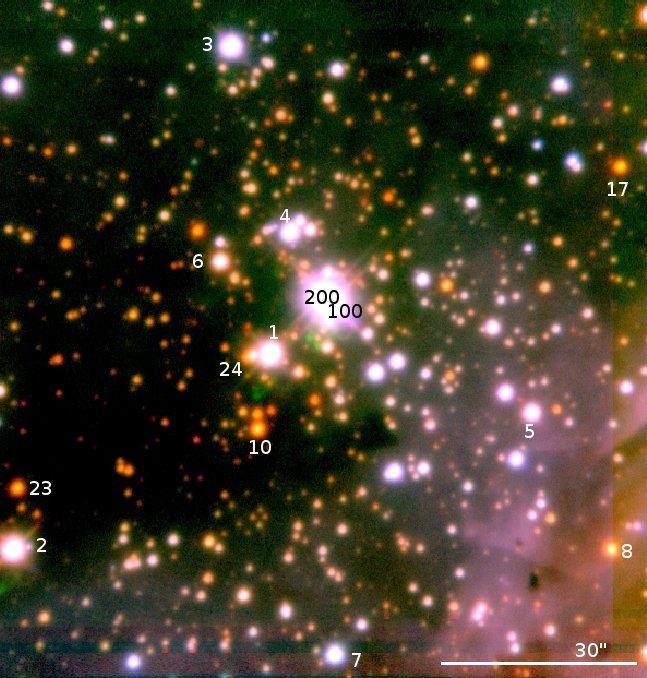}
\end{minipage} \hfill
\begin{minipage}[b]{0.47\linewidth}
\includegraphics[height=10cm,width=\linewidth]{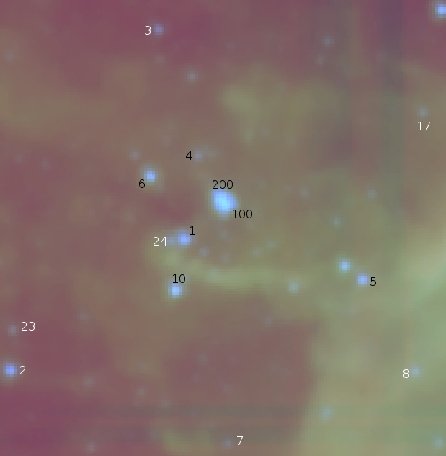}
\end{minipage}
\caption{{ Color images of G15.0-0.7 (M17). Left: $JHK_{s}$ color image. Right: IRAC-{\it Spitzer} color image. In both images, size is $\approx$ 1.5 arcmin on a side.}}
\label{fig:M17-color}
\end{figure*}

\begin{figure*}
\begin{minipage}[b]{0.48\linewidth}
\includegraphics[height=10cm,width=\linewidth]{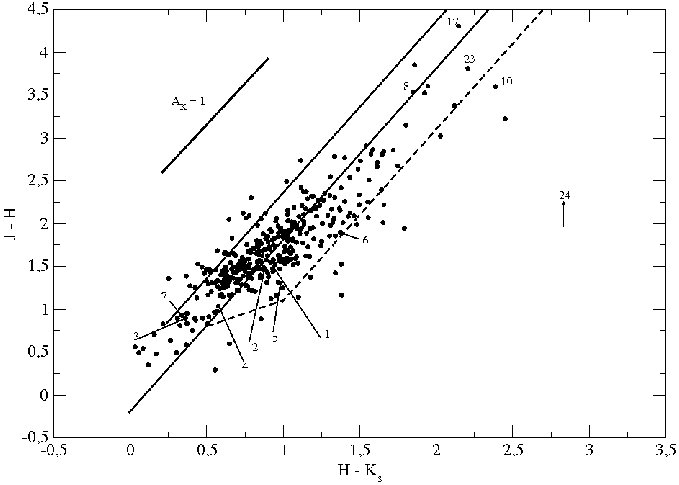}
\end{minipage} \hfill
\begin{minipage}[b]{0.49\linewidth}
\includegraphics[height=10cm,width=\linewidth]{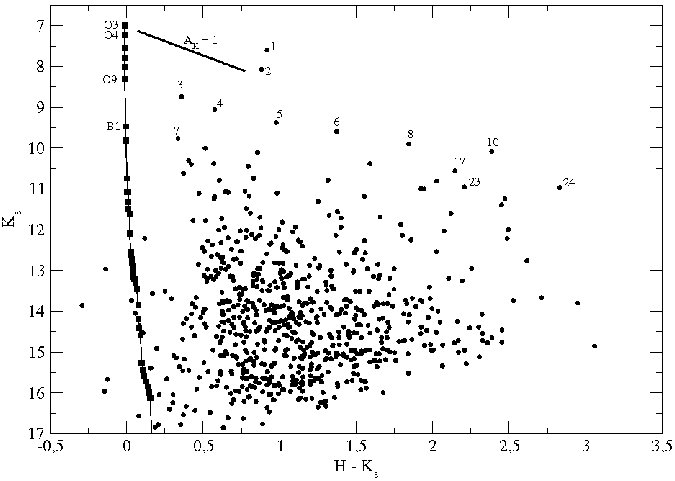}
\end{minipage}
\caption{{ H\,{\sc{ii}} region G15.0-0.7 (M17). Left: Color-color diagram (C-C). Right: Color-magnitude diagram (C-M). Object \#24 was not detected at $J$-band, so we have used a limiting magnitude $J$ = 16.0 mag. In agreement with the work of \citet{Hanson97}, objects \#1 and \#2, if de-reddened indicate a smaller distance than the kinematic results.}}
\label{fig:M17-CMD}
\end{figure*}


\begin{figure*}
\begin{minipage}[b]{0.47\linewidth}
\includegraphics[height=10cm,width=\linewidth]{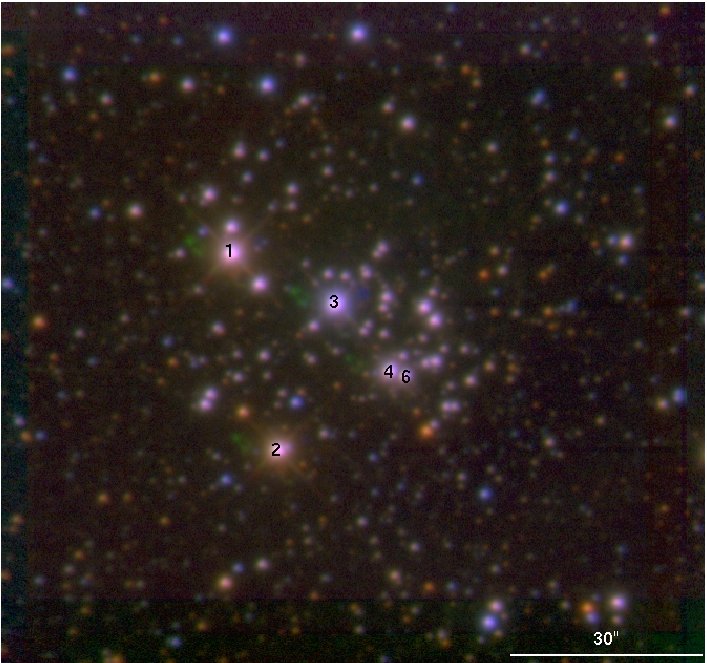}
\end{minipage} \hfill
\begin{minipage}[b]{0.47\linewidth}
\includegraphics[height=10cm,width=\linewidth]{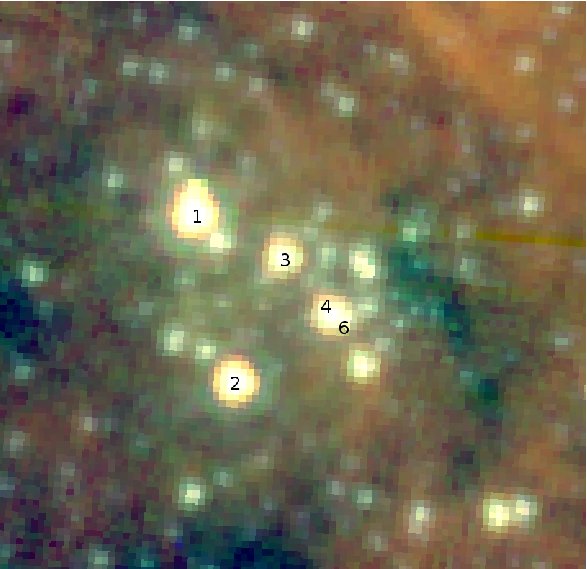}
\end{minipage}
\caption{{ Color images of G22.7-0.4. Left: $JHK_{s}$ color image. Right: IRAC-{\it Spitzer} color image. The size is $\approx$ 2.0 arcmin on a side. No nebulosity was detected in the near infrared, but in the {\it Spitzer} image we see weak emission which is not clearly connected with the star cluster.}}
\label{fig:W41-color}
\end{figure*}

\begin{figure*}
\begin{minipage}[b]{0.48\linewidth}
\includegraphics[height=10cm,width=\linewidth]{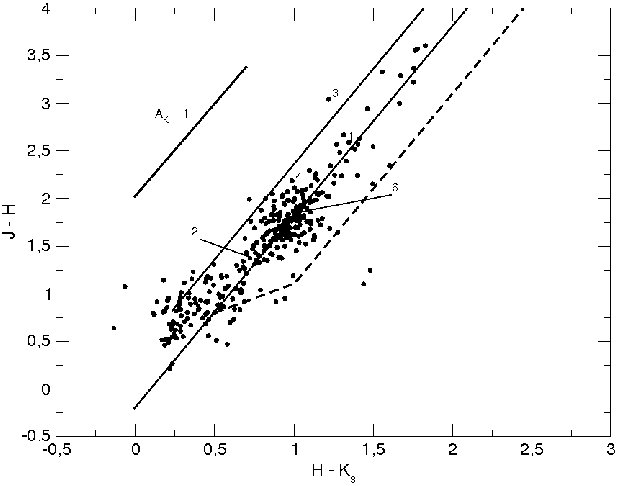}
\end{minipage} \hfill
\begin{minipage}[b]{0.49\linewidth}
\includegraphics[height=10cm,width=\linewidth]{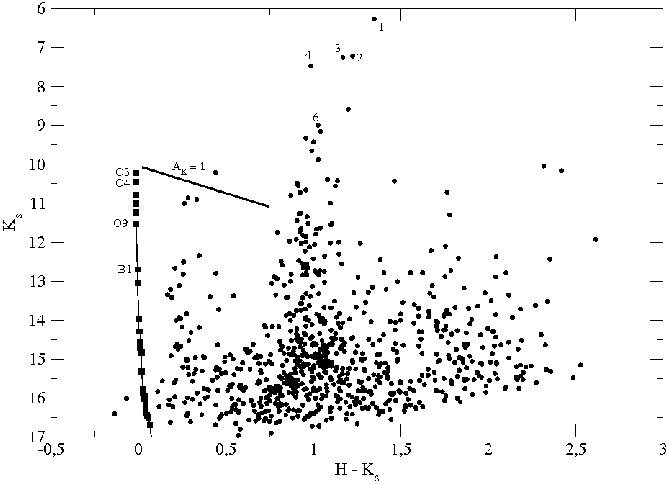}
\end{minipage}
\caption{{ H\,{\sc{ii}} region G22.7-0.4. Left: Color-color diagram (C-C). Right: Color-magnitude diagram (C-M). In the C-M diagram, the kinematic distance appears larger than what would be expected from the photometry. The tip of the main sequence is fainter than the brightest cluster members.}}
\label{fig:W41-CMD}
\end{figure*}


\begin{figure*}
\begin{minipage}[b]{0.47\linewidth}
\includegraphics[height=10cm,width=\linewidth]{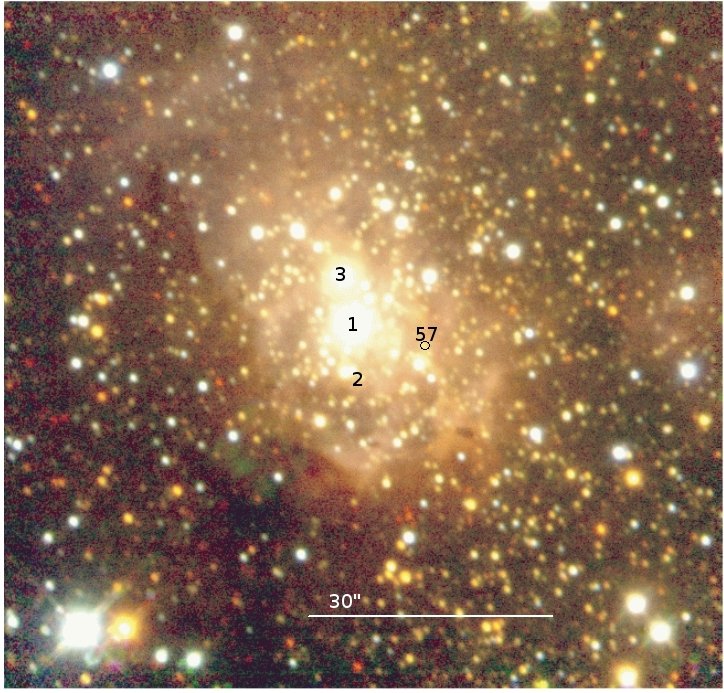}
\end{minipage} \hfill
\begin{minipage}[b]{0.47\linewidth}
\includegraphics[height=10cm,width=\linewidth]{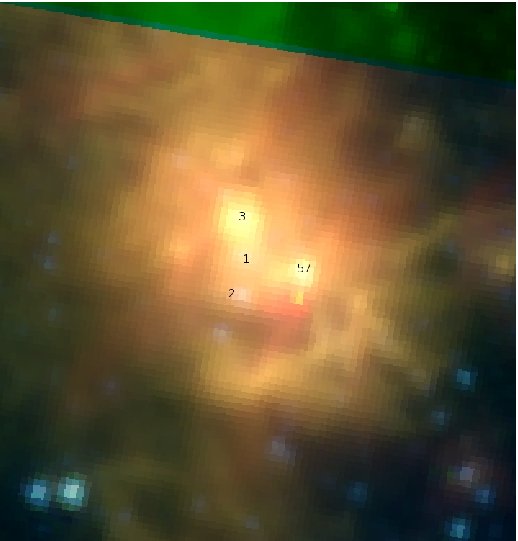}
\end{minipage}
\caption{{ Color images of G25.4-0.2 (W42). Left: $JHK_{s}$ color image, reproduced from \citet{Blum00}. Right: IRAC-{\it Spitzer} color image. In both images, the size is $\approx$ 1.5 arcmin on a side.}}
\label{fig:W42-color}
\end{figure*}

\begin{figure*}
\begin{minipage}[b]{0.48\linewidth}
\includegraphics[height=10cm,width=\linewidth]{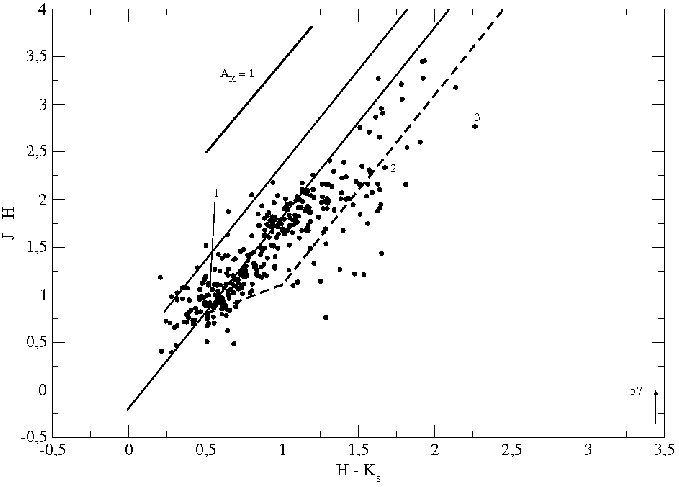}
\end{minipage} \hfill
\begin{minipage}[b]{0.49\linewidth}
\includegraphics[height=10cm,width=\linewidth]{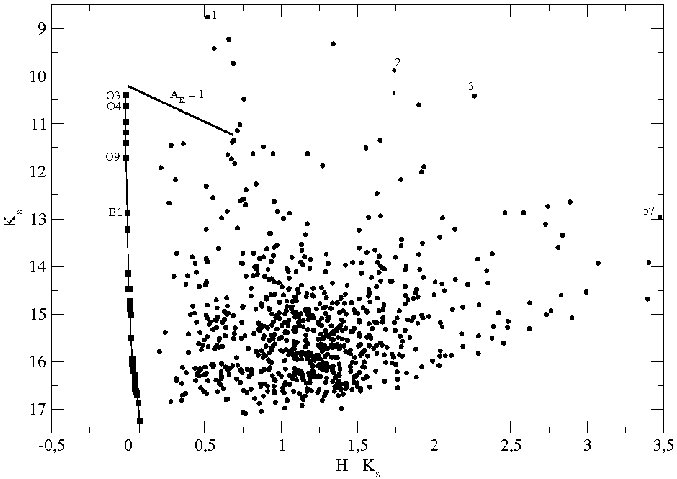}
\end{minipage}
\caption{{ H\,{\sc{ii}} region G25.4-0.2 (W42). Left: Color-color diagram (C-C). Right: Color-magnitude diagram (C-M). \citet{Blum00} used objects \#1, \#2 and \#3 to obtain a spectrophotometric distance of 2.2 kpc, while the (far) kinematic distance is 11.5 kpc \citep{Russeil03}. The main sequence position (C-M diagram) is for the kinematic distance, and we see that the bright objcts of W42 are brighter than the tip of the main sequence, which points to a smaller distance. Object \#57 was not detected in the $J$-band and we used a limiting magnitude of $J$ = 16.5 mag.}}
\label{fig:W42-CMD}
\end{figure*}


\begin{figure*}
\begin{minipage}[b]{0.47\linewidth}
\includegraphics[height=10cm,width=\linewidth]{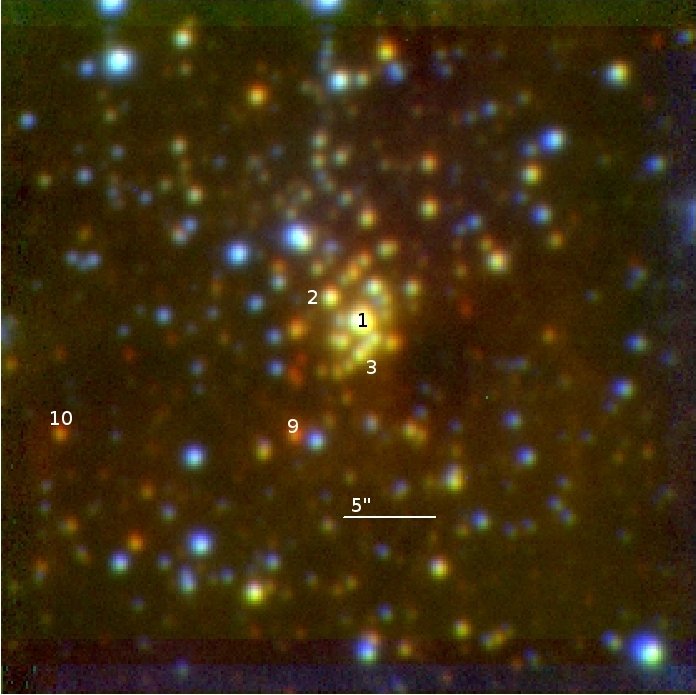}
\end{minipage} \hfill
\begin{minipage}[b]{0.47\linewidth}
\includegraphics[height=10cm,width=\linewidth]{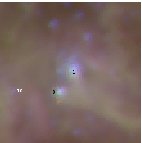}
\end{minipage}
\caption{{ Color images of G30.8-0.2 (W43). Left: $JHK_{s}$ color image. Right: IRAC-{\it Spitzer} color image. In both images, the size is $\approx$ 0.83 arcmin on a side.}}
\label{fig:W43-color}
\end{figure*}

\begin{figure*}
\begin{minipage}[b]{0.48\linewidth}
\includegraphics[height=10cm,width=\linewidth]{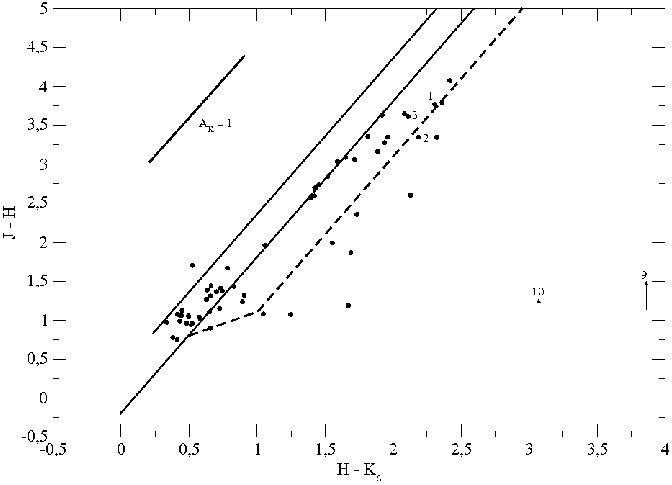}
\end{minipage} \hfill
\begin{minipage}[b]{0.49\linewidth}
\includegraphics[height=10cm,width=\linewidth]{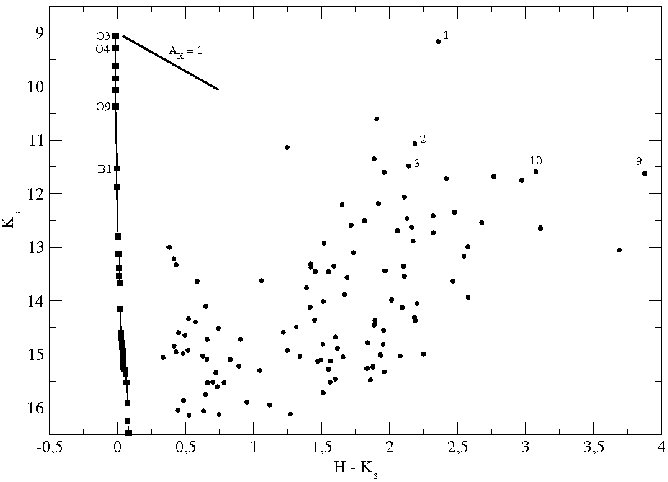}
\end{minipage}
\caption{{ H\,{\sc{ii}} region G30.8-0.2 (W43). Left: Color-color diagram (C-C). Right: Color-magnitude diagram (C-M). Objects \#1, \#2 and \#3 were used to determine the distance from spectrophotometric parallax \citep[4.3 kpc,][]{Blum99}. With the photometry alone, object \#1 if de-reddened will be brighter than the tip of the main sequence, which also indicates a smaller distance. The adopted $J$-band magnitude for objects \#9 and \#10 is $J$ = 17.0 mag.}}
\label{fig:W43-CMD}
\end{figure*}


\begin{figure*}
\begin{minipage}[b]{0.47\linewidth}
\includegraphics[height=09cm,width=\linewidth]{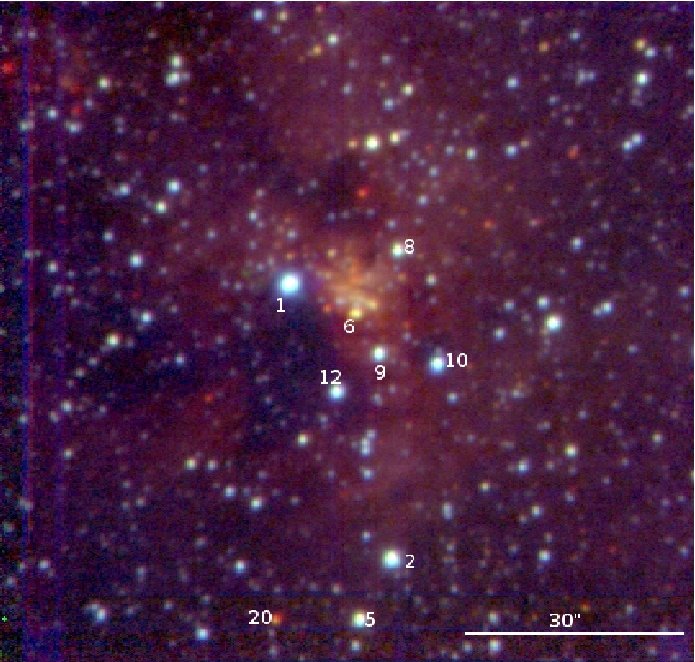}
\end{minipage} \hfill
\begin{minipage}[b]{0.47\linewidth}
\includegraphics[height=09cm,width=\linewidth]{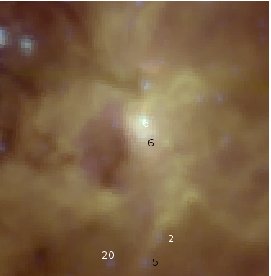}
\end{minipage}
\caption{{ Color images of G45.5+0.1 (K47). Left: $JHK_{s}$ color image. Right: IRAC-{\it Spitzer} color image. In both images, the size is $\approx$ 1.5 arcmin on a side.}}
\label{fig:K47-color}
\end{figure*}

\begin{figure*}
\begin{minipage}[b]{0.48\linewidth}
\includegraphics[height=10cm,width=\linewidth]{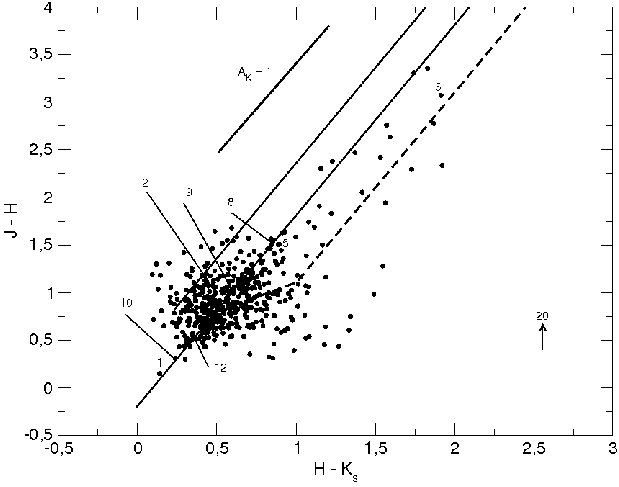}
\end{minipage} \hfill
\begin{minipage}[b]{0.49\linewidth}
\includegraphics[height=10cm,width=\linewidth]{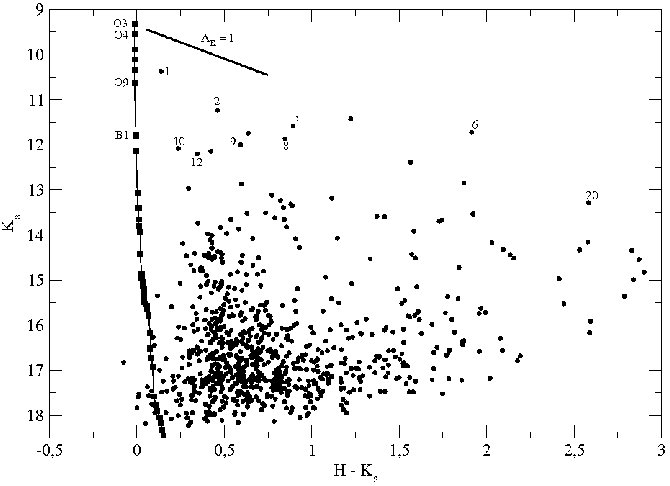}
\end{minipage}
\caption{{ H\,{\sc{ii}} region G45.5+0.1 (K47). Left: Color-color diagram (C-C). Right: Color-magnitude diagram (C-M). Due to the absence of a star cluster, the distance analyses is inconclusive. The limiting magnitude in the $J$-band is $16.5$ mag.}}
\label{fig:K47-CMD}
\end{figure*}

\clearpage

\begin{figure*}
\begin{minipage}[b]{0.47\linewidth}
\includegraphics[height=09cm,width=09cm]{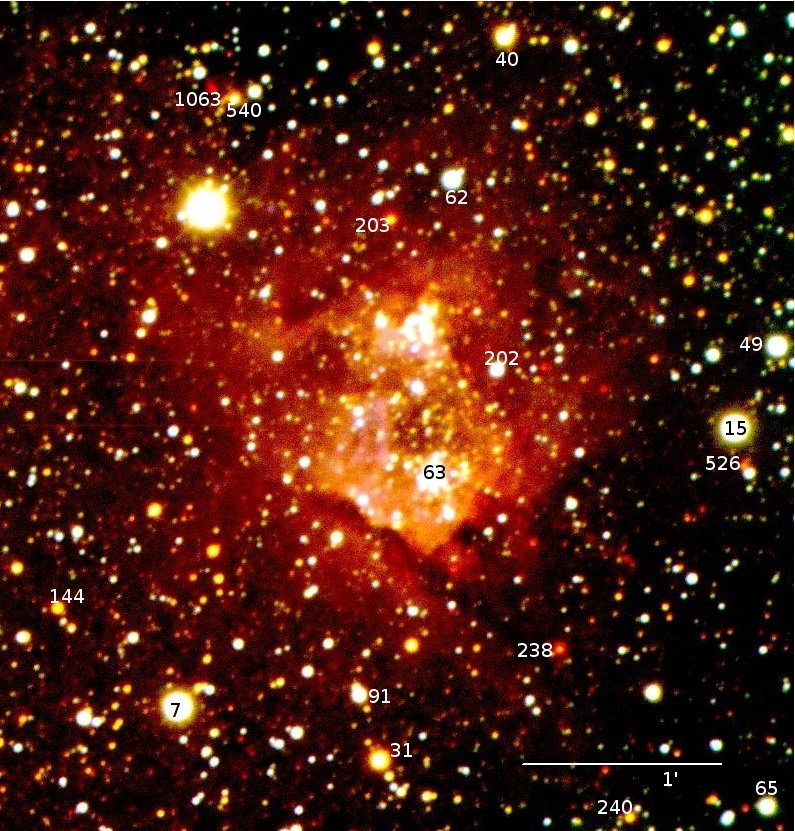}
\end{minipage} \hfill
\begin{minipage}[b]{0.47\linewidth}
\includegraphics[height=09cm,width=09cm]{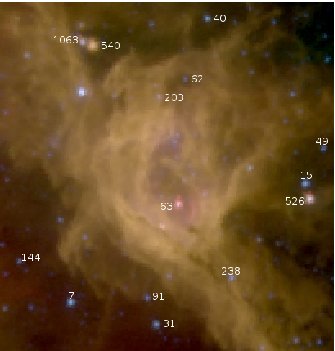}
\end{minipage}
\caption{{ Color images of G48.9-0.3 (W51). Left: $JHK_{s}$ color image. Right: IRAC-{\it Spitzer} color image. In both images, the size is $\approx$ 3.3 arcmin on a side.}}
\label{fig:W51-color}
\end{figure*}

\begin{figure*}
\begin{minipage}[b]{0.48\linewidth}
\includegraphics[height=10cm,width=\linewidth]{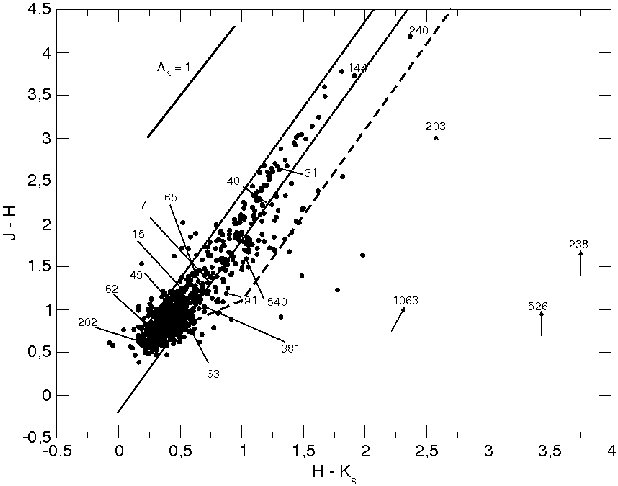}
\end{minipage} \hfill
\begin{minipage}[b]{0.49\linewidth}
\includegraphics[height=10cm,width=\linewidth]{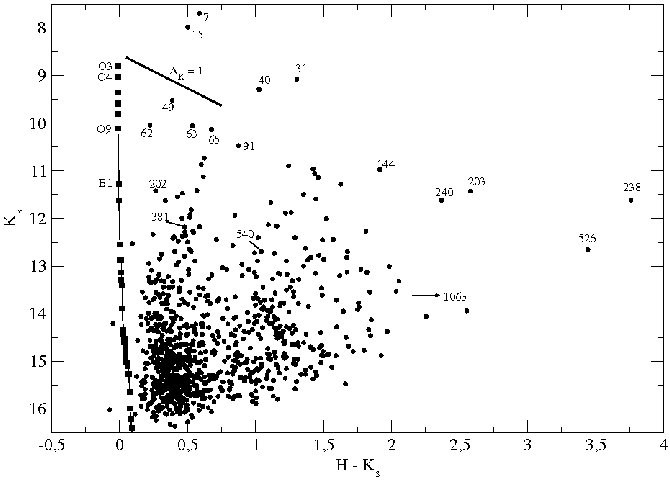}
\end{minipage}
\caption{{ H\,{\sc{ii}} region G48.9-0.3 (W51). Left: Color-color diagram (C-C). Right: Color-magnitude diagram (C-M). The brightest objects are brighter than the tip of the main sequence line, indicating a distance smaller than the kinematic results. Limiting magnitudes for this region (adopted for the objects not detected): $J$ = 17.0 and $H$ = 15.5 mag.}}
\label{fig:W51-CMD}
\end{figure*}

\clearpage

\begin{figure*}
\begin{minipage}[b]{0.47\linewidth}
\includegraphics[height=07.5cm,width=09cm]{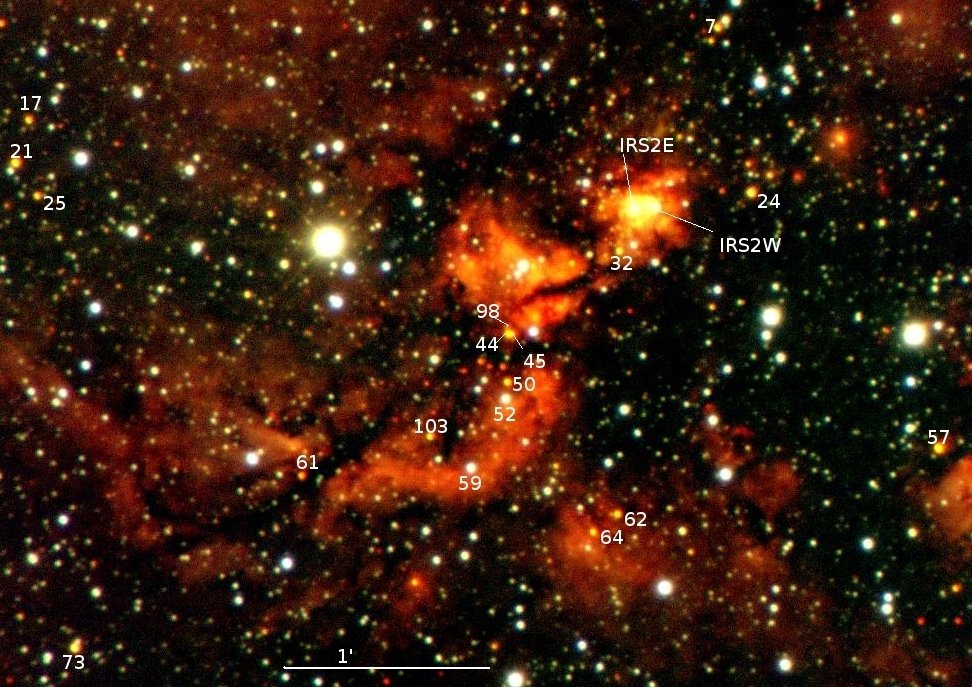}
\end{minipage} \hfill
\begin{minipage}[b]{0.47\linewidth}
\includegraphics[height=07.5cm,width=09cm]{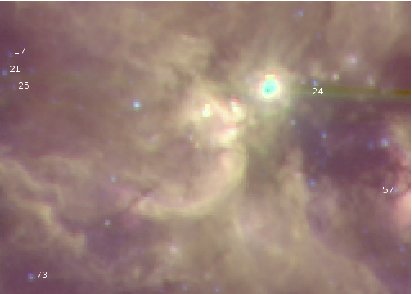}
\end{minipage}
\caption{{ Color images of G49.5-0.4 (W51A). Left: $JHK_{s}$ color image, reproduced from Figu\^eredo et al. (2005). Right: IRAC-{\it Spitzer} color image. In both images, the FOV is $\approx$ 4.0 $\times$ 3.5 arcmin.}}
\label{fig:W51A-color}
\end{figure*}

\begin{figure*}
\begin{minipage}[b]{0.48\linewidth}
\includegraphics[height=09cm,width=\linewidth]{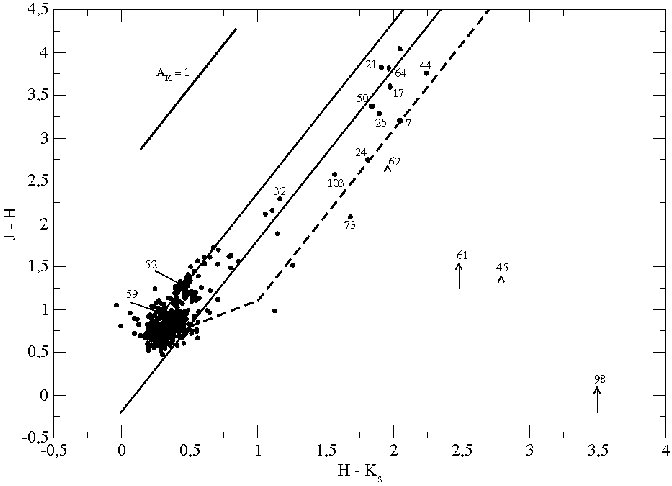}
\end{minipage} \hfill
\begin{minipage}[b]{0.49\linewidth}
\includegraphics[height=09cm,width=\linewidth]{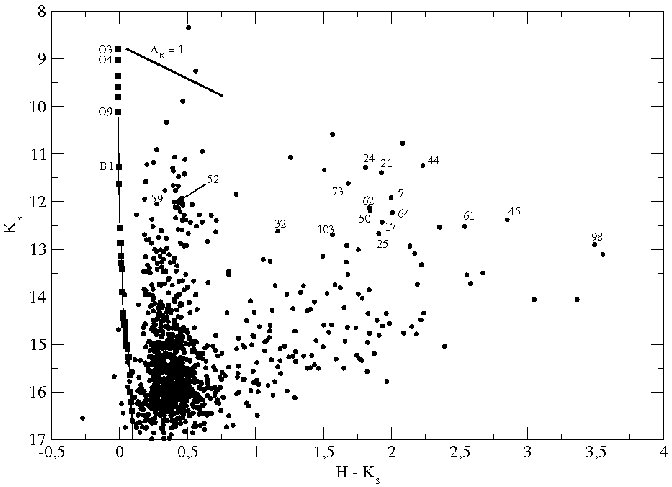}
\end{minipage}
\caption{{ H\,{\sc{ii}} region G49.5-0.4 (W51A). Left: Color-color diagram (C-C). Right: Color-magnitude diagram (C-M). The adopted kinematic distance leads to a main sequence whose tip is brighter than the stars of this region. There are some stars brighter than the tip of the main sequence, but they are foreground objects. The kinematic distance is 5.5 kpc \citep{Russeil03} which is similar to the value found by Barbosa et al. (2008) from an analyses of a $K-$band spectrum of a source associated with the UCH\,{sc{ii}} W51d; see text. The limiting magnitude in $J$-band is 16.5 mag.}}
\label{fig:W51A-CMD}
\end{figure*}


\begin{figure*}
\begin{minipage}[b]{0.47\linewidth}
\includegraphics[height=09cm,width=09cm]{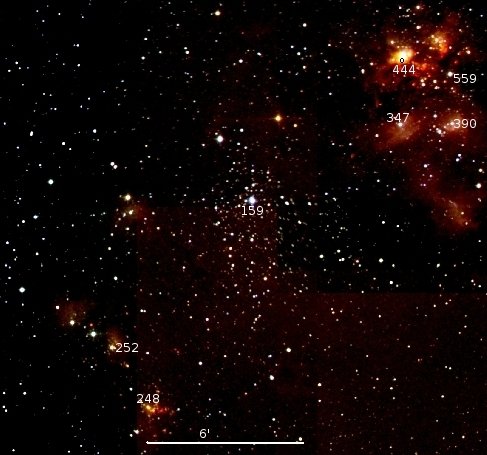}
\end{minipage} \hfill
\begin{minipage}[b]{0.47\linewidth}
\includegraphics[height=09cm,width=09cm]{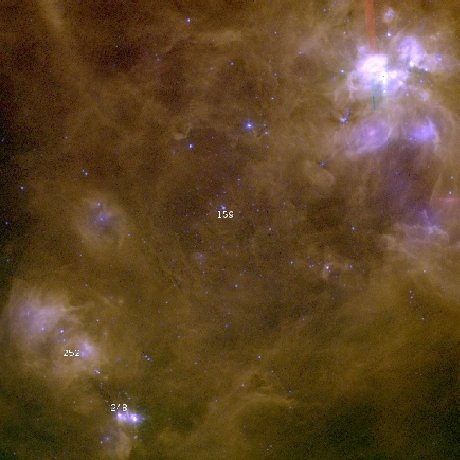}
\end{minipage}
\caption{{ Color images of G133.7+1.2 (W3). Left: $JHK_{s}$ 2MASS mosaic color image. Right: IRAC-{\it Spitzer} color image. In both images, the sizes are $\approx$ 18.0 arcmin on a side.}}
\label{fig:W3-color}
\end{figure*}

\begin{figure*}
\begin{minipage}[b]{0.48\linewidth}
\includegraphics[height=09cm,width=\linewidth]{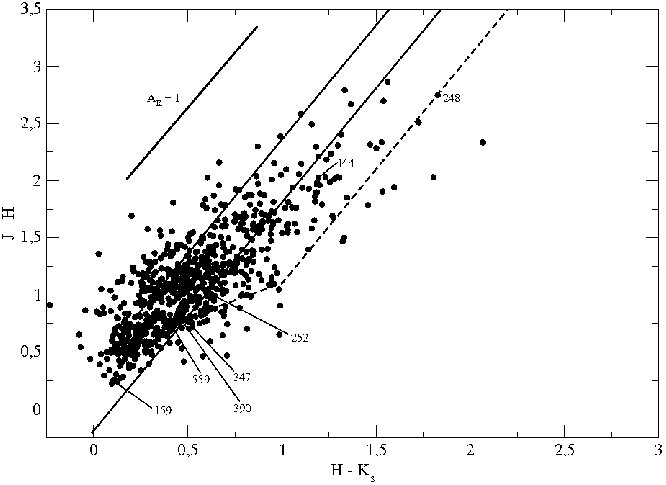}
\end{minipage} \hfill
\begin{minipage}[b]{0.49\linewidth}
\includegraphics[height=09cm,width=\linewidth]{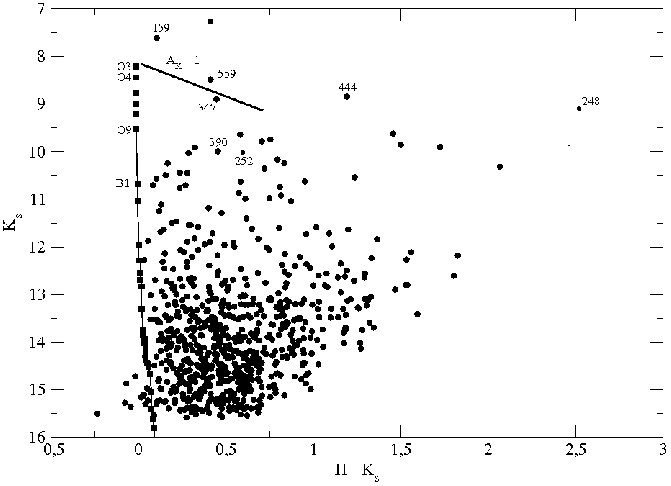}
\end{minipage}
\caption{{ H\,{\sc{ii}} region G133.7+1.2 (W3). Left: Color-color diagram (C-C). Right: Color-magnitude diagram (C-M). Here, we have used 2MASS photometric data to identify the O-type stars and derive a spectrophotometric distance of 1.85 kpc (Navarete et al. in preparation). This result (closer) is in agreement with that of \citet{Xu06}, who derived a trigonometric parallax distance of 1.95 kpc. The kinematic distance to this region is 4.2 kpc \citep{Russeil03}.}}
\label{fig:W3-CMD}
\end{figure*}


\begin{figure*}
\begin{minipage}[b]{0.47\linewidth}
\includegraphics[height=10cm,width=09cm]{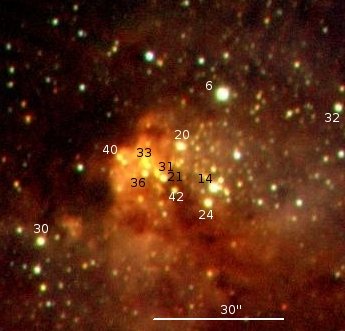}
\end{minipage} \hfill
\begin{minipage}[b]{0.47\linewidth}
\includegraphics[height=10cm,width=09cm]{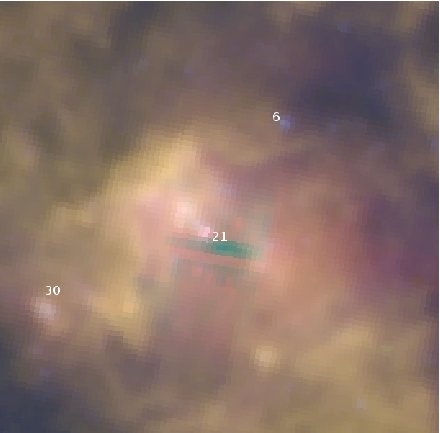}
\end{minipage}
\caption{{ Color images of G274.0-1.1 (RCW42). Left: $JHK_{s}$ color image. Right: IRAC-{\it Spitzer} color image. The size is $\approx$ 1.5 arcmin on a side. In the $JHK_{s}$ color image we can see a deeply embedded cluster of stars.}}
\label{fig:G274-color}
\end{figure*}

\begin{figure*}
\begin{minipage}[b]{0.48\linewidth}
\includegraphics[height=10cm,width=\linewidth]{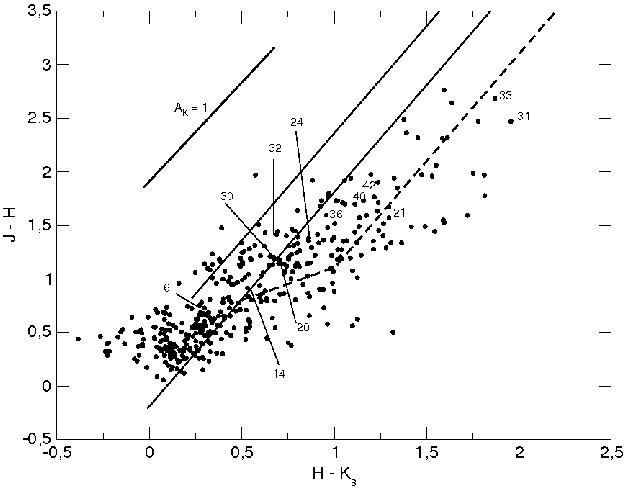}
\end{minipage} \hfill
\begin{minipage}[b]{0.49\linewidth}
\includegraphics[height=10cm,width=\linewidth]{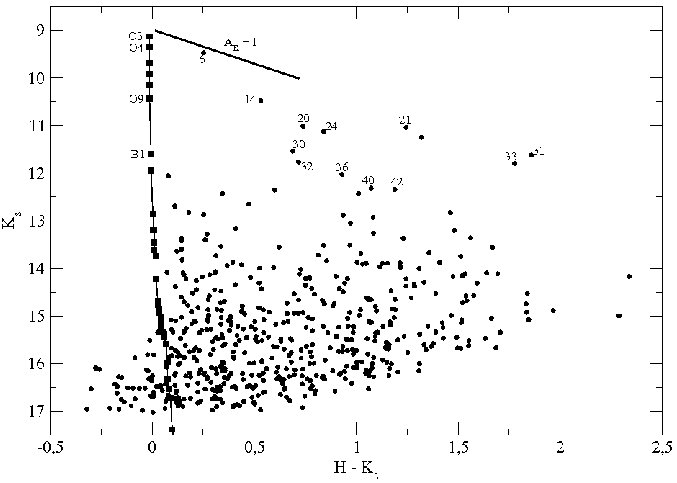}
\end{minipage}
\caption{{ H\,{\sc{ii}} region G274.0-1.1 (RCW42). Left: Color-color diagram (C-C). Right: Color-magnitude diagram (C-M). The C-M diagram appears consistent with the adopted kinematic distance (agreement).}}
\label{fig:G274-CMD}
\end{figure*}


\begin{figure*}
\begin{minipage}[b]{0.47\linewidth}
\includegraphics[height=10cm,width=\linewidth]{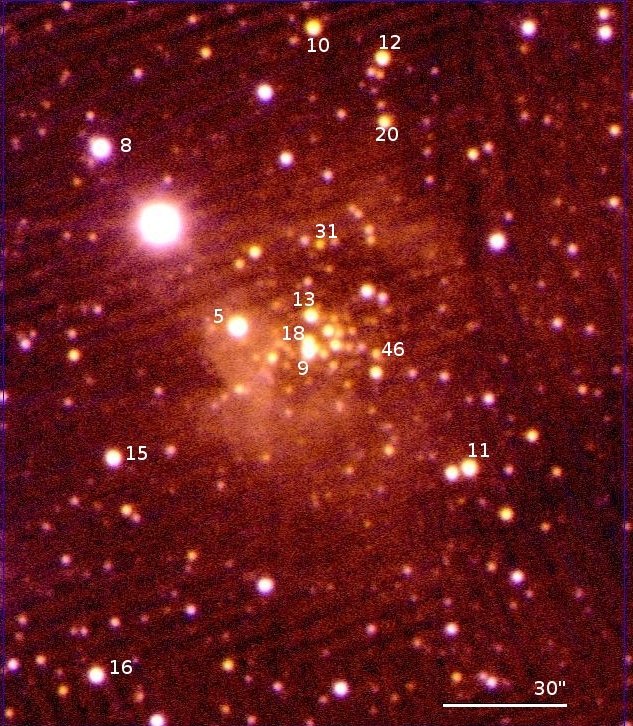}
\end{minipage} \hfill
\begin{minipage}[b]{0.46\linewidth}
\includegraphics[height=10cm,width=\linewidth]{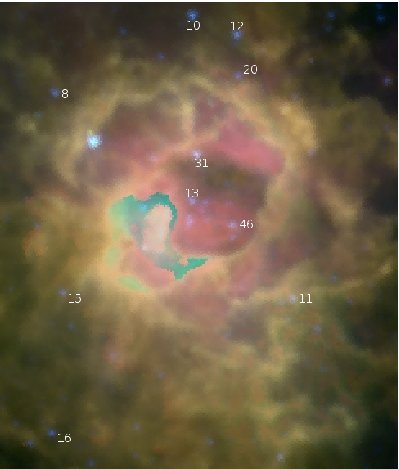}
\end{minipage}
\caption{{ Color images of G282.0-1.2 (RCW46). Left: $JHK_{s}$ color image. Right: IRAC-{\it Spitzer} color image. In both images, the size is $\approx$ 3.0 arcmin on a side.}}
\label{fig:G282-color}
\end{figure*}

\begin{figure*}
\begin{minipage}[b]{0.48\linewidth}
\includegraphics[height=10cm,width=\linewidth]{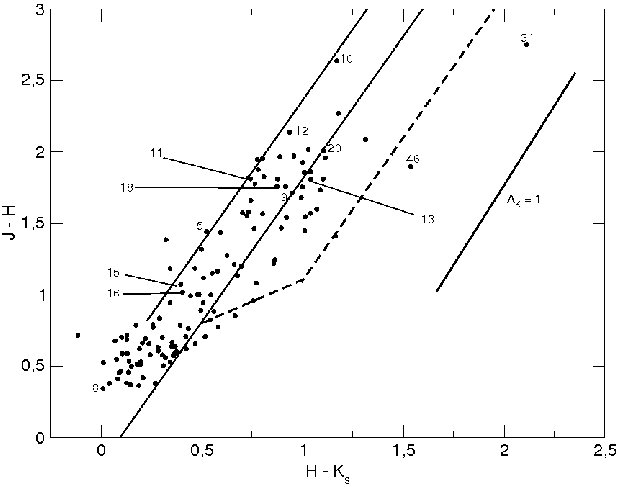}
\end{minipage} \hfill
\begin{minipage}[b]{0.49\linewidth}
\includegraphics[height=10cm,width=\linewidth]{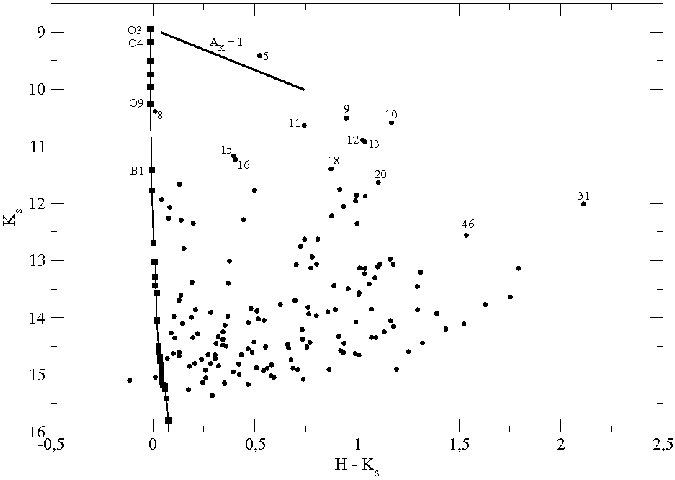}
\end{minipage}
\caption{{ H\,{\sc{ii}} region G282.0-1.2 (RCW46). Left: Color-color diagram (C-C). Right: Color-magnitude diagram (C-M). The tip of the main sequence line is brighter than the objects of this region, which indicates this adopted kinematic distance may be correct (agreement).}}
\label{fig:G282-CMD}
\end{figure*}


\begin{figure*}
\begin{minipage}[b]{0.47\linewidth}
\includegraphics[height=09cm,width=09cm]{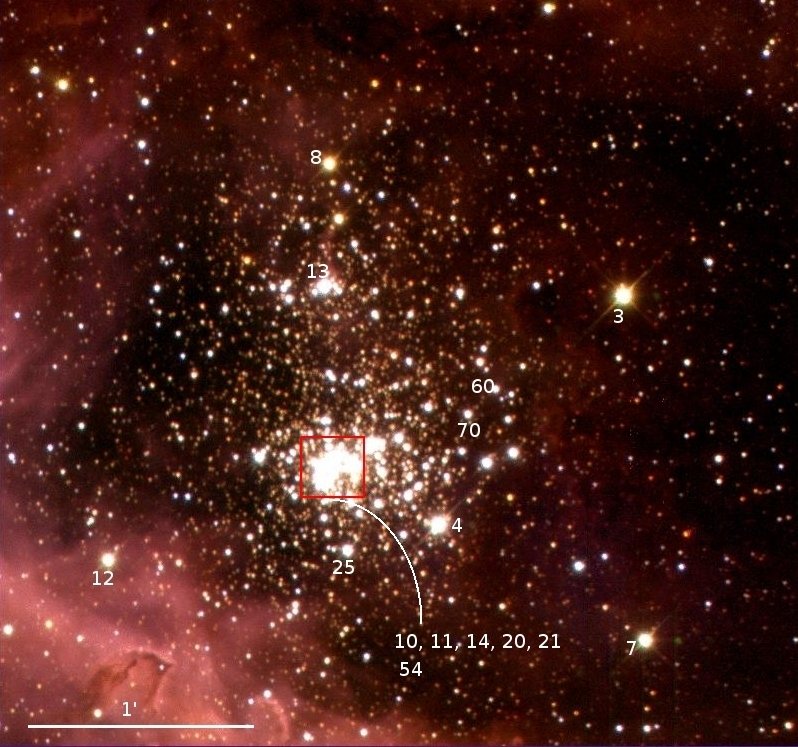}
\end{minipage} \hfill
\begin{minipage}[b]{0.47\linewidth}
\includegraphics[height=09cm,width=09cm]{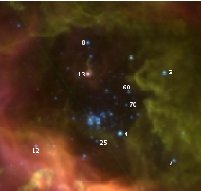}
\end{minipage}
\caption{{ Color images of G284.3-0.3 (NGC3247). Left: $JHK_{s}$ color image. Right: IRAC-{\it Spitzer} color image. In both images, the size is $\approx$ 3.3 arcmin on a side. Some of the objects in the crowded cluster (red box) are indicated in the near infrared image.}}
\label{fig:G284-color}
\end{figure*}

\begin{figure*}
\begin{minipage}[b]{0.48\linewidth}
\includegraphics[height=10cm,width=\linewidth]{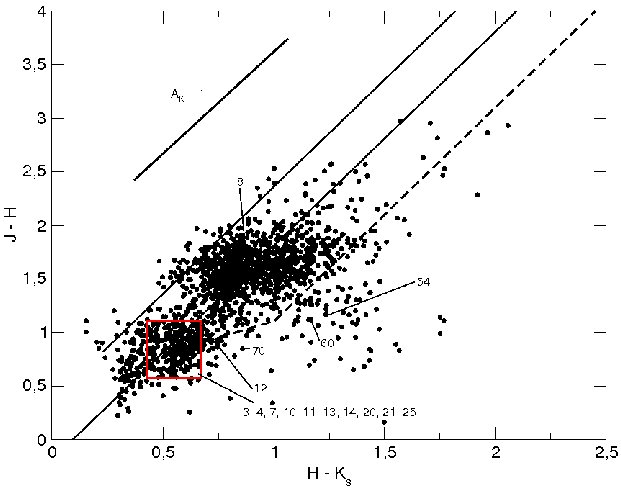}
\end{minipage} \hfill
\begin{minipage}[b]{0.49\linewidth}
\includegraphics[height=10cm,width=\linewidth]{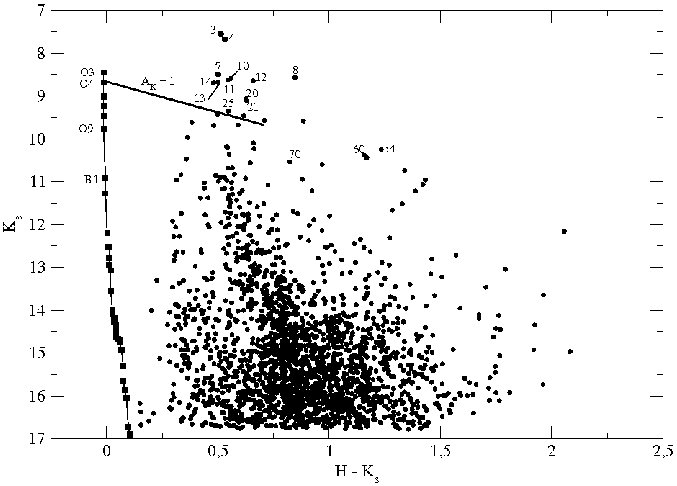}
\end{minipage}
\caption{{ H\,{\sc{ii}} region G284.3-0.3 (NGC3247). Left: Color-color diagram (C-C). Right: Color-magnitude diagram (C-M). The cluster members are around $H - K_{s}$ = 0.6. Our photometry suggests this region may be closer than the adopted distance of 4.7 kpc \citep{Russeil03}.}}
\label{fig:G284-CMD}
\end{figure*}

\clearpage

\begin{figure*}
\begin{minipage}[b]{0.47\linewidth}
\includegraphics[height=09cm,width=09cm]{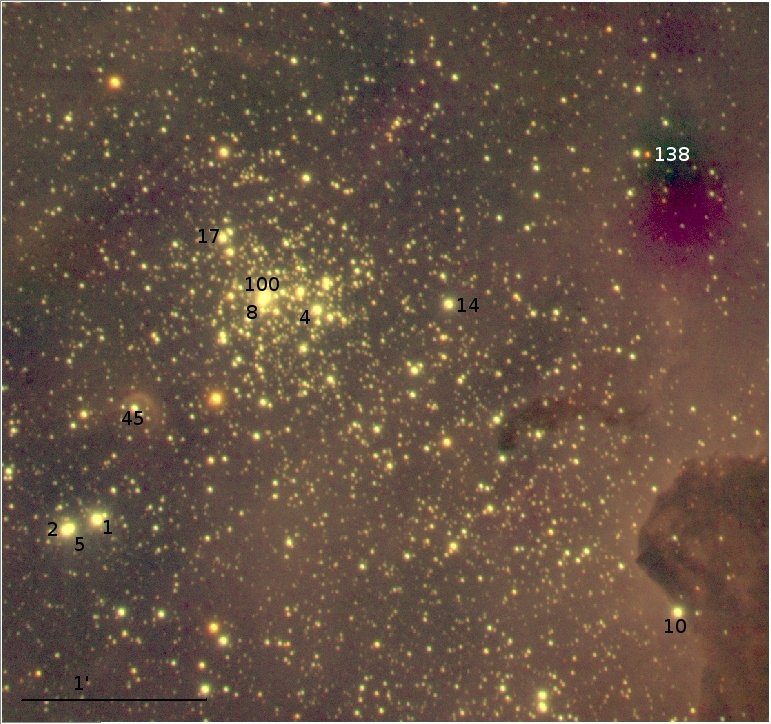}
\end{minipage} \hfill
\begin{minipage}[b]{0.47\linewidth}
\includegraphics[height=09cm,width=09cm]{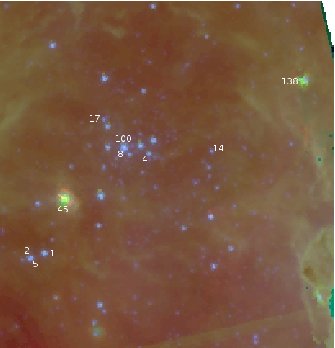}
\end{minipage}
\caption{{ Color images of G287.4-0.6 (NGC3372). Left: $JHK_{s}$ color image. Right: IRAC-{\it Spitzer} color image. In both images, the size is $\approx$ 4.0 arcmin on a side.}}
\label{fig:G287-color}
\end{figure*}

\begin{figure*}
\begin{minipage}[b]{0.48\linewidth}
\includegraphics[height=10cm,width=\linewidth]{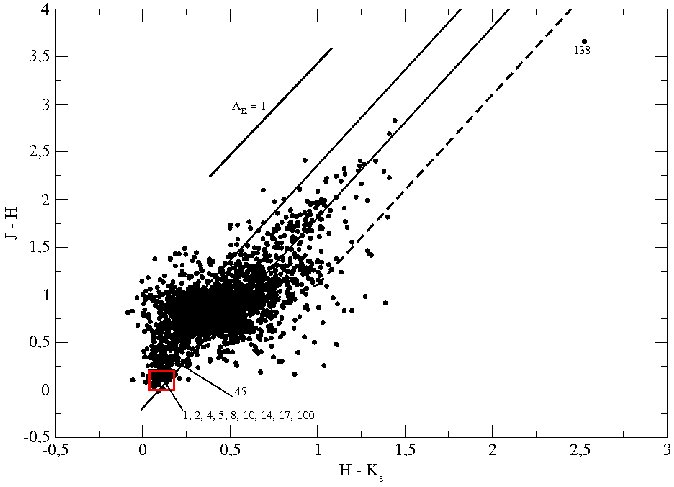}
\end{minipage} \hfill
\begin{minipage}[b]{0.49\linewidth}
\includegraphics[height=10cm,width=\linewidth]{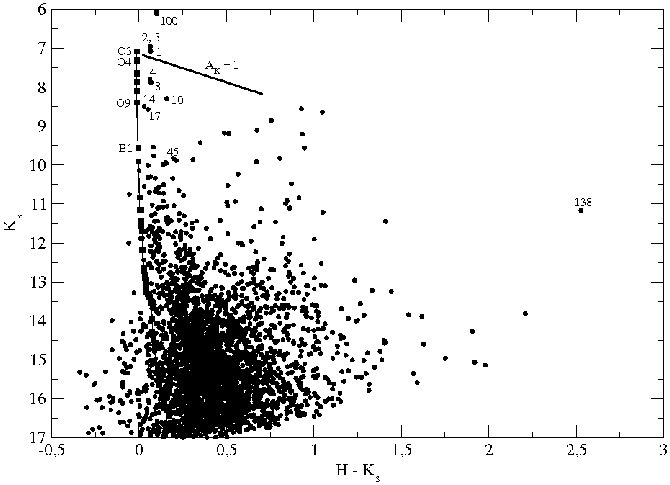}
\end{minipage}
\caption{{ H\,{\sc{ii}} region G287.4-0.6 (NGC3372). Left: Color-color diagram (C-C). Right: Color-magnitude diagram (C-M). The kinematic distance leads to a main sequence which is fainter than expected based on the photometry, closer distance.}}
\label{fig:G287-CMD}
\end{figure*}


\begin{figure*}
\begin{minipage}[b]{0.47\linewidth}
\includegraphics[height=09cm,width=09cm]{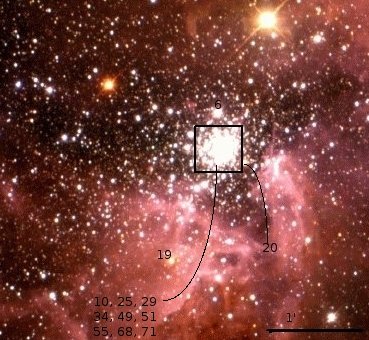}
\end{minipage} \hfill
\begin{minipage}[b]{0.47\linewidth}
\includegraphics[height=09cm,width=09cm]{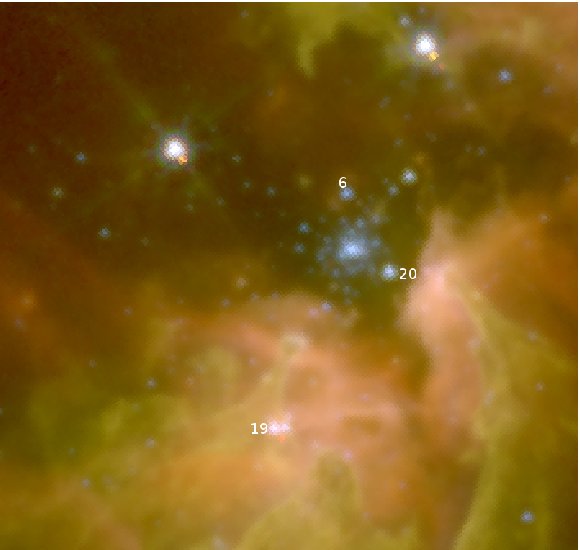}
\end{minipage}
\caption{{ Color images of G291.6-0.5 (NGC3603). Left: $JHK_{s}$ color image, except objects \#6, \#19 and \#20, the objects numbered in the diagrams are inside the black box. Right: IRAC-{\it Spitzer} image. In both images, the FOV is $\approx$ 4.0 $\times$ 5.0 arcmin.}}
\label{fig:NGC3603-color}
\end{figure*}

\begin{figure*}
\begin{minipage}[b]{0.48\linewidth}
\includegraphics[height=10cm,width=\linewidth]{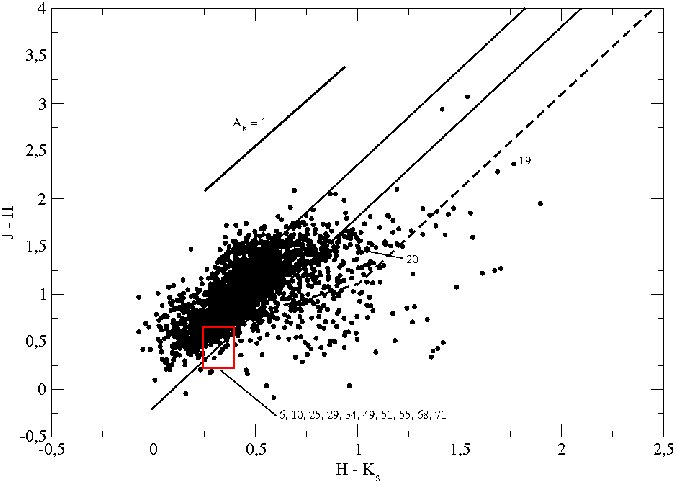}
\end{minipage} \hfill
\begin{minipage}[b]{0.49\linewidth}
\includegraphics[height=10cm,width=\linewidth]{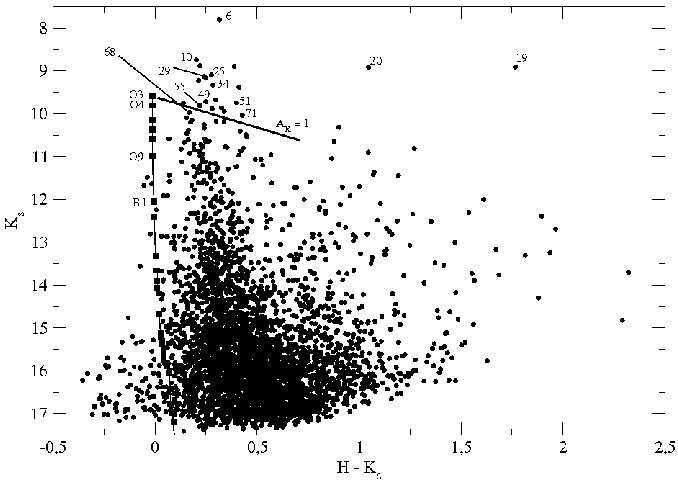}
\end{minipage}
\caption{{ H\,{\sc{ii}} region G291.6-0.5 (NGC3603). Left: Color-color diagram (C-C). Right: Color-magnitude diagram (C-M). In the center of the $JHK_{s}$ color image, we see a crowded cluster of stars. The members of this cluster are indicated in the black box in the $JHK_{s}$ image and in the red box of the C-C diagram. The disagreement between the photometry and the tip of the main sequence line is clear, this region is closer than the kinematic results. }}
\label{fig:NGC3603-CMD}
\end{figure*}

\clearpage


\begin{figure*}
\begin{minipage}[b]{0.47\linewidth}
\includegraphics[height=10cm,width=\linewidth]{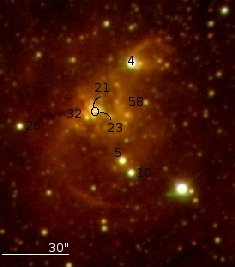}
\end{minipage} \hfill
\begin{minipage}[b]{0.47\linewidth}
\includegraphics[height=10cm,width=\linewidth]{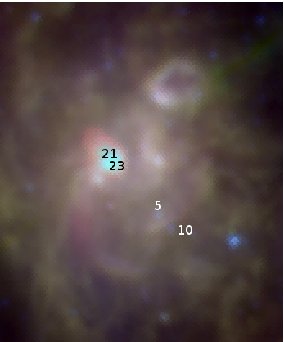}
\end{minipage}
\caption{{ Color images of G298.2-0.3. Left: $JHK_{s}$ color image. Right: IRAC-{\it Spitzer} image. In both images, the bottom size is $\approx$ 2.0 arcmin. The photometry suggests this region might be closer than the adopted distance of 7.9 kpc \citep{Russeil03}.}}
\label{fig:G298-2-color}
\end{figure*}

\begin{figure*}
\begin{minipage}[b]{0.48\linewidth}
\includegraphics[height=10cm,width=\linewidth]{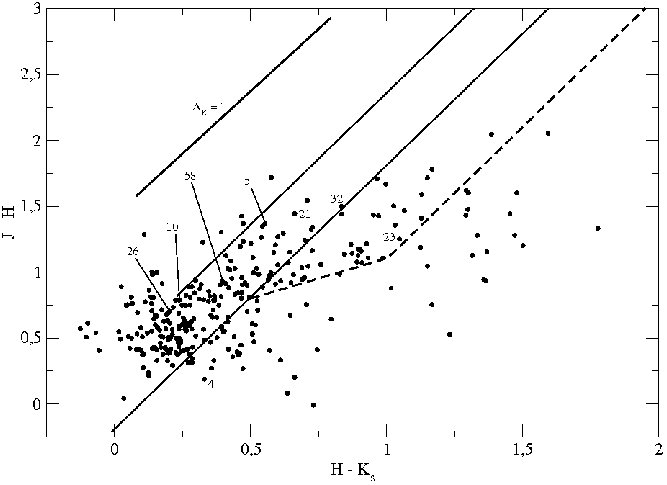}
\end{minipage} \hfill
\begin{minipage}[b]{0.49\linewidth}
\includegraphics[height=10cm,width=\linewidth]{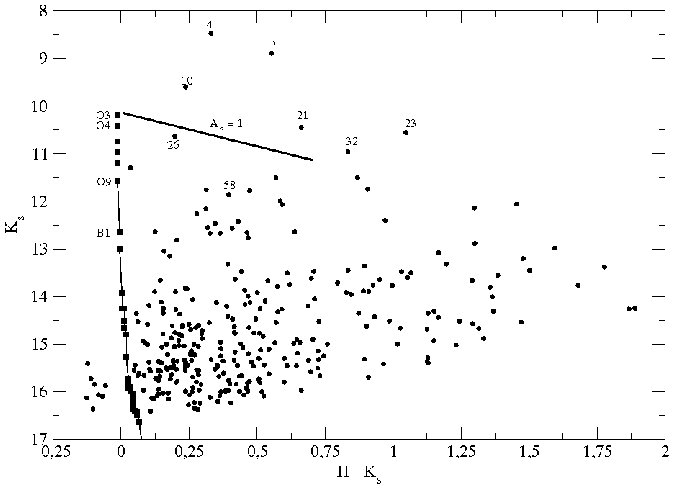}
\end{minipage}
\caption{{ H\,{\sc{ii}} region G298.2-0.3. Left: Color-color diagram (C-C). Right: Color-magnitude diagrama (C-M). Since there is a few objects detected in this region, the distance analyses is inconclusive.}}
\label{fig:G298-2-CMD}
\end{figure*}

\clearpage

\begin{figure*}
\begin{minipage}[b]{0.47\linewidth}
\includegraphics[height=09cm,width=\linewidth]{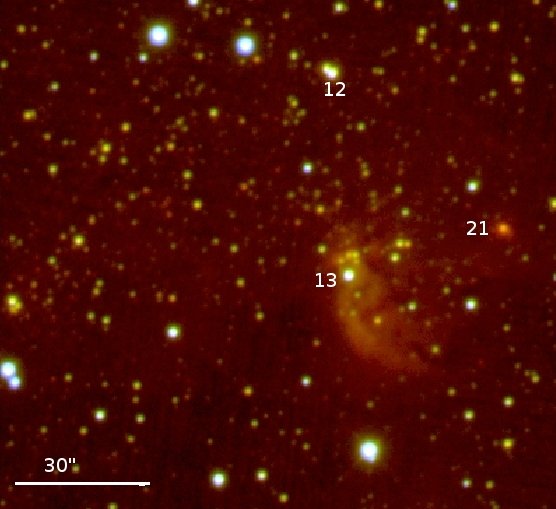}
\end{minipage} \hfill
\begin{minipage}[b]{0.47\linewidth}
\includegraphics[height=09cm,width=\linewidth]{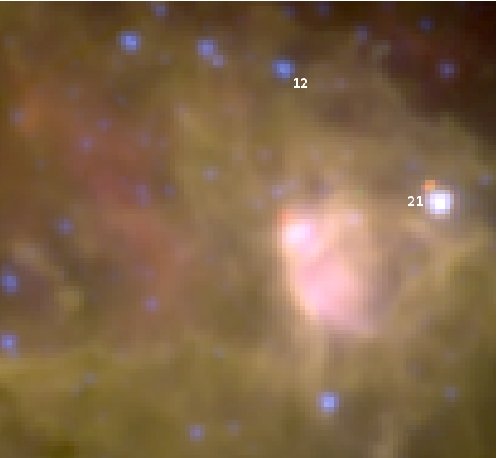}
\end{minipage}
\caption{{ Color images of G298.9-0.4. Left: $JHK_{s}$ color image. Right: IRAC-{\it Spitzer} image. In both images, the bottom edge is $\approx$ 2.0 arcmin.}}
\label{fig:G298-color}
\end{figure*}

\begin{figure*}
\begin{minipage}[b]{0.48\linewidth}
\includegraphics[height=10cm,width=\linewidth]{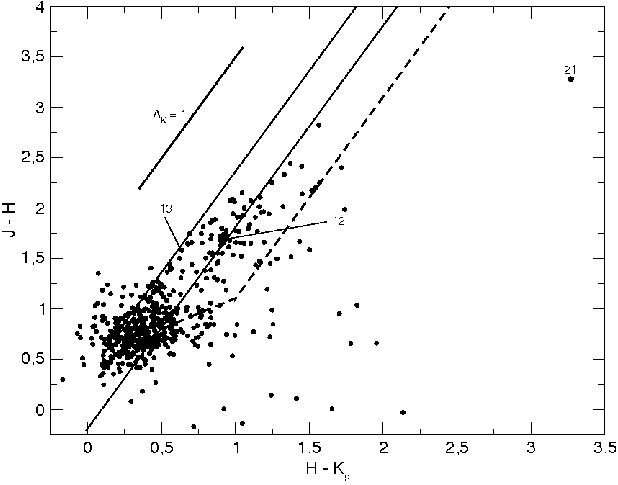}
\end{minipage} \hfill
\begin{minipage}[b]{0.49\linewidth}
\includegraphics[height=10cm,width=\linewidth]{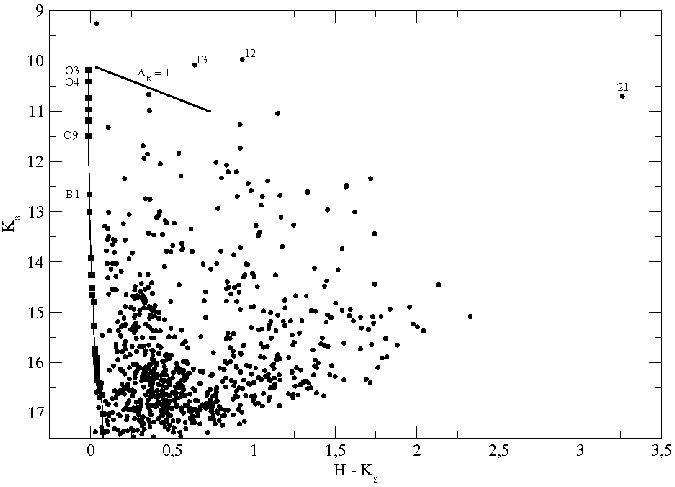}
\end{minipage}
\caption{{ H\,{\sc{ii}} region G298.9-0.4. Left: Color-color diagram (C-C). Right: Color-magnitude diagram (C-M). 
Since there is not a detected star cluster in this region, the distance analyses is inconclusive.}}
\label{fig:G298-CMD}
\end{figure*}


\begin{figure*}
\begin{minipage}[b]{0.47\linewidth}
\includegraphics[height=08.5cm,width=09cm]{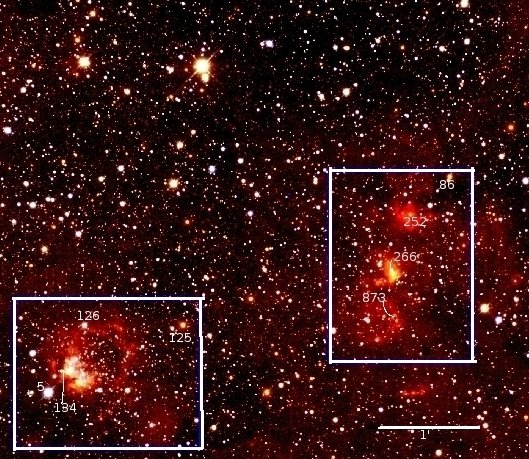}
\end{minipage} \hfill
\begin{minipage}[b]{0.47\linewidth}
\includegraphics[height=08.5cm,width=09cm]{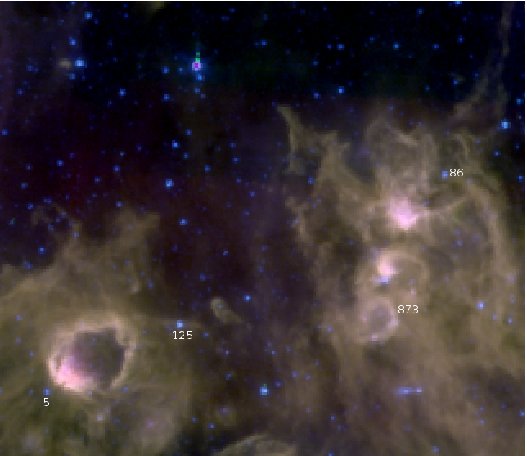}
\end{minipage}
\caption{{ Color images of G305.2+0.0. Left: $JHK_{s}$ color image. Right: IRAC-{\it Spitzer} image, the {\it Spitzer} image is saturated in some regions (the blue zones). In both images, each side has $\approx$ 5.5 arcmin.}}
\label{fig:G305-0-color}
\end{figure*}

\begin{figure*}
\begin{minipage}[b]{0.48\linewidth}
\includegraphics[height=10cm,width=\linewidth]{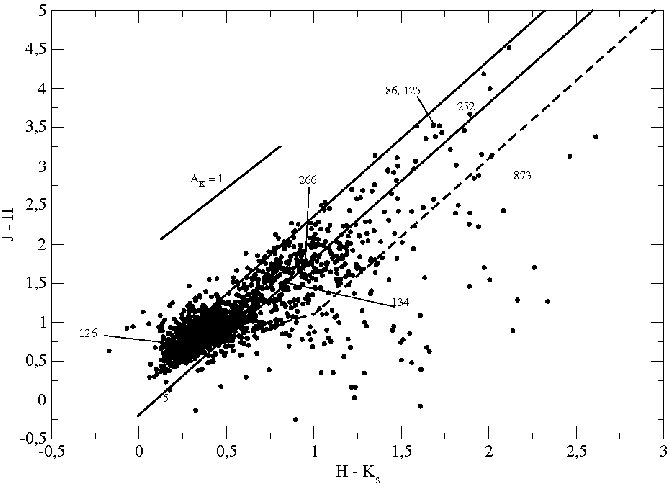}
\end{minipage} \hfill
\begin{minipage}[b]{0.49\linewidth}
\includegraphics[height=10cm,width=\linewidth]{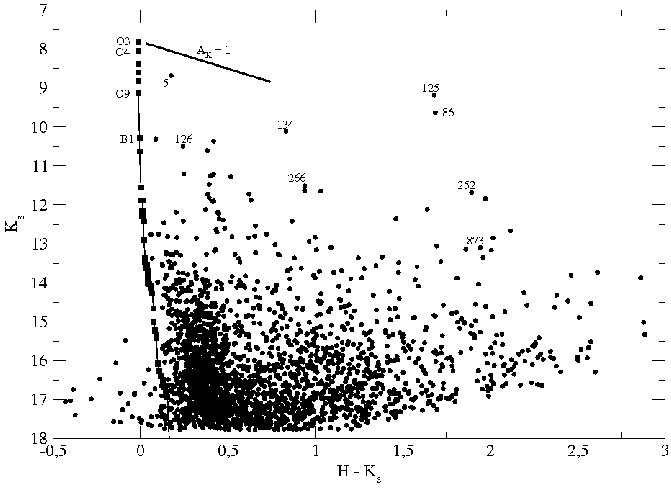}
\end{minipage}
\caption{{ H\,{\sc{ii}} region G305.2+0.0. Left: Color-color diagram (C-C). Right: Color-magnitude diagram (C-M). The photometry was taken from the regions indicated with the two white rectangles in the $JHK_{s}$ color image. The adopted kinematic distance of 3.5 kpc \citep{Russeil03} seems to be consistent with the photometry, but the small quantity of detected stars makes this distance analysis inconclusive.}}
\label{fig:G305-0-CMD}
\end{figure*}


\begin{figure*}
\begin{minipage}[b]{0.47\linewidth}
\includegraphics[height=10cm,width=09cm]{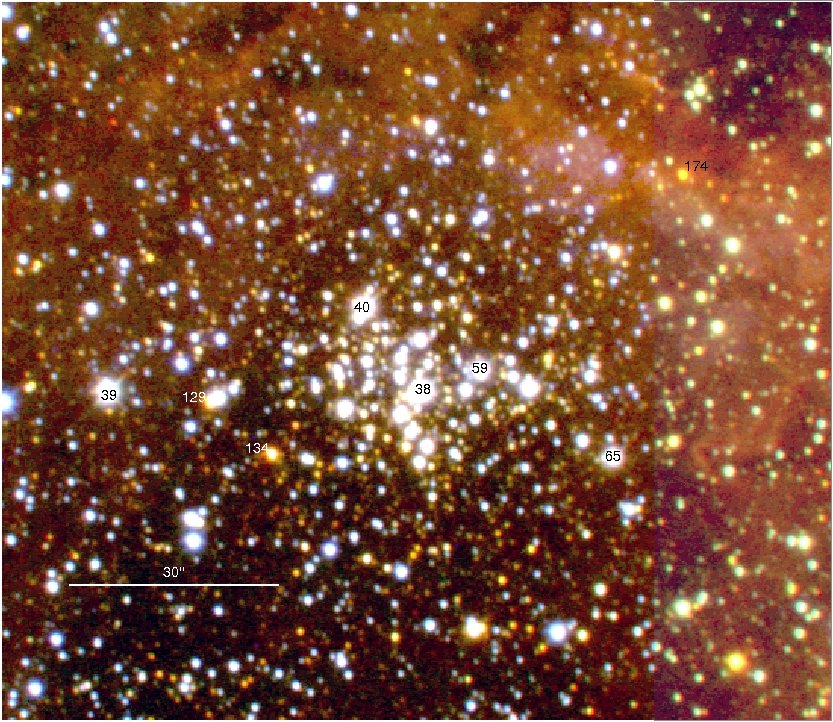}
\end{minipage} \hfill
\begin{minipage}[b]{0.47\linewidth}
\includegraphics[height=10cm,width=09cm]{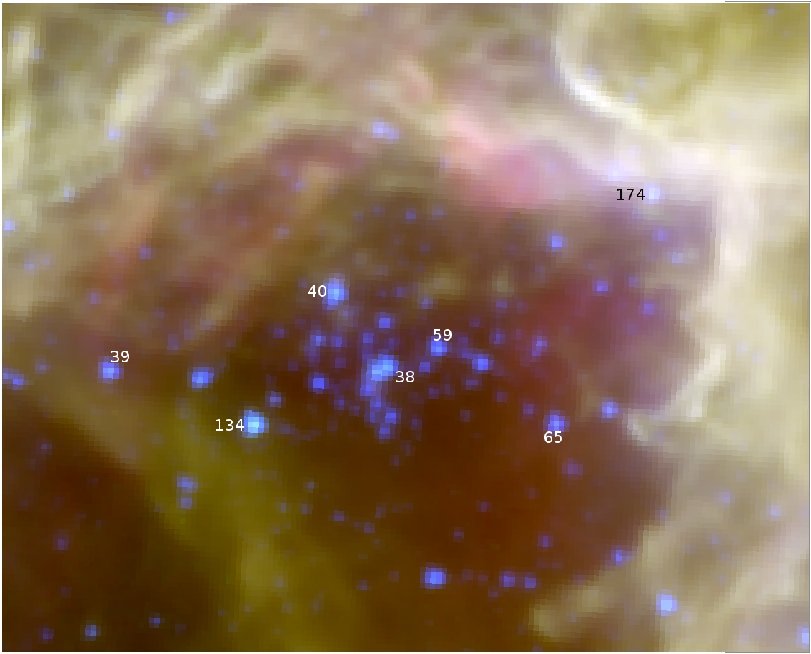}
\end{minipage}
\caption{{ Color images of G305.2+0.2. Left: $JHK_{s}$ color image. Right: IRAC-{\it Spitzer} image. In both images, each side has size of $\approx$ 2.0 arcmin.}}
\label{fig:G305-color}
\end{figure*}

\begin{figure*}
\begin{minipage}[b]{0.48\linewidth}
\includegraphics[height=10cm,width=\linewidth]{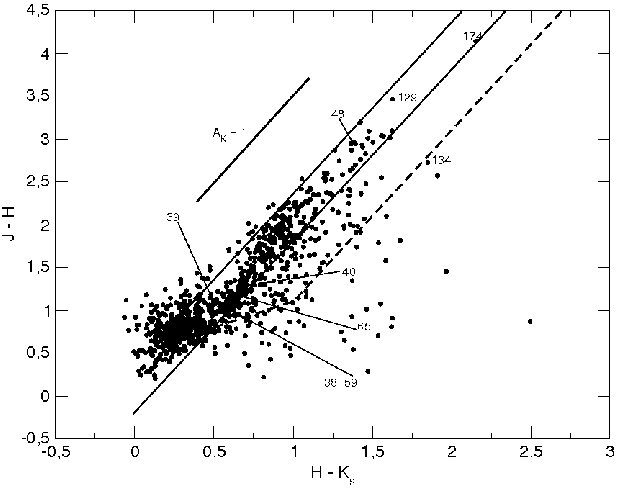}
\end{minipage} \hfill
\begin{minipage}[b]{0.49\linewidth}
\includegraphics[height=10cm,width=\linewidth]{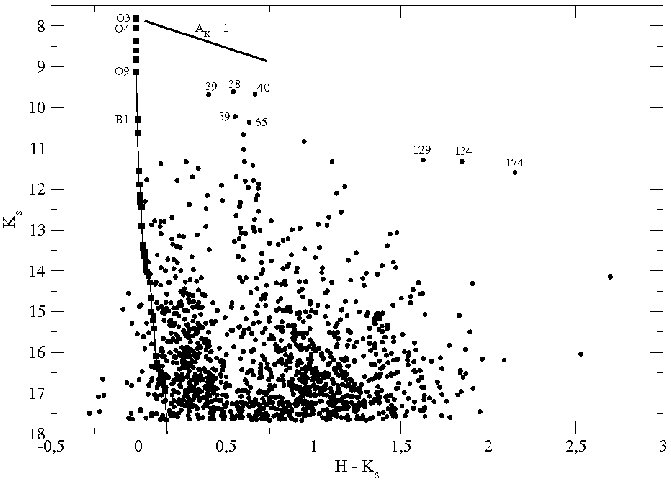}
\end{minipage}
\caption{{ H\,{\sc{ii}} region G305.2+0.2. Left: Color-color diagram (C-C). Right: Color-magnitude diagram (C-M). The adopted kinematic distance of 3.5 kpc \citep{Russeil03} appears consistent with the photometry.}}
\label{fig:G305-CMD}
\end{figure*}


\begin{figure*}
\begin{minipage}[b]{0.47\linewidth}
\includegraphics[height=10cm,width=\linewidth]{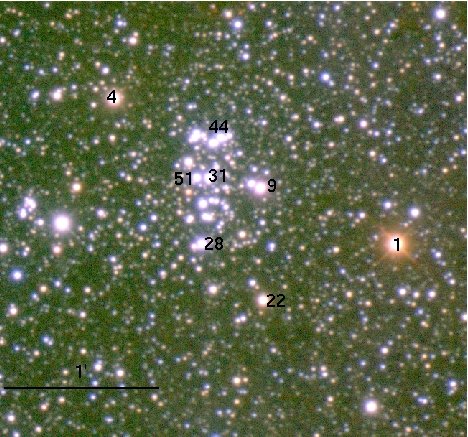}
\end{minipage} \hfill
\begin{minipage}[b]{0.47\linewidth}
\includegraphics[height=10cm,width=\linewidth]{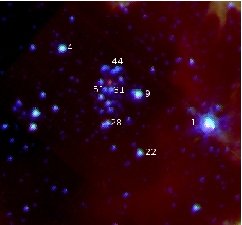}
\end{minipage}
\caption{{ Color images of G308.7+0.6. Left: $JHK_{s}$ color image. Right: IRAC-{\it Spitzer} image. In both images, the size is $\approx$ 2.5 arcmin on a side. In this small field we do not see nebulosity, but in the larger field $\approx$ 8 arcmin given by the ISPI image (not shown), weak nebulosity surrounds the star cluster.}}
\label{fig:G308-color}
\end{figure*}

\begin{figure*}
\begin{minipage}[b]{0.48\linewidth}
\includegraphics[height=10cm,width=\linewidth]{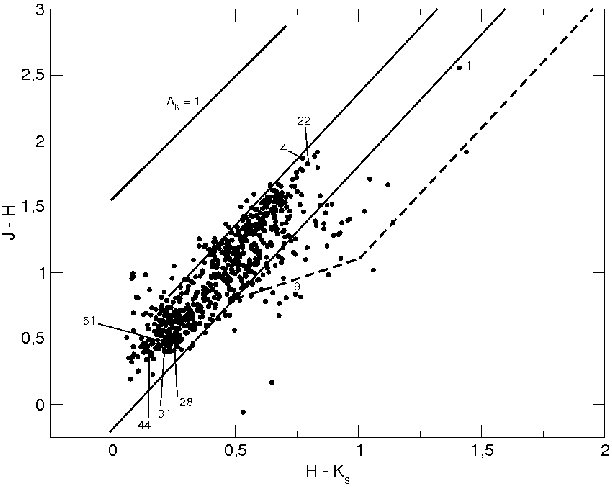}
\end{minipage} \hfill
\begin{minipage}[b]{0.49\linewidth}
\includegraphics[height=10cm,width=\linewidth]{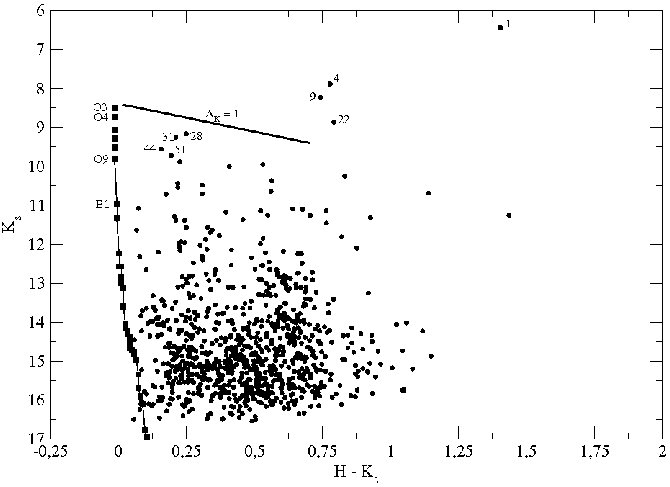}
\end{minipage}
\caption{{ H\,{\sc{ii}} region G308.7+0.6. Left: Color-color diagram (C-C). Right: Color-magnitude diagrama (C-M). The main sequence seems to be correct.}}
\label{fig:G308-CMD}
\end{figure*}


\begin{figure*}
\begin{minipage}[b]{0.47\linewidth}
\includegraphics[height=08cm,width=09cm]{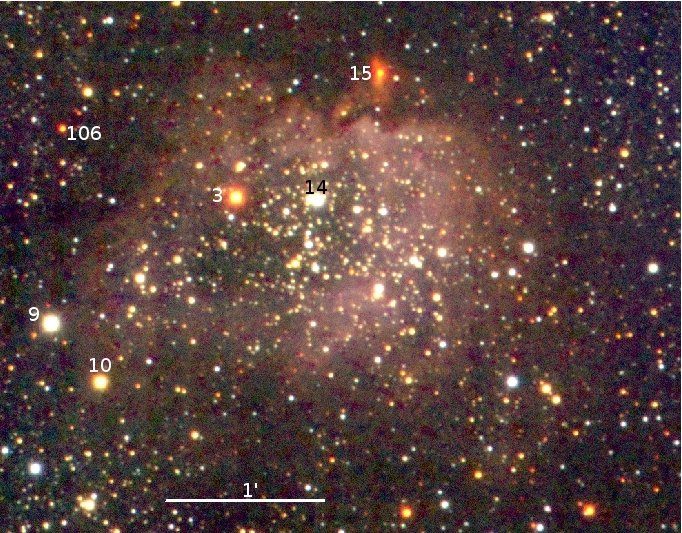}
\end{minipage} \hfill
\begin{minipage}[b]{0.47\linewidth}
\includegraphics[height=08cm,width=09cm]{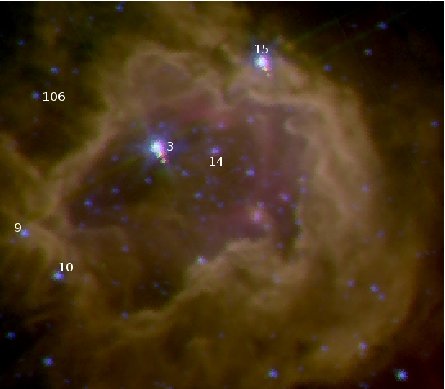}
\end{minipage}
\caption{{ Color images of G320.1+0.8 (RCW87). Left: $JHK_{s}$ color image. Right: IRAC-{\it Spitzer} image. In both images, the size is $\approx$ 4.0 arcmin on a side.}}
\label{fig:G320-color}
\end{figure*}

\vspace{4cm}

\begin{figure*}
\begin{minipage}[b]{0.48\linewidth}
\includegraphics[height=09cm,width=\linewidth]{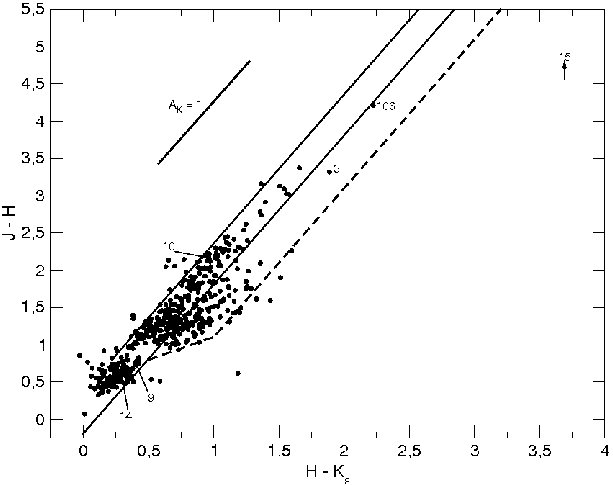}
\end{minipage} \hfill
\begin{minipage}[b]{0.49\linewidth}
\includegraphics[height=09cm,width=\linewidth]{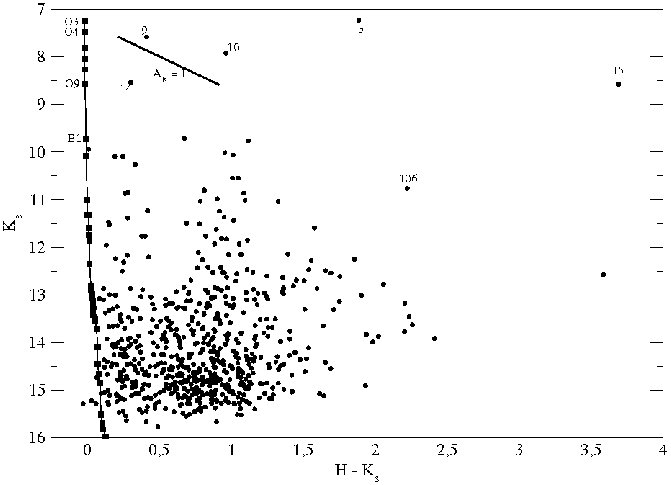}
\end{minipage}
\caption{{ H\,{\sc{ii}} region G320.1+0.8 (RCW87). Left: Color-color diagram (C-C). Right: Color-magnitude diagram (C-M). A very embedded H\,{\sc{ii}} region with strong nebulosity. The magnitude limit in the $J$-band is 17.0 mag. The brightest stars of the cluster are fainter than the tip of the main sequence at the adopted kinematic distance \citep[$d$ = 2.7 kpc][]{Russeil03}. But, due to the difficulty with the membership identification, the distance analysis is inconclusive.}}
\label{fig:G320-CMD}
\end{figure*}


\begin{figure*}
\begin{minipage}[b]{0.47\linewidth}
\includegraphics[height=08.5cm,width=09cm]{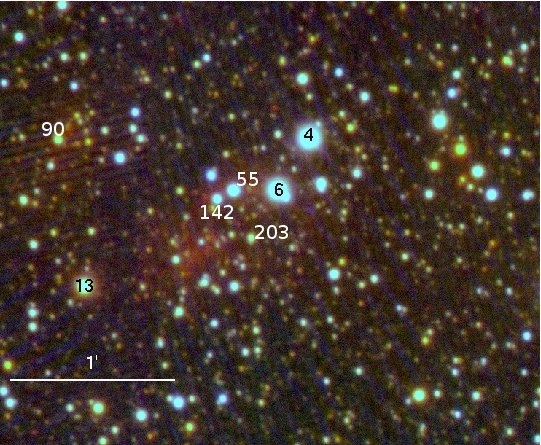}
\end{minipage} \hfill
\begin{minipage}[b]{0.47\linewidth}
\includegraphics[height=08.5cm,width=09cm]{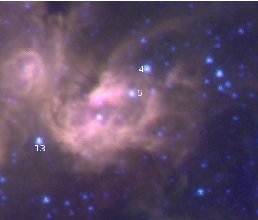}
\end{minipage}
\caption{{Color images of G320.3-0.2. Left: $JHK_{s}$ color image. Right: IRAC-{\it Spitzer} image. In both images, the size is $\approx$ 3.0 arcmin on a side.}}
\label{fig:G320-2-color}
\end{figure*}

\begin{figure*}
\begin{minipage}[b]{0.48\linewidth}
\includegraphics[height=09.5cm,width=\linewidth]{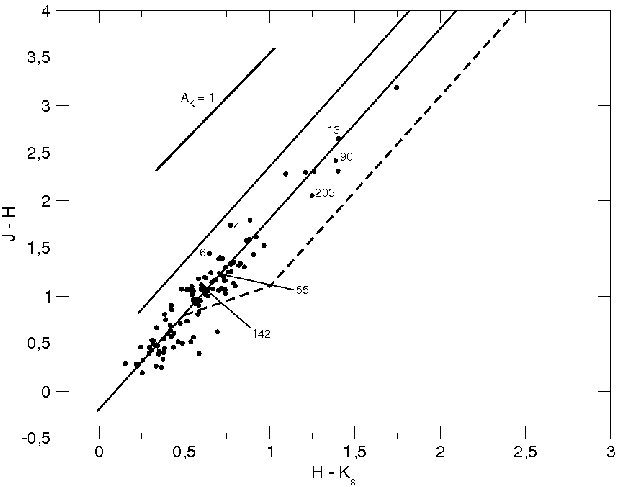}
\end{minipage} \hfill
\begin{minipage}[b]{0.49\linewidth}
\includegraphics[height=09.5cm,width=\linewidth]{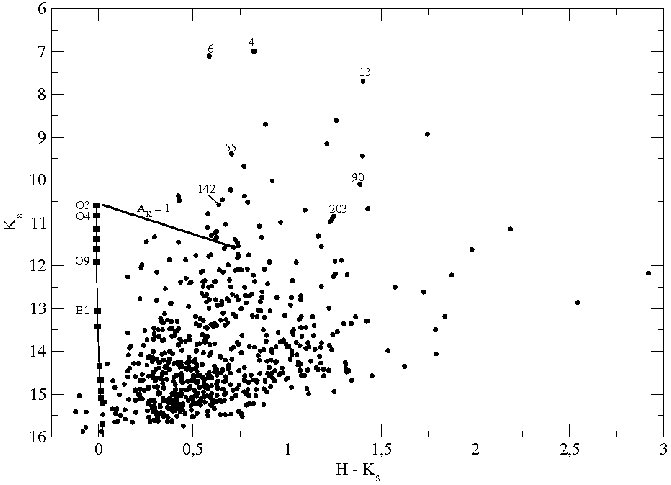}
\end{minipage}
\caption{{ H\,{\sc{ii}} region G320.3-0.2. Left: Color-color diagram (C-C). Right: Color-magnitude diagram (C-M). An evolved H\,{\sc{ii}} region, but with nebulosity in the {\it Spitzer} image. Due to the absence of a star cluster, the distance analysis is inconclusive.}}
\label{fig:G320-2-CMD}
\end{figure*}


\begin{figure*}
\begin{minipage}[b]{0.47\linewidth}
\includegraphics[height=10cm,width=\linewidth]{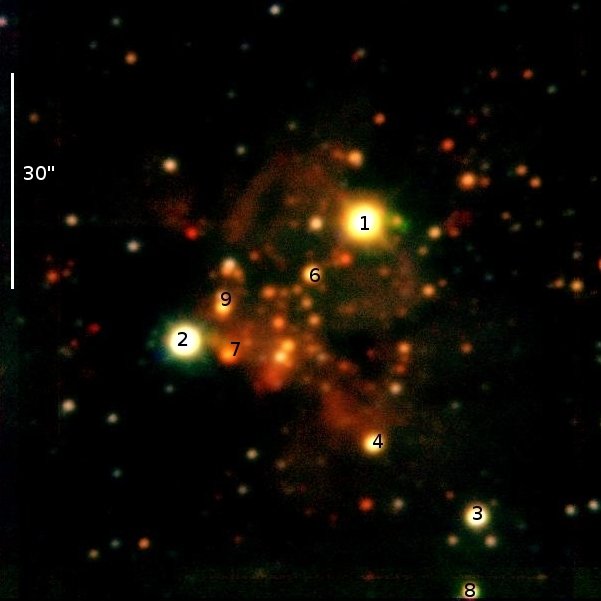}
\end{minipage} \hfill
\begin{minipage}[b]{0.47\linewidth}
\includegraphics[height=10cm,width=\linewidth]{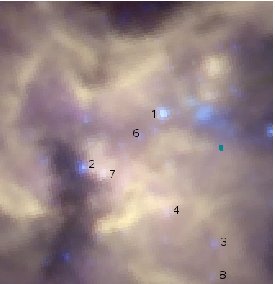}
\end{minipage}
\caption{{ Color images of G322.2+0.6 (RCW92). Left: $JHK_{s}$ is red. Right: IRAC-{\it Spitzer} image. In both images, the FOV is $\approx$ 1.0 $\times$ 1.5 arcmin.}}
\label{fig:G322-color}
\end{figure*}

\begin{figure*}
\begin{minipage}[b]{0.48\linewidth}
\includegraphics[height=10cm,width=\linewidth]{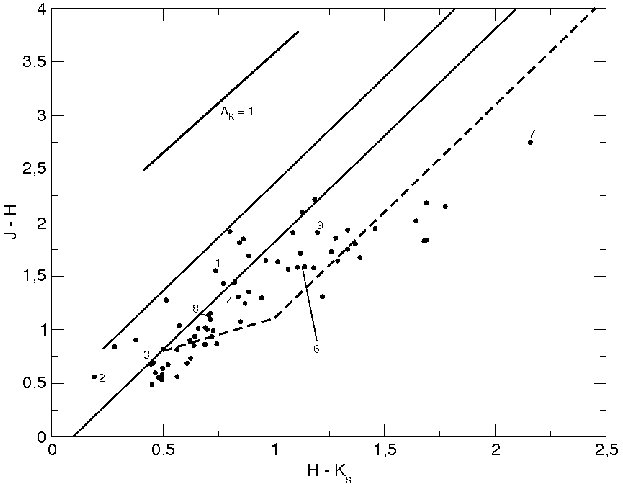}
\end{minipage} \hfill
\begin{minipage}[b]{0.49\linewidth}
\includegraphics[height=10cm,width=\linewidth]{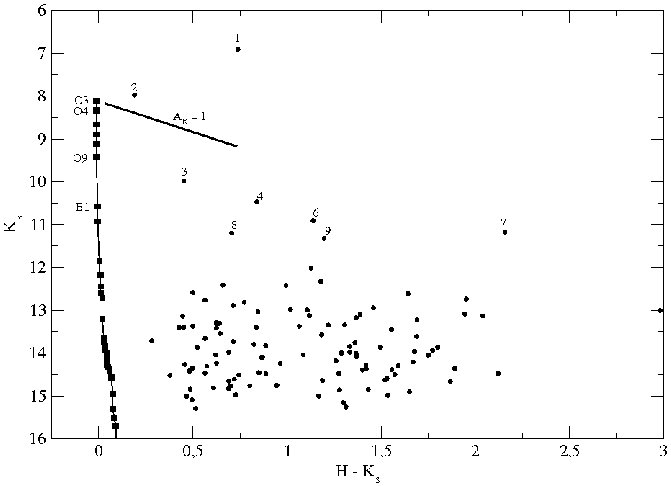}
\end{minipage}
\caption{{ H\,{\sc{ii}} region G322.2+0.6 (RCW92). Left: Color-color diagram (C-C). Right: Color-magnitude diagram (C-M). A very obscured H\,{\sc{ii}} region with the brightest objects apparently in the foreground and a 'hidden cluster' seen in the {\it Spitzer} image. Due to the small number of detected objects, the distance analysis is inconclusive.}}
\label{fig:G322-CMD}
\end{figure*}

\clearpage

\begin{figure*}
\begin{minipage}[b]{0.47\linewidth}
\includegraphics[height=08cm,width=09cm]{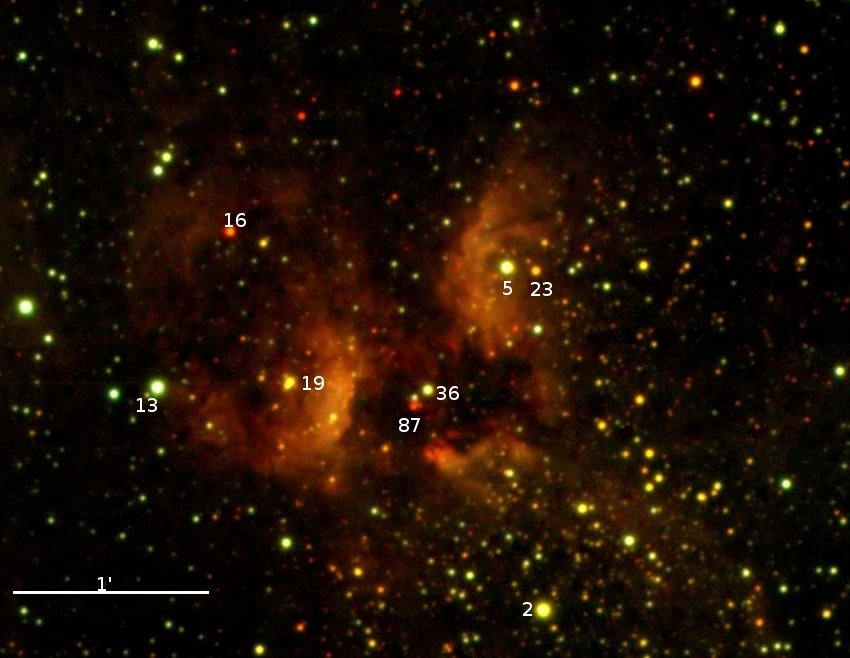}
\end{minipage} \hfill
\begin{minipage}[b]{0.47\linewidth}
\includegraphics[height=08cm,width=09cm]{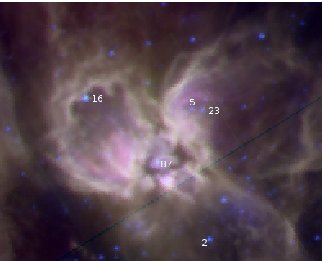}
\end{minipage}
\caption{{ Color images of G327.3-0.5 (RCW97). Left: $JHK_{s}$ color image. Right: IRAC-{\it Spitzer} image. In both images, the size is $\approx$ 3.5 arcmin on a side.}}
\label{fig:RCW97-color}
\end{figure*}

\begin{figure*}
\begin{minipage}[b]{0.48\linewidth}
\includegraphics[height=09.5cm,width=\linewidth]{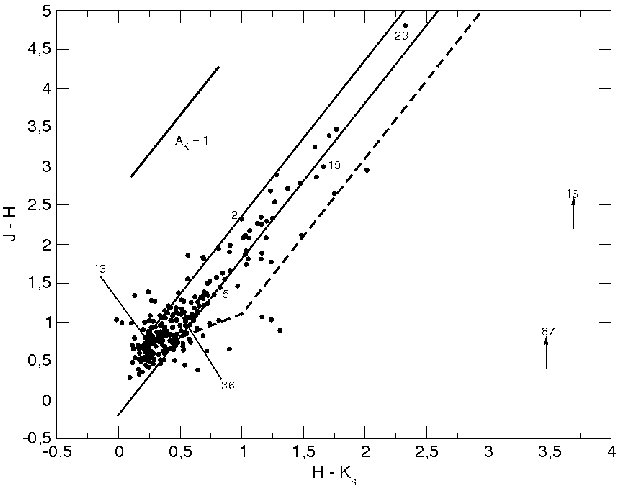}
\end{minipage} \hfill
\begin{minipage}[b]{0.49\linewidth}
\includegraphics[height=09.5cm,width=\linewidth]{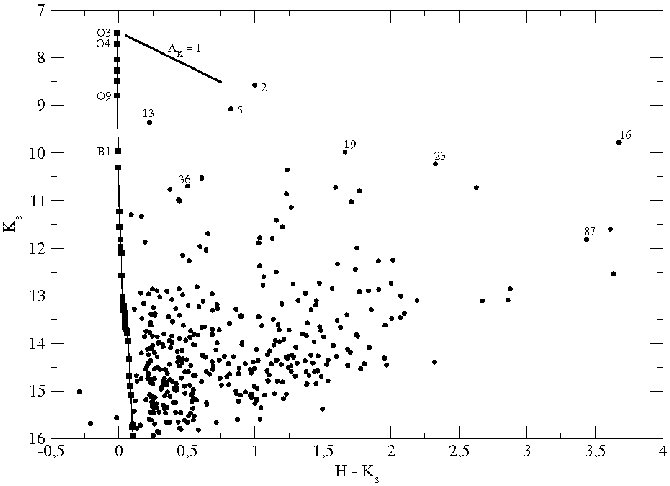}
\end{minipage}
\caption{{ H\,{\sc{ii}} region G327.3-0.5 (RCW97). Left: Color-color diagram (C-C), most of the objects seem to be foreground.  Right: Color-magnitude diagrama (C-M). The limiting magnitude is $J$ = 16.0 mag. There are few objects associated with this region, but their photometry seems to be in agreement with the kinematic distance. }}
\label{fig:RCW97-CMD}
\end{figure*}


\begin{figure*}
\begin{minipage}[b]{0.47\linewidth}
\includegraphics[height=10cm,width=\linewidth]{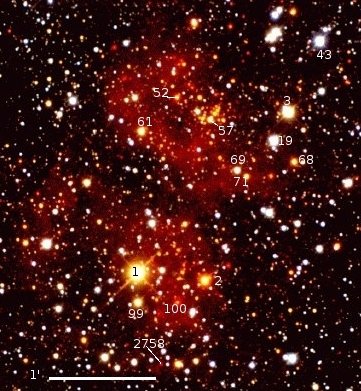}
\end{minipage} \hfill
\begin{minipage}[b]{0.47\linewidth}
\includegraphics[height=10cm,width=\linewidth]{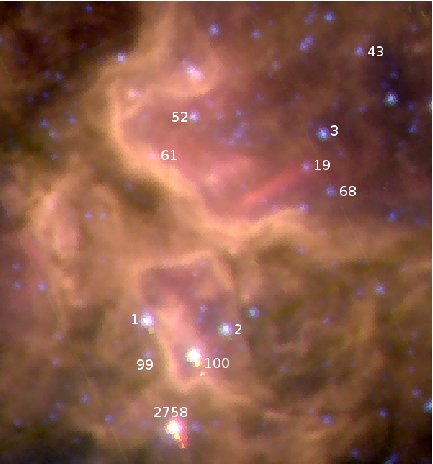}
\end{minipage}
\caption{{ Color images of G331.5-0.1. Left: $JHK_{s}$ color image. Right: IRAC-{\it Spitzer} image. In both images, the FOV is $\approx$ 3.0 $\times$ 4.0 arcmin.}}
\label{fig:G331-5-color}
\end{figure*}

\begin{figure*}
\begin{minipage}[b]{0.48\linewidth}
\includegraphics[height=10cm,width=\linewidth]{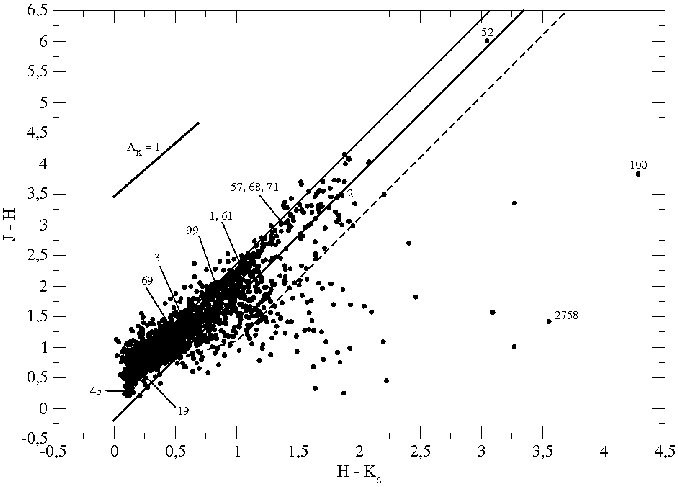}
\end{minipage} \hfill
\begin{minipage}[b]{0.49\linewidth}
\includegraphics[height=10cm,width=\linewidth]{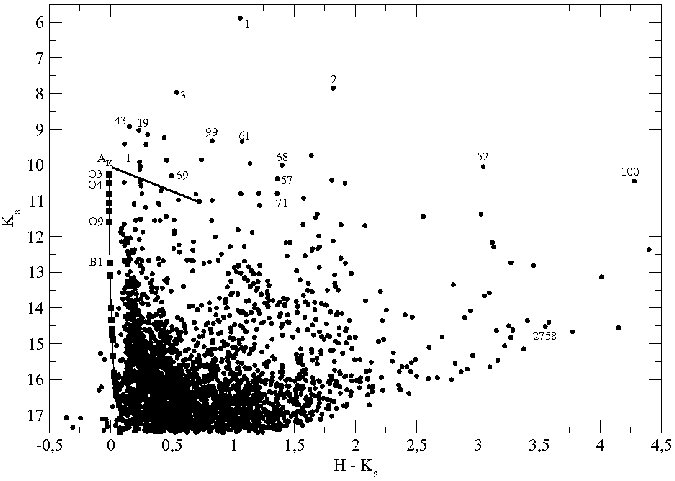}
\end{minipage}
\caption{{H\,{\sc{ii}} region G331.5-0.1. Left: Color-color diagram (C-C). Right: Color-magnitude diagram (C-M). Objects \#1, \#2 and \#3 are saturated in our images; we have used 2MASS photometric data for these objects. The adopted distance for the main sequence location is 10.8 kpc \citep{Russeil03} which appears too far relative to our photometry.}}
\label{fig:G331-5-CMD}
\end{figure*}


\begin{figure*}
\begin{minipage}[b]{0.47\linewidth}
\includegraphics[height=08cm,width=09cm]{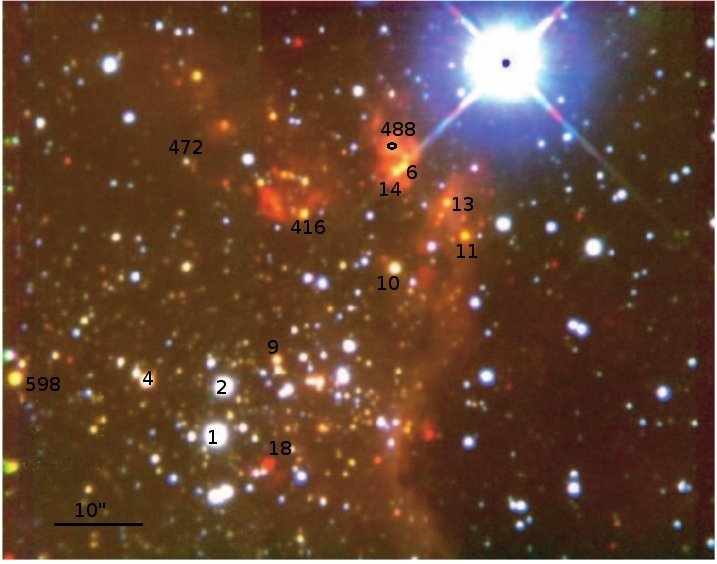}
\end{minipage} \hfill
\begin{minipage}[b]{0.47\linewidth}
\includegraphics[height=08cm,width=09cm]{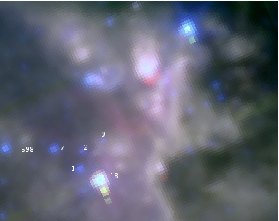}
\end{minipage}
\caption{{ Color images of G333.1-0.4. Left: $JHK_{s}$ color image, reproduced from Figuer\^edo et al. (2005). Right: IRAC-{\it Spitzer} image. In both images, the size is $\approx$ 1.5 arcmin on a side.}}
\label{fig:G333-1-color}
\end{figure*}

\begin{figure*}
\begin{minipage}[b]{0.48\linewidth}
\includegraphics[height=09.5cm,width=\linewidth]{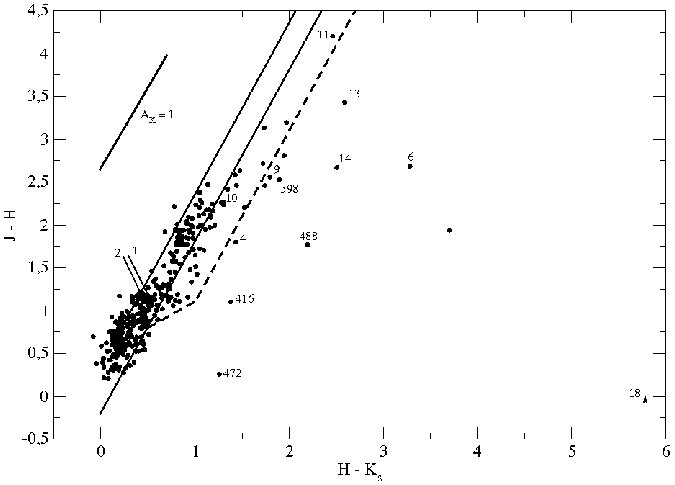}
\end{minipage} \hfill
\begin{minipage}[b]{0.49\linewidth}
\includegraphics[height=09.5cm,width=\linewidth]{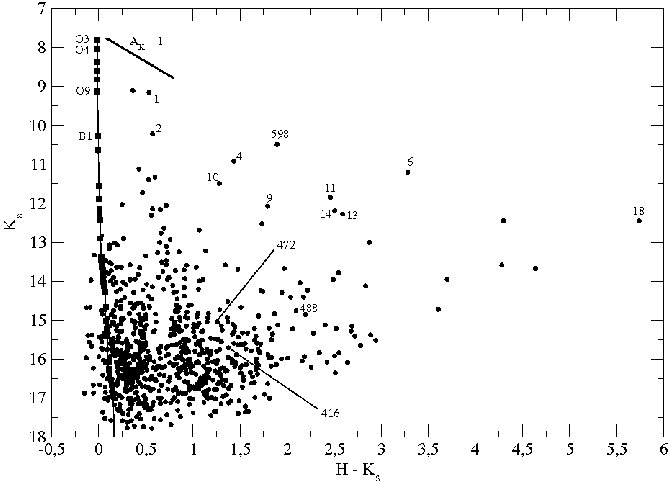}
\end{minipage}
\caption{{H\,{\sc{ii}} region G333.1-0.4. Left: Color-color diagram (C-C). Right: Color-magnitude diagram (C-M). Figuer\^edo et al. (2005) derived a spectrophotometric distance of 2.6 kpc to this region using $K$-band spectra of the stars \#1 and \#2, while the kinematic distance is 3.5 kpc. The limiting magnitude in the $J$-band is 18.0 mag.}}
\label{fig:G333-1-CMD}
\end{figure*}


\clearpage
\begin{figure*}
\begin{minipage}[b]{0.47\linewidth}
\includegraphics[height=12cm,width=\linewidth]{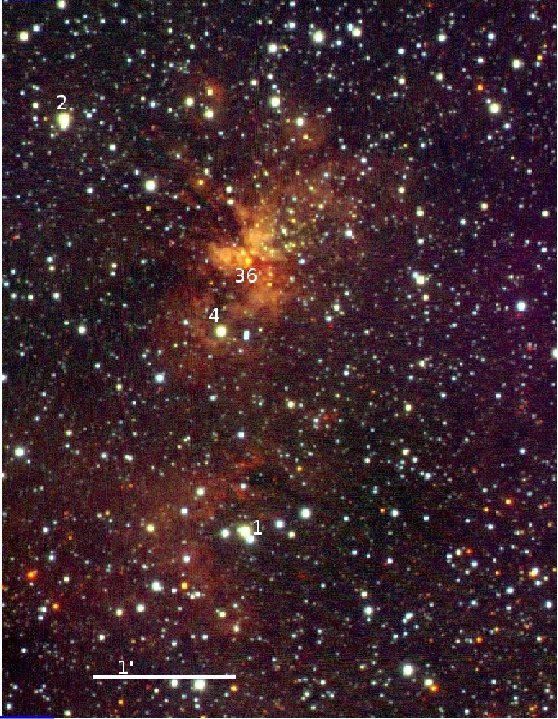}
\end{minipage} \hfill
\begin{minipage}[b]{0.47\linewidth}
\includegraphics[height=12cm,width=\linewidth]{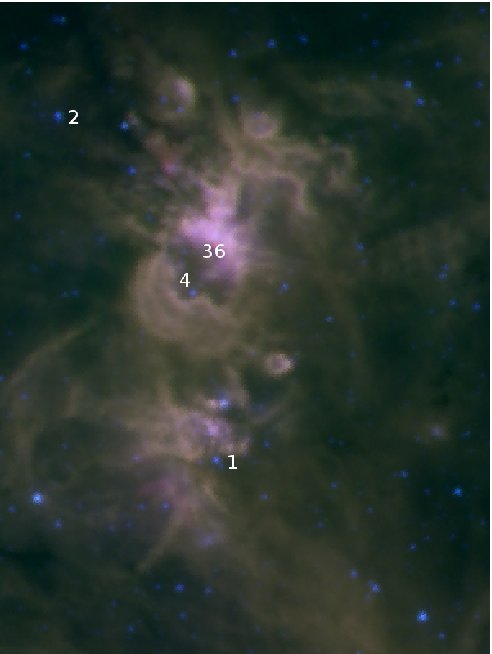}
\end{minipage}
\caption{{ Color images of G333.3-0.4. Left: $JHK_{s}$ color image. Right: IRAC-{\it Spitzer} image. In both images, the FOV is $\approx$ 3.7 $\times$ 5 arcmin.}}
\label{fig:G333-color}
\end{figure*}

\begin{figure*}
\begin{minipage}[b]{0.48\linewidth}
\includegraphics[height=08cm,width=\linewidth]{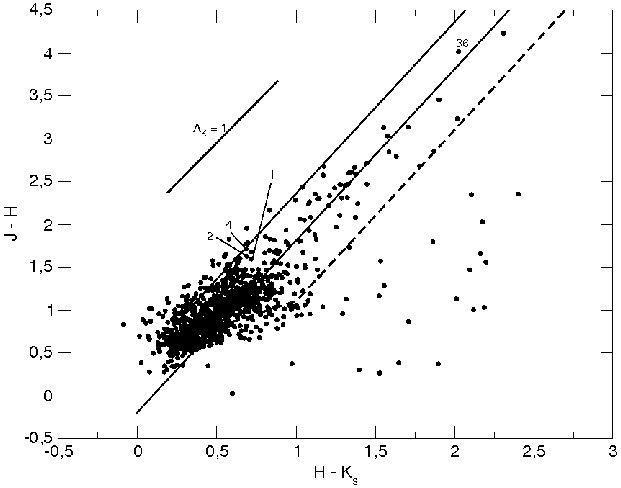}
\end{minipage} \hfill
\begin{minipage}[b]{0.49\linewidth}
\includegraphics[height=08cm,width=\linewidth]{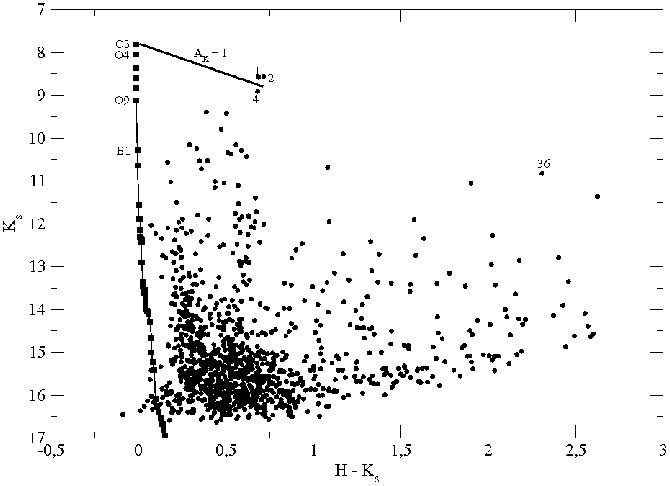}
\end{minipage}
\caption{{H\,{\sc{ii}} region G333.3-0.4. Left: Color-color diagram (C-C). Right: Color-magnitude diagram (C-M). The associated cluster appears to still be heavily buried. The absence of a cluster makes the analysis of the distance inconclusive.}}
\label{fig:G333-CMD}
\end{figure*}


\begin{figure*}
\begin{minipage}[b]{0.47\linewidth}
\includegraphics[height=09cm,width=09cm]{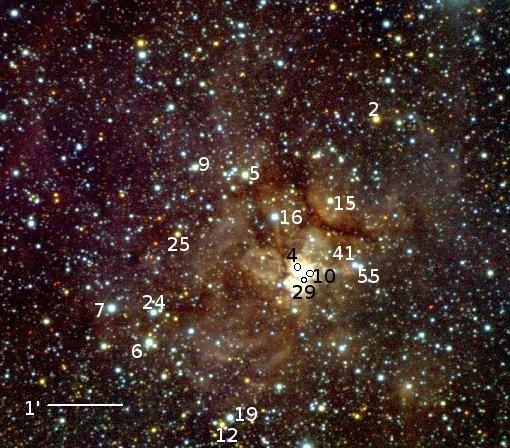}
\end{minipage} \hfill
\begin{minipage}[b]{0.47\linewidth}
\includegraphics[height=09cm,width=09cm]{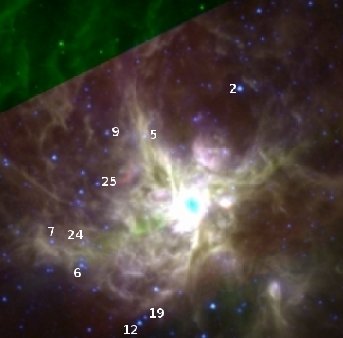}
\end{minipage}
\caption{{ Color images of G333.6-0.2. Left: $JHK_{s}$ color image. Right: IRAC-{\it Spitzer} image. In both images, the size is $\approx$ 5.0 arcmin on a side.}}
\label{fig:G333-2-color}
\end{figure*}

\begin{figure*}
\begin{minipage}[b]{0.48\linewidth}
\includegraphics[height=10cm,width=\linewidth]{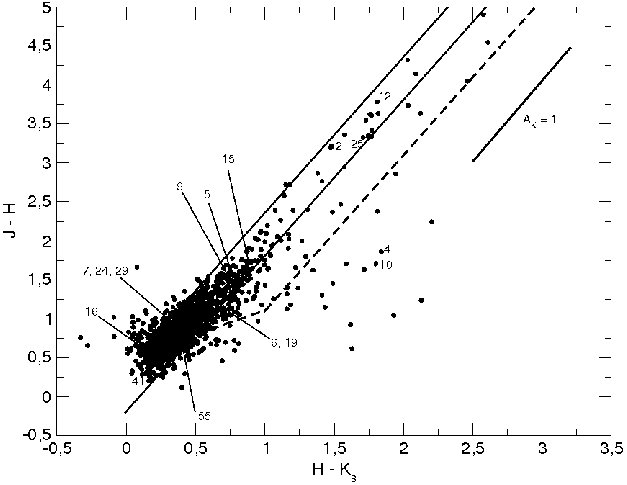}
\end{minipage} \hfill
\begin{minipage}[b]{0.49\linewidth}
\includegraphics[height=10cm,width=\linewidth]{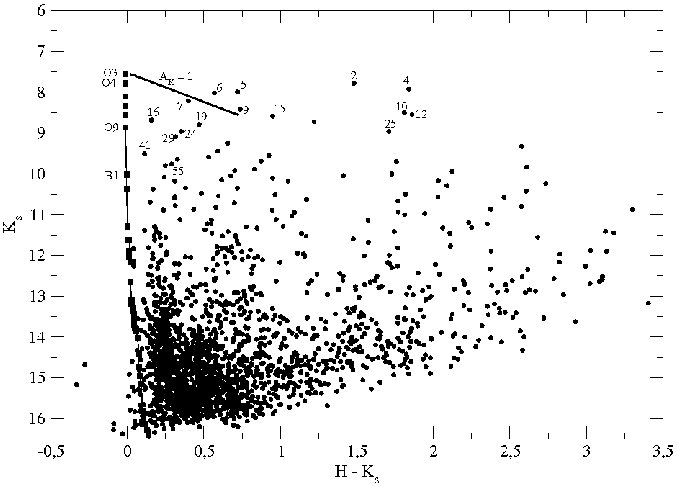}
\end{minipage}
\caption{{H\,{\sc{ii}} region G333.6-0.2. Left: Color-color diagram (C-C). Right: Color-magnitude diagrama (C-M). The detection of few stars associated with this H\,{\sc{ii}} region makes the analysis of the distance inconclusive.}}
\label{fig:G333-2-CMD}
\end{figure*}


\begin{figure*}
\begin{minipage}[b]{0.47\linewidth}
\includegraphics[height=08.5cm,width=09cm]{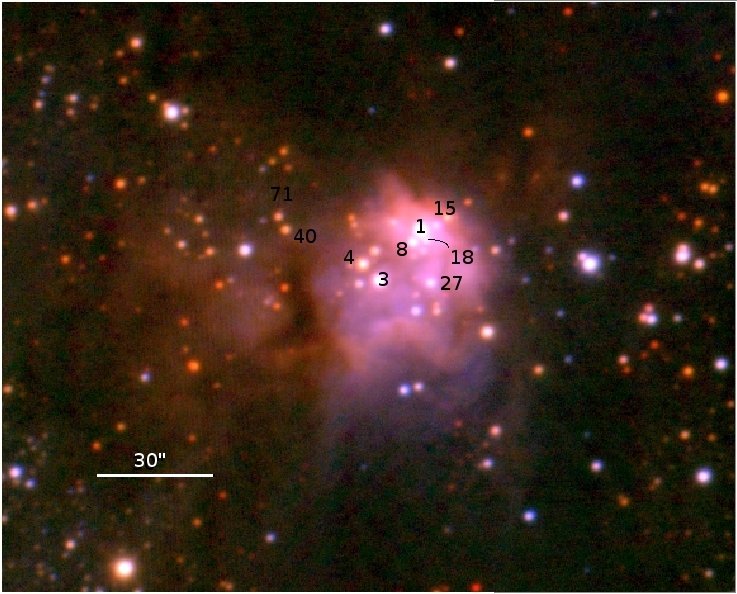}
\end{minipage} \hfill
\begin{minipage}[b]{0.47\linewidth}
\includegraphics[height=08.5cm,width=09cm]{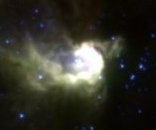}
\end{minipage}
\caption{{ Color images of G336.5-1.5 (RCW108). Left: $JHK_{s}$ color image. Right: IRAC-{\it Spitzer} image. In both images, the size is $\approx$ 3.0 arcmin on a side.}}
\label{fig:G336-color}
\end{figure*}

\begin{figure*}
\begin{minipage}[b]{0.48\linewidth}
\includegraphics[height=09cm,width=\linewidth]{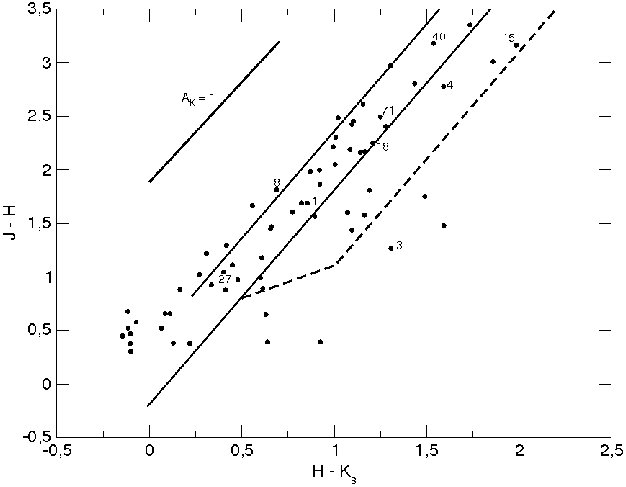}
\end{minipage} \hfill
\begin{minipage}[b]{0.49\linewidth}
\includegraphics[height=09cm,width=\linewidth]{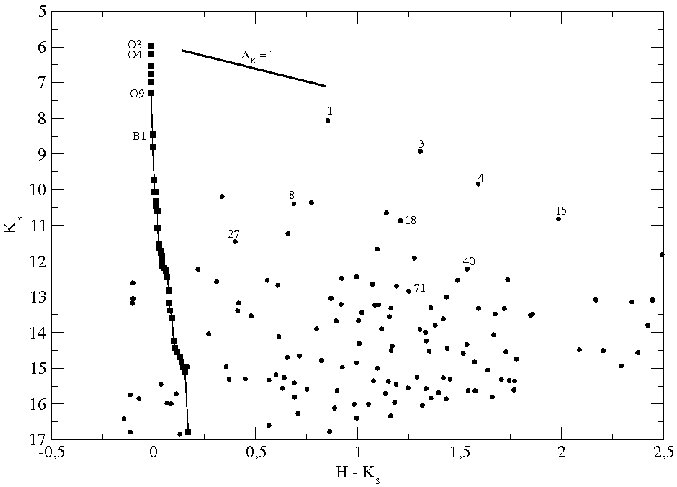}
\end{minipage}
\caption{{ H\,{\sc{ii}} region G336.5-1.5 (RCW108). Left: Color-color diagram (C-C). Right: Color-magnitude diagram (C-M). Due to the strong nebulosity, only a few stars were detected in this region but their photometry seems to be in agreement with the kinematic distance.}}
\label{fig:G336-CMD}
\end{figure*}


\begin{figure*}
\begin{minipage}[b]{0.47\linewidth}
\includegraphics[height=09cm,width=09cm]{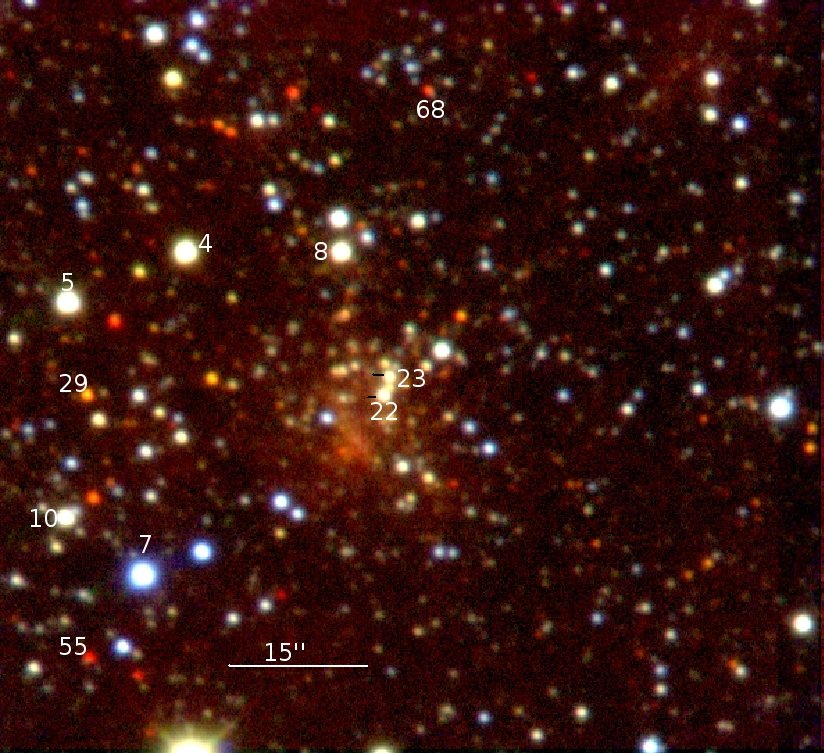}
\end{minipage} \hfill
\begin{minipage}[b]{0.47\linewidth}
\includegraphics[height=09cm,width=09cm]{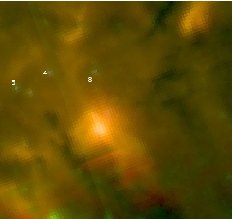}
\end{minipage}
\caption{{ Color images of G336.8-0.0. Left: $JHK_{s}$ color image. Right: IRAC-{\it Spitzer} image. In both images, the size is $\approx$ 1.8 arcmin on a side.}}
\label{fig:G336-2-color}
\end{figure*}

\begin{figure*}
\begin{minipage}[b]{0.48\linewidth}
\includegraphics[height=10cm,width=\linewidth]{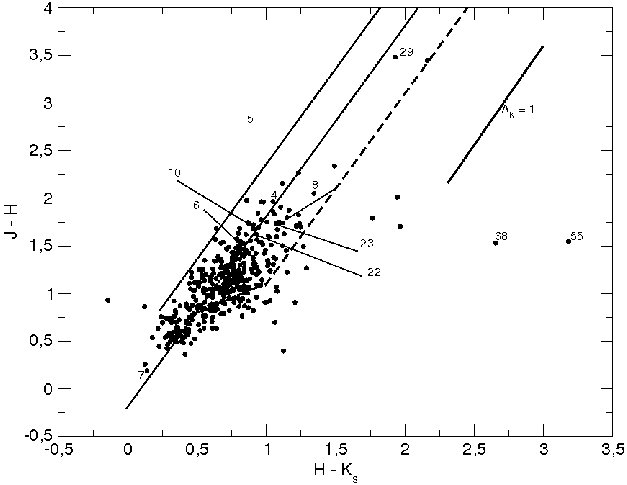}
\end{minipage} \hfill
\begin{minipage}[b]{0.49\linewidth}
\includegraphics[height=10cm,width=\linewidth]{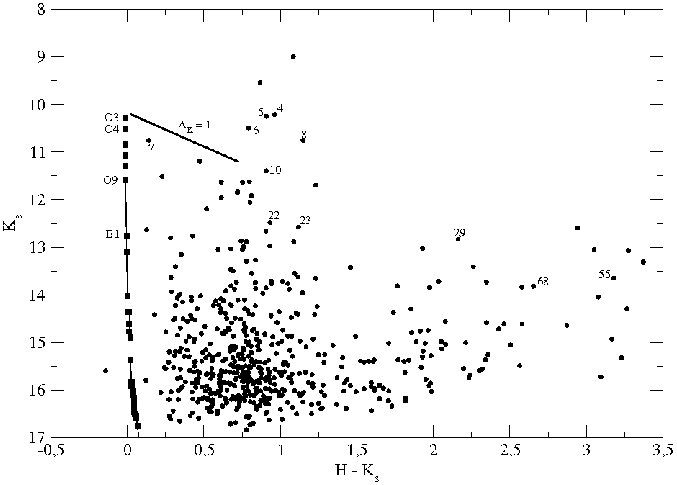}
\end{minipage}
\caption{{ H\,{\sc{ii}} region G336.8-0.0. Left: Color-color diagram (C-C). Right: Color-magnitude diagram (C-M). The presence of a star cluster is not clear, but if the objects \#22 and \#23 are associated with the nebulosity, the kinematic distance is in agreement with the photometry.}}
\label{fig:G336-2-CMD}
\end{figure*}


\begin{figure*}
\begin{minipage}[b]{0.47\linewidth}
\includegraphics[height=09cm,width=09cm]{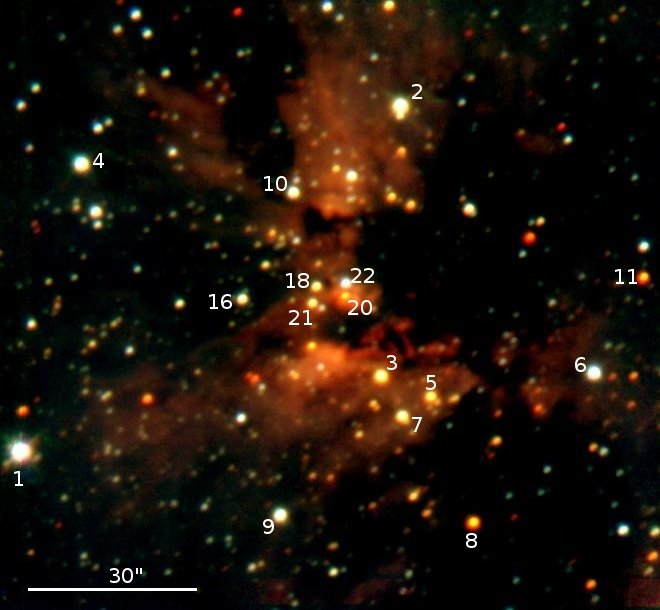}
\end{minipage} \hfill
\begin{minipage}[b]{0.47\linewidth}
\includegraphics[height=09cm,width=09cm]{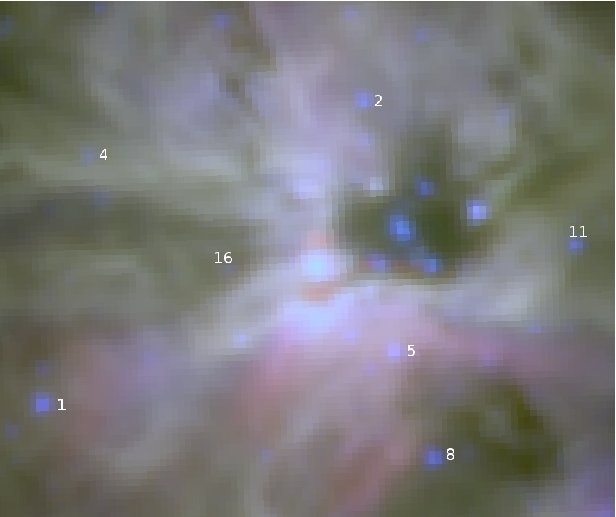}
\end{minipage}
\caption{{ Color images of G348.7-1.0 (RCW122). Left: $JHK_{s}$ color image. Right: IRAC-{\it Spitzer} image. In both images, the size is $\approx$ 1.8 arcmin on a side. Highly obscured stars behind the dark cloud appear in the {\it Spitzer} image.}}
\label{fig:RCW122-color}
\end{figure*}

\begin{figure*}
\begin{minipage}[b]{0.48\linewidth}
\includegraphics[height=10cm,width=\linewidth]{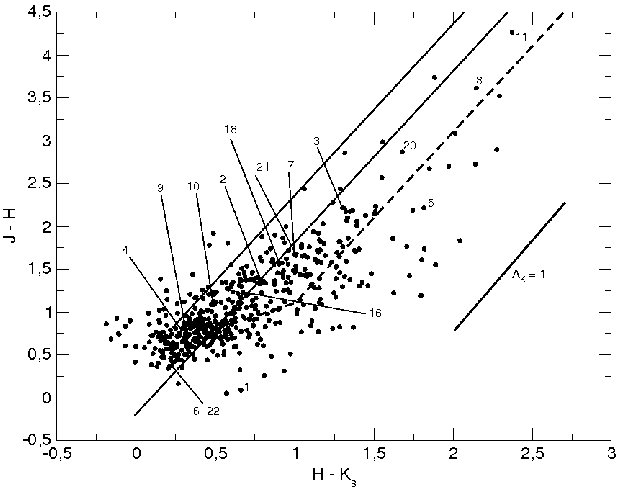}
\end{minipage} \hfill
\begin{minipage}[b]{0.49\linewidth}
\includegraphics[height=10cm,width=\linewidth]{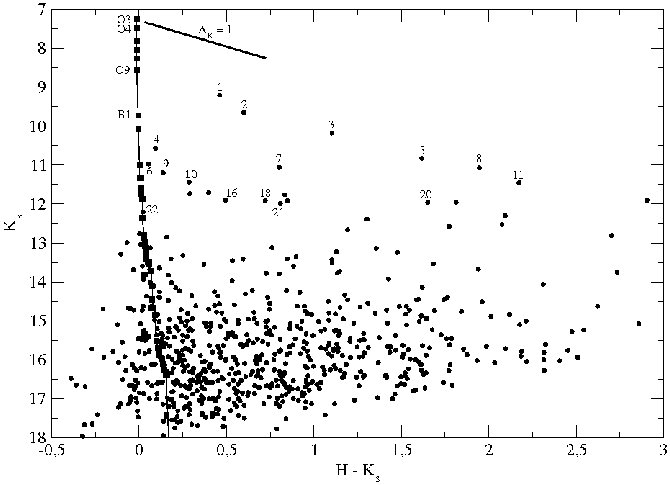}
\end{minipage}
\caption{{ H\,{\sc{ii}} region G348.7-1.0 (RCW122). Left: Color-color diagram (C-C). Right: Color-magnitude diagram (C-M). Most of the stars are likely foreground. But all the objects, including the candidates (for intance, \#18, \#20 and \#22) are fainter than the tip of the main sequence line, which indicates the kinematic distance may be in agreement with the photometry.}}
\label{fig:RCW122-CMD}
\end{figure*}


\begin{figure*}
\begin{minipage}[b]{0.47\linewidth}
\includegraphics[height=08cm,width=09cm]{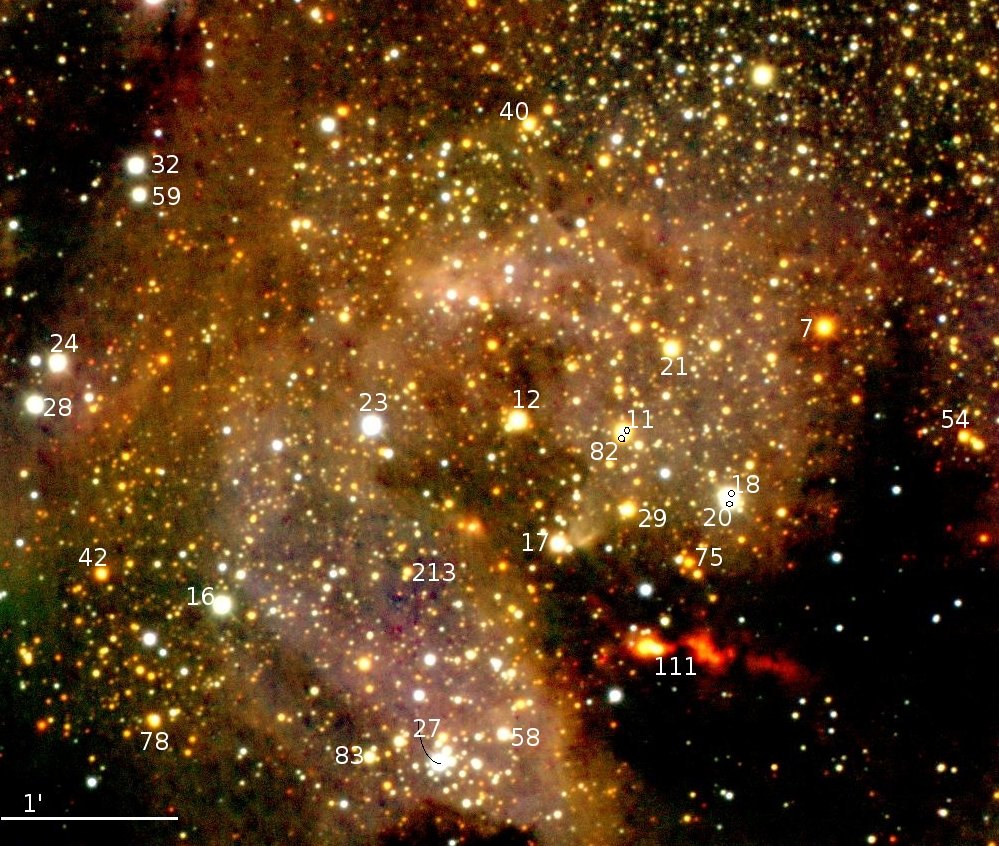}
\end{minipage} \hfill
\begin{minipage}[b]{0.47\linewidth}
\includegraphics[height=08cm,width=09cm]{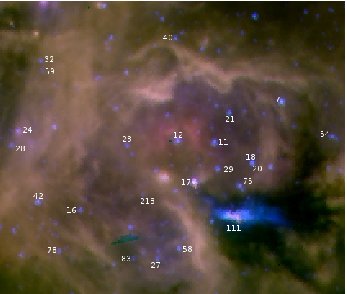}
\end{minipage}
\caption{{ Color images of G351.2+0.7. Left: $JHK_{s}$ image. Right: IRAC-{\it Spitzer} image. In both images, the size is $\approx$ 5.0 arcmin on a side.}}
\label{fig:G351-color}
\end{figure*}

\begin{figure*}
\begin{minipage}[b]{0.48\linewidth}
\includegraphics[height=09cm,width=\linewidth]{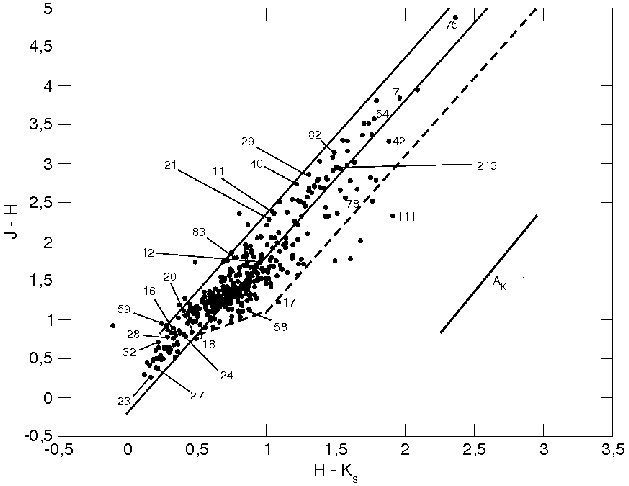}
\end{minipage} \hfill
\begin{minipage}[b]{0.49\linewidth}
\includegraphics[height=09cm,width=\linewidth]{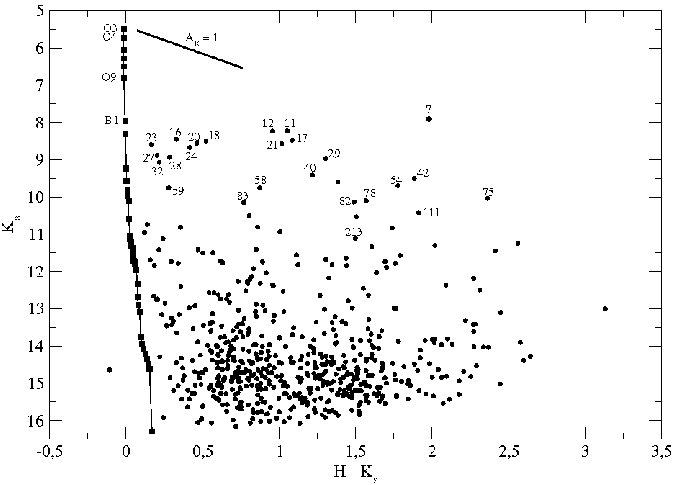}
\end{minipage}
\caption{{ H\,{\sc{ii}} region G351.2+0.7. Left: Color-color diagram (C-C). Right: Color-magnitude diagram (C-M). The kinematic distance seems to be smaller than what is predicted by the photometry. The main sequence line position requires few ionizing sources, which is in disagreement with the strong nebulosity in the images. The cluster likely is further away than the kinematic distance of 1.2 kpc.}}
\label{fig:G351-CMD}
\end{figure*}


\begin{figure*}
\begin{minipage}[b]{0.47\linewidth}
\includegraphics[height=08cm,width=09cm]{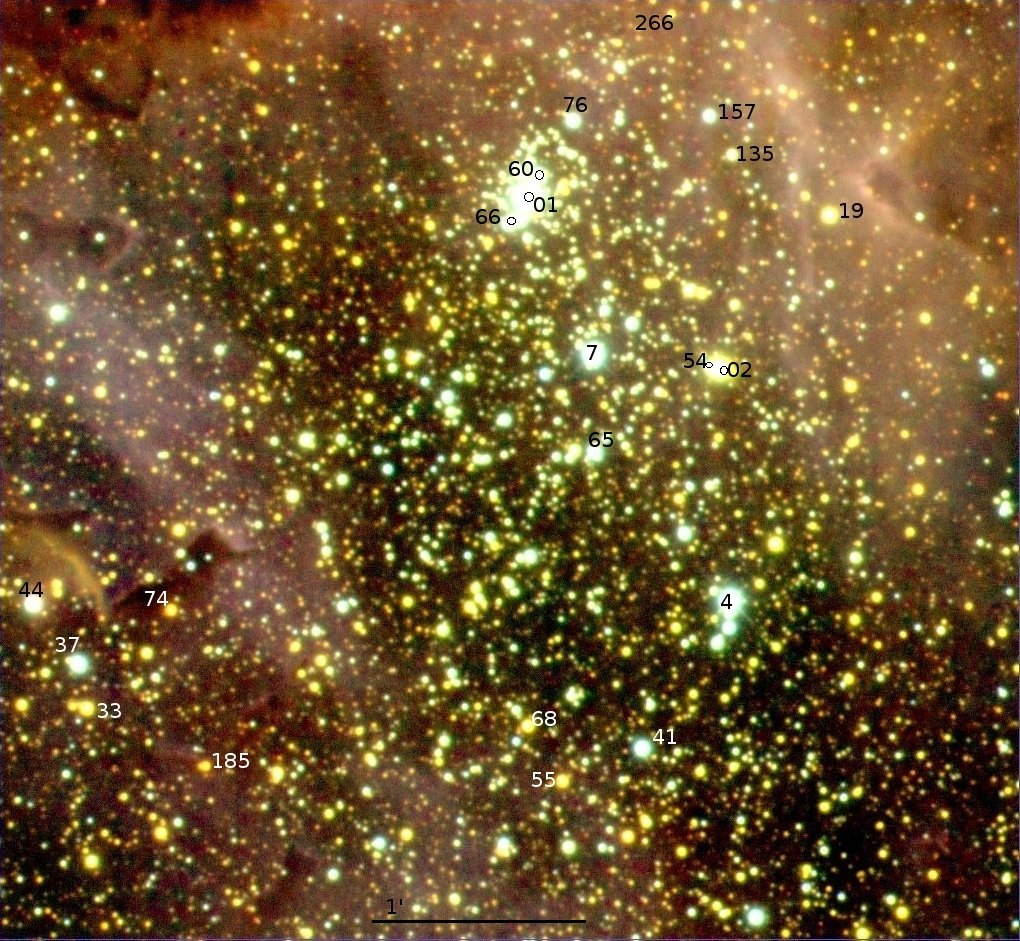}
\end{minipage} \hfill
\begin{minipage}[b]{0.47\linewidth}
\includegraphics[height=08cm,width=09cm]{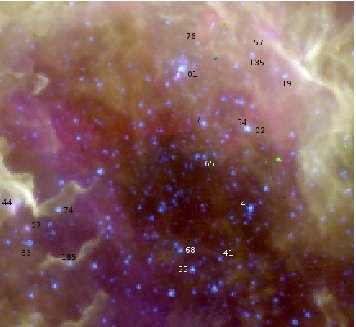}
\end{minipage}
\caption{{ Color images of G353.2+0.6 (RCW131). Left: $JHK_{s}$ color image. Right: IRAC-{\it Spitzer} image. In both images, the size is $\approx$ 4.0 arcmin on a side.}}
\label{fig:G353-color}
\end{figure*}

\begin{figure*}
\begin{minipage}[b]{0.48\linewidth}
\includegraphics[height=09cm,width=\linewidth]{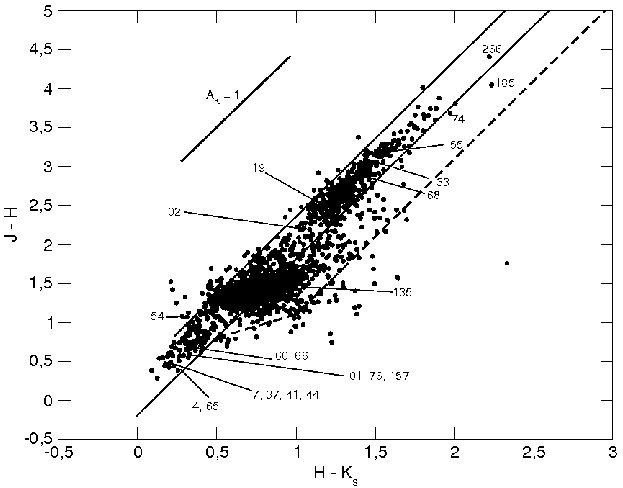}
\end{minipage} \hfill
\begin{minipage}[b]{0.49\linewidth}
\includegraphics[height=09cm,width=\linewidth]{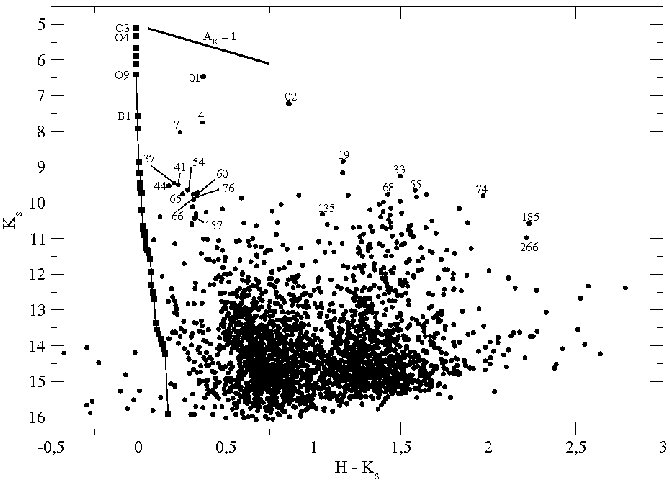}
\end{minipage}
\caption{{ H\,{\sc{ii}} region G353.2+0.6 (RCW131). Left: Color-color diagram (C-C). Right: Color-magnitude diagram (C-M). Both diagrams indicate, at least, three groups of objects. The first is a group of objects almost not affected by interstellar reddening. The second is grouped at $H - K_{s}$ $\approx$ 0.75. The last one is grouped at $H - K_{s}$ $\approx$ 1.5. The main sequence position requires few ionizing sources, which is in disagreement with the strong nebulosity in the images. The cluster likely is further away than the kinematic distance of 1.0 kpc.}}
\label{fig:G353-CMD}
\end{figure*}

\label{lastpage}

\end{document}